\documentstyle[12pt,subfigure,epsfig,amssymb]{article}
\begin{document}
%
\newcommand{\bq}{\begin{equation}}
\newcommand{\eq}{\end{equation}}
\newcommand{\ba}{\begin{eqnarray}}
\newcommand{\ea}{\end{eqnarray}}
%
\newcommand{\mco}{\multicolumn}
\newcommand{\ds}      {\mbox{$ {\mathrm d} \sigma                          $}}
\newcommand{\cotw}     {\mbox{$ \cot\theta_{\mathrm W}                      $}}
\newcommand{\tanw}     {\mbox{$ \tan\theta_{\mathrm W}                      $}}
\newcommand{\sinw}     {\mbox{$ \sin\theta_{\mathrm W}                      $}}
\newcommand{\cosw}     {\mbox{$ \cos\theta_{\mathrm W}                      $}}
\newcommand{\ee}      {\mbox{$ e^+    e^-                                   $}}
\newcommand{\W}       {\mbox{$ W                                            $}}
\newcommand{\Ws}      {\mbox{$W$s                                            }}
\newcommand{\Wp}      {\mbox{$ W^+                                          $}}
\newcommand{\Wm}      {\mbox{$ W^-                                         $}}
\newcommand{\WW}      {\mbox{$ W W                                         $}}
\newcommand{\WpWm}      {\mbox{$ W^+ W^-                                    $}}
\newcommand{\eeWW}    {\mbox{$ \ee \rightarrow \WpWm                         $}}
\newcommand{\eeffff}    {\mbox{$ \ee \rightarrow f_1 f_2 f_3 f_4           $}}
\newcommand{\jjlv}    {\mbox{$ j j \ell \nu                                $}}
\newcommand{\jjenu}   {\mbox{$ j j e \nu                                   $}}
\newcommand{\jjmunu}  {\mbox{$ j j \mu \nu                                 $}}
\newcommand{\jjtaunu} {\mbox{$ j j \tau \nu                                $}}
\newcommand{\ZZ}      {\mbox{$ Z Z                                         $}}
\newcommand{\jjjj}    {\mbox{$ j j j j                                     $}}
\newcommand{\lvlv}    {\mbox{$ \ell \nu \ell \nu                           $}}
\newcommand{\dz}      {\mbox{$ \delta g_Z                                  $}}
\newcommand{\kg}       {\mbox{$ \kappa_{\gamma}                             $}}
\newcommand{\kz}       {\mbox{$ \kappa_Z                                    $}}
\newcommand{\lm}       {\mbox{$ \lambda_{\gamma}                            $}}
\newcommand{\lz}       {\mbox{$ \lambda_Z                                   $}}
\newcommand{\xg}       {\mbox{$ x_{\gamma}                                  $}}
\newcommand{\xz}       {\mbox{$ x_Z                                         $}}
\newcommand{\yg}       {\mbox{$ y_{\gamma}                                  $}}
\newcommand{\yz}       {\mbox{$ y_Z                                         $}}
\newcommand{\aWphi}   {\mbox{$\alpha_{W\phi}                               $}}
\newcommand{\aBphi}   {\mbox{$\alpha_{B\phi}                               $}}
\newcommand{\aW}      {\mbox{$\alpha_{W}                                   $}}
\newcommand{\ctw}     {\mbox{$ \cos\theta                                  $}}
\newcommand{\thw}     {\mbox{$ \theta                                      $}}
\newcommand{\thetaa}  {\mbox{$ \theta_1                                    $}}
\newcommand{\phia}    {\mbox{$ \phi_1                                      $}}
\newcommand{\cta}     {\mbox{$ \cos\theta_1                                $}}
\newcommand{\thetab}  {\mbox{$ \theta_2                                    $}}
\newcommand{\phib}    {\mbox{$ \phi_2                                      $}}
\newcommand{\ctb}     {\mbox{$ \cos\theta_2                                $}}
\newcommand{\ctl}     {\mbox{$ \cos\theta_l                                $}}
\newcommand{\phil}    {\mbox{$ \phi_l                                      $}}
\newcommand{\ctj}     {\mbox{$ \cos\theta_j                                $}}
\newcommand{\phij}    {\mbox{$ \phi_j                                      $}}
\newcommand{\ctja}     {\mbox{$ \cos\theta_{j_1}                           $}}
\newcommand{\phija}    {\mbox{$ \phi_{j_1}                                 $}}
\newcommand{\ctjb}     {\mbox{$ \cos\theta_{j_2}                           $}}
\newcommand{\phijb}    {\mbox{$ \phi_{j_2}                                 $}}
\newcommand{\fold}     {\mbox{$ _{\mathrm{folded}}                          $}}
%
%
\def\ap#1#2#3   {{\em Ann. Phys. (NY)} {\bf#1} (#2) #3}
\def\apj#1#2#3  {{\em Astrophys. J.} {\bf#1} (#2) #3}
\def\apjl#1#2#3 {{\em Astrophys. J. Lett.} {\bf#1} (#2) #3}
\def\app#1#2#3  {{\em Acta. Phys. Pol.} {\bf#1} (#2) #3}
\def\ar#1#2#3   {{\em Ann. Rev. Nucl. Part. Sci.} {\bf#1} (#2) #3}
\def\cpc#1#2#3  {{\em Computer Phys. Comm.} {\bf#1} (#2) #3}
\def\err#1#2#3  {{\it Erratum} {\bf#1} (#2) #3}
\def\ib#1#2#3   {{\it ibid.} {\bf#1} (#2) #3}
\def\jmp#1#2#3  {{\em J. Math. Phys.} {\bf#1} (#2) #3}
\def\ijmp#1#2#3 {{\em Int. J. Mod. Phys.} {\bf#1} (#2) #3}
\def\jetp#1#2#3 {{\em JETP Lett.} {\bf#1} (#2) #3}
\def\jpg#1#2#3  {{\em J. Phys. G.} {\bf#1} (#2) #3}
\def\mpl#1#2#3  {{\em Mod. Phys. Lett.} {\bf#1} (#2) #3}
\def\nat#1#2#3  {{\em Nature (London)} {\bf#1} (#2) #3}
\def\nc#1#2#3   {{\em Nuovo Cim.} {\bf#1} (#2) #3}
\def\nim#1#2#3  {{\em Nucl. Instr. Meth.} {\bf#1} (#2) #3}
\def\np#1#2#3   {{\em Nucl. Phys.} {\bf#1} (#2) #3}
\def\pcps#1#2#3 {{\em Proc. Cam. Phil. Soc.} {\bf#1} (#2) #3}
\def\pr#1#2#3   {{\em Phys. Rev.} {\bf#1} (#2) #3}
\def\pl#1#2#3   {{\em Phys. Lett.} {\bf#1} (#2) #3}
\def\prep#1#2#3 {{\em Phys. Rep.} {\bf#1} (#2) #3}
\def\pl#1#2#3   {{\em Phys. Lett.} {\bf#1} (#2) #3}
\def\prep#1#2#3 {{\em Phys. Rep.} {\bf#1} (#2) #3}
\def\prev#1#2#3 {{\em Phys. Rev.} {\bf#1} (#2) #3}
\def\prl#1#2#3  {{\em Phys. Rev. Lett.} {\bf#1} (#2) #3}
\def\prs#1#2#3  {{\em Proc. Roy. Soc.} {\bf#1} (#2) #3}
\def\ptp#1#2#3  {{\em Prog. Th. Phys.} {\bf#1} (#2) #3}
\def\ps#1#2#3   {{\em Physica Scripta} {\bf#1} (#2) #3}
\def\rmp#1#2#3  {{\em Rev. Mod. Phys.} {\bf#1} (#2) #3}
\def\rpp#1#2#3  {{\em Rep. Prog. Phys.} {\bf#1} (#2) #3}
\def\sjnp#1#2#3 {{\em Sov. J. Nucl. Phys.} {\bf#1} (#2) #3}
\def\spj#1#2#3  {{\em Sov. Phys. JEPT} {\bf#1} (#2) #3}
\def\spu#1#2#3  {{\em Sov. Phys.-Usp.} {\bf#1} (#2) #3}
\def\zp#1#2#3   {{\em Zeit. Phys.} {\bf#1} (#2) #3}
%
 
\def\ie{{\it i.e.\/}}
\def\eg{{\it e.g.\/}}
\def\etc{{\it etc.\/}}
\def\etal{{\it et.al.\/}}
%
 
\def\df{\mathrel{:=}}
\def\fd{\mathrel{=:}}
 

\def\hf{{1\over 2}}
\def\nth#1{{1\over #1}}
\def\sfrac#1#2{{\scriptstyle {#1} \over {#2}}}  
\def\stack#1#2{\buildrel{#1}\over{#2}}
\def\dd#1#2{{{d #1} \over {d #2}}}  
\def\ppartial#1#2{{{\partial #1} \over {\partial #2}}}
\def\grad{\nabla}
\def\Bsl{\hbox{/\kern-.6700em$B$}} 
\def\Dsl{\hbox{/\kern-.6700em$D$}} 
\def\Wsl{\hbox{/\kern-.6700em$W$}} 

\def\roughly#1{\mathrel{\raise.3ex
    \hbox{$#1$\kern-.75em\lower1ex\hbox{$\sim$}}}}
\def\lsim{\roughly<}
\def\gsim{\roughly>}
 
\def\bv#1{{\bf #1}}
\def\scr#1{{\cal #1}}
\def\op#1{{\widehat #1}}
\def\twi{\widetilde}
\def\tw#1{\tilde{#1}}
\def\ol#1{\overline{#1}}
 
\def\L{ {\cal L }}
\def\O{ {\cal O }}
\def\sw{s_W}
\def\cw{c_W}
\def\swd{s^2_W}
\def\cwd{c^2_W}
\def\ed{e^2}
\def\mwd{M_W^2}
\def\mw{M_W}
\def\lw{\lambda_W}
\def\rd{\sqrt2}

\def\ep#1#2{(\epsilon_{#1}\epsilon_{#2})}
\def\dh#1{ {1\over D_H({#1})} }
\def\nh#1{  D_H({#1}) }
\def\co{\biggm[}
\def\cf{\biggm]}

\begin{center}
{\Large \bf Triple Gauge Boson Couplings } 
\end{center}
\begin{center}
{\it Conveners: } G. Gounaris, J.-L. Kneur and D. Zeppenfeld 
\end{center}
{\it Working group:} 
Z. Ajaltouni, A. Arhrib, G. Bella, F. Berends,
M. Bilenky, A. Blondel, \\ 
J. Busenitz, D. Charlton, D Choudhury,
P. Clarke, J. E. Conboy, M. Diehl, 
D. Fassouliotis, \\ 
J.-M. Fr\`ere, C. Georgiopoulos, M. Gibbs, 
M. Gr\"unewald, 
J. B. Hansen, C. Hartmann, \\ B. N. Jin, J. Jousset,  J. Kalinowski, 
M. Kocian, 
A. Lahanas, J. Layssac, E. Lieb, C. Markou,\\ C. Matteuzzi, P. M\"attig, 
J. M. Moreno, G. Moultaka, A. Nippe, J. Orloff, C. G. Papadopoulos, \\ 
J. Paschalis, C. Petridou, 
H. Phillips, 
F. Podlyski, M. Pohl, F. M. Renard, J.-M. Rossignol,\\ 
R. Rylko, 
R. L. Sekulin, A. van Sighem, E. Simopoulou,
A. Skillman, V. Spanos, A. Tonazzo, \\ 
M. Tytgat, S. Tzamarias,
C. Verzegnassi, N.~D.~Vlachos, E. Zevgolatakos \\ 

\vspace{1 truecm}  
%
%

{\bf 1. Introduction} 

{\bf 2. Parametrization, models and present bounds on TGC} 

{\bf 3. The W pair production process} 

{\bf 4. Statistical techniques for TGC determination } 

{\bf 5. Precision of TGC determination at LEP2: generator level studies} 

{\bf 6. Analysis of the \jjenu\ and \jjmunu\ final states} 

{\bf 7. Analysis of the \jjtaunu\ final state} 

{\bf 8. Analysis of the \jjjj\ final state } 

{\bf 9. Analysis of the \lvlv\ final state} 

{\bf 10. Other anomalous couplings and other channels}

{\bf 11. Conclusions}
\newpage 
\section{Introduction}\label{TGCintro}
Present measurements of the vector boson-fermion couplings at LEP
and SLC  accurately confirm the Standard Model (SM)
 predictions at the 0.1~--~1\%
level \cite{LEP1}, which may readily be considered to be  
evidence
for the gauge boson nature of the W and the Z. Nevertheless the most
crucial consequence of the $SU(2) \times U(1)$ gauge theory, namely the
specific form of the non-Abelian
self-couplings of the W, Z and photon, remains poorly measured to date. 
A direct and more accurate measurement of the trilinear 
self-couplings is possible via pair production of
electroweak bosons in present and future collider experiments
($W^+W^-$ at LEP2, $W\gamma$,
$WZ$ and $W^+W^-$ at hadron colliders).  \par

The major goal of such experiments at LEP2 will be to corroborate
the SM predictions. If sufficient accuracy is reached, such
measurements can be used to probe New Physics (NP) 
in the bosonic sector. This possibility raises a number of
other questions. What are the expected sizes of such effects
in definite models of NP? What type of specifically bosonic NP
contributions could have escaped detection in other experiments, 
e.g. at LEP1? 
Are there significant constraints from low-energy measurements?
Although we shall address these questions, the aim of this report is mostly
to elaborate on a detailed phenomenological strategy
for the direct measurement of the self-couplings at LEP2, which should 
allow their determination from data with the greatest possible accuracy. 
\section{Parametrization, Models and Present Bounds on TGC}\label{TGCeffparam}
We shall restrict ourselves to Triple Gauge boson Couplings
(TGC) in most of the report (possibilities to test quartic couplings
at LEP2 are extremely limited).
Analogous to the introduction
of arbitrary vector and axial-vector couplings $g_V$ and $g_A$ of the
gauge bosons to fermions, the measurements of the TGC can be made
quantitative by introducing a more general WWV vertex. We thus start
with a parametrization in terms of a purely phenomenological effective
Lagrangian~\footnote{We use $\epsilon^{0123}=1$.}
 \cite{GaGo79,Haetal87} [$V \equiv \gamma$ or $Z$]
\ba \label{LeffWWV}
i{\cal L}_{eff}^{WWV} & = & g_{WWV}\, \Bigl[ g_1^V V^\mu \left(
W^{-}_{\mu\nu}W^{+\nu}-W^{+}_{\mu\nu}W^{-\nu}\right) +
\kappa_V\,  W^{+}_{\mu}W^{-}_{\nu}V^{\mu\nu} + \\ \nonumber
& & {\lambda_V\over m_W^2}\,
V^{\mu\nu}W^{+\rho}_{\!\!\nu}W^-_{\rho\mu}
+ig_5^V\varepsilon_{\mu\nu\rho\sigma}\left(
(\partial^\rho W^{- \mu})W^{+\nu} -
W^{- \mu}(\partial^{\rho}W^{+\nu}) \right) V^{\sigma} \\ \nonumber
& & +ig_4^V W^{-}_{\mu}W^+_{\nu} (\partial^\mu V^\nu+\partial^\nu V^\mu)
-\frac{\tilde \kappa_V}{2} W^{-}_{\mu}W^+_{\nu}
\varepsilon^{\mu\nu\rho\sigma}
V_{\rho\sigma} -{\tilde \lambda_V\over {2 m_W^2}}\,
W^{-}_{\rho\mu}{W^{+\mu}}_{\nu}\varepsilon^{\nu\rho\alpha\beta}
V_{\alpha\beta}
\Bigr]\; ,
\ea
which gives  the most general
Lorentz invariant $WWV$ vertex  observable  in
processes where the vector bosons couple to effectively massless
fermions.
Here the overall couplings are defined as $g_{WW\gamma}=e$ and
$g_{WWZ}= e \cot\theta_W$, $W_{\mu\nu}=\partial_\mu W_\nu - \partial_\nu
W_\mu$, and $V_{\mu\nu}=\partial_\mu V_\nu - \partial_\nu V_\mu$.
For on-shell photons, $g_1^\gamma(q^2=0)=1$ and $g_5^\gamma(q^2=0)=0$
are fixed by electromagnetic gauge invariance
\footnote{For $q^2 \neq 0$ deviations due to form factor effects are
always possible, see section \ref{TGCzprime} below in this connection.}
Within
the SM, at tree level, the couplings are given by $g_1^Z = g_1^\gamma =
\kappa_Z = \kappa_\gamma = 1$, with all other couplings in (\ref{LeffWWV})
vanishing.
Terms with higher derivatives in (\ref{LeffWWV})
are equivalent to a dependence of the couplings
on the vector boson momenta and thus merely lead to a form-factor
behaviour of them. We also note that
$g_1^V$, $\kappa_V$ and $\lambda_V$ conserve $C$ and $P$
separately, while $g_5^V$ violates $C$ and $P$ but conserves $CP$.
Finally $g_4^V$, $\tilde\kappa_V$ and $\tilde\lambda_V$
parameterize a possible CP violation in the bosonic sector,
which will not be much studied in this report, as it may be
considered a more remote possibility for LEP2 studies
\footnote{Data on the neutron electric dipole moment allow observable
effects of e.g. $\tilde\kappa_\gamma$ at LEP2 only if fine tuning at the
$10^{-3}$ level is accepted~\cite{ndipole}.}. However, there exist
definite and simple means to test for such CP violation, see section 
\ref{TGCWpairpheno}. 
The $C$ and $P$
conserving terms in ${\cal L}_{eff}^{WW\gamma}$
correspond to the lowest order terms in a multipole expansion of the
$W-$photon interactions: the charge $Q_W$, the magnetic dipole
moment $\mu_W$ and the electric quadrupole moment $q_W$  of the
$W^+$~\cite{Ar69}:
\bq \label{EQ:multipole}
Q_W ~ = ~ e g_1^\gamma\; ~~,~~
\mu_W ~ = ~ {e \over 2m_W}
\left(g_1^\gamma + \kappa_\gamma + \lambda_\gamma\right) \; , ~~
q_W ~ = ~ -{e \over m_W^2} \left(\kappa_\gamma-\lambda_\gamma\right) \; .
\eq
For practical purposes it is convenient to introduce
deviations from the (tree-level) SM as
\ba
\label{deviation}
& \Delta g_1^Z \equiv (g_1^Z - 1)  ~~\equiv \tan\theta_W \delta_Z ~,~~~
 \Delta \kappa_\gamma \equiv
(\kappa_\gamma-1) ~~\equiv \xg ~, \\ \nonumber ~~~ & \Delta\kappa_Z
\equiv (\kappa_Z-1) ~~\equiv \tan\theta_W (x_Z +\delta_Z)~, \\ \nonumber
~& \lambda_\gamma ~~\equiv y_\gamma,
  ~~~\lambda_Z      ~~\equiv \tan\theta_W\, y_Z\; .
\ea
For completeness (and easy comparison) the correspondence of the most
studied C and P conserving parameters has also been given for another
equivalent set ($\delta_Z,\;x_V,\;y_V$), which was used in some recent
analyses~\cite{Bietal93,Se94}.
\subsection{Gauge-invariant Parametrization of TGC}\label{TGCgaugeinv}
Any of the interaction terms in (\ref{LeffWWV}) can be rendered
$SU(2)\times U(1)$ gauge invariant by adding to it interactions involving
additional gauge bosons~\cite{KuReSc87}, and/or additional Would Be Goldstone
Bosons (WBGBs)  and the physical Higgs
(if it exists)\cite{Ruetal92,Haetal93,GoRe93}.
However, one needs to consider $SU(2)\times U(1)$
gauge invariant operators of high dimension in order to reproduce all
couplings in (\ref{LeffWWV}). For example, if the Higgs particle
exists one needs to consider operators of dimension up to $d=12$.
Depending on the NP dynamics,
such operators could be generated  at the NP mass scale
$\Lambda_{NP}$, with a strength which is generally
suppressed by factors like
$(m_W/\Lambda_{NP})^{d-4}$ or $(\sqrt{s}/\Lambda_{NP})^{d-4}$
\cite{dyn1, dyn2}.
Accordingly, the gauge invariance
requirement {\it alone} does not provide any constraint on the form of
possible interactions. Rather it is a low energy approximation,
the neglect of operators of dimension greater
than 4 or 6, which leads to relations among the various TGCs.

Such relations among TGCs are highly desirable,
given the somewhat limited statistics accessible at LEP2.
They  were first derived in ~\cite{MaScSc86,KuReSc87} by
imposing approximate global $SU(2)$
symmetry conditions on the phenomenological Lagrangian (\ref{LeffWWV}).
In the next subsection we present them following  an
approach based on $SU(2)\times U(1)$ gauge
invariance and dimensional considerations.
The connection to the approach based on
``global $SU(2)$'' symmetry will be discussed at the end.

In order to write down all allowed operators of a given dimensionality
one must first identify the low energy degrees of freedom
participating in NP. We
assume that these include {\it only} the $SU(2)\times U(1)$
gauge fields and the remnants of the
spontaneous breaking of the gauge symmetry, the WBGBs
that exist already in the standard model.
If a relatively light Higgs boson is assumed to exist,
then NP is described in terms of a direct extension
of the ordinary SM formalism; \ie\@ using a {\it linear}
realization of the symmetry.
On the other hand, if the Higgs
is absent from the spectrum (or, equivalently for our purpose, if it is
sufficiently heavy), then the effective Lagrangian should be
expressed using a nonlinear realization of the symmetry.
\subsubsection{Linear Realization}\label{TGClr}
In addition to a Higgs doublet field $\Phi$, 
the building blocks of the gauge-invariant
operators are the
covariant derivatives of the Higgs field, $D_\mu \Phi$,
and the non-Abelian
field strength tensors $\hat{W}_{\mu\nu} = W_{\mu\nu} -g W_\mu \times 
W_\nu $ and $B_{\mu\nu}$ of the
$SU(2)_L$ and $U(1)_Y$ gauge fields respectively.\par
 
Considering  CP-conserving interactions of dimension $d=6$,
11~independent operators can
be constructed~\cite{BuWy86,Ruetal92,Haetal93}. Four of these operators
affect the gauge boson propagators at tree
level~\cite{GrWi91} and as a result their coefficients are
severely constrained by present low energy data \cite{Ruetal92,Haetal93}.
Another subset of these operators generates anomalous Higgs
couplings and will be discussed in section \ref{TGChghz} below.
Here we consider the three remaining operators
which do not affect the gauge boson propagators at
tree-level, but give rise to deviations in the C and P-conserving TGC.
Denoting the corresponding couplings as  {\bf \aWphi},
{\bf \aBphi}, and {\bf \aW}, the TGC inducing effective Lagrangian is
written as
\bq
\label{eq:alphas}
{\cal L}^{TGC}_{d=6} =  ig^\prime {\aBphi \over m^2_W}\,
 (D_\mu\Phi)^\dagger B^{\mu\nu}(D_\nu\Phi)\,
+\, i g {\aWphi \over m^2_W} \,
(D_\mu\Phi)^\dagger \vec \tau \cdot \vec{ \hat{W}}^{\mu\nu}
(D_\nu\Phi)\,
+g {\aW \over {6 m^2_W}}\,
\vec{ \hat{W}}^{\mu}_{\phantom{\mu}\nu}
\cdot (\vec{ \hat{W}}^\nu_{\phantom{\nu}\rho}
\times \vec{ \hat{W}}^{\rho}_{\phantom{\rho}\mu}) \,,
\eq
with $g$, $g^\prime$ the $SU(2)_L$ and $U(1)_Y$ couplings respectively.
Replacing the Higgs doublet field by its vacuum expectation value,
$\Phi^T\to (0, v/\sqrt{2})$, yields nonvanishing anomalous TGCs in
(\ref{LeffWWV}),
\bq
\label{tgcd6}
\Delta g_1^Z = \frac{\aWphi}{c^2_W}\;, \qquad
\Delta\kappa_\gamma =  -\frac{c^2_W}{s^2_W}(
\Delta\kappa_Z -\Delta g_1^Z ) =  \aWphi +\aBphi \;, \qquad
\lambda_\gamma = \lambda_Z  =  \aW \; , 
\eq
where $s_W \equiv \sin \theta_W$, $c_W \equiv \cos\theta_W $.
 The normalization of the dimension 6 operators in (\ref{eq:alphas}) has
been chosen such that the coefficients $\alpha_i$ correspond directly to
$\Delta\kappa_\gamma$ and $\lambda_\gamma$. It should be noted that, as the
NP scale $\Lambda_{NP}$ is increased, the $\alpha_i$ are expected to decrease
as $(m_W/\Lambda_{NP})^2$.

This scaling behaviour can be quantified to some extent by invoking 
(tree-level) unitarity constraints~\cite{BaZe88, GoRe94, uni2, dyn2}. 
A constant anomalous TGC
leads to a rapid growth of vector boson pair production cross-sections with
energy, saturating the unitarity limit at $\sqrt{s}=\Lambda_U$.
A larger value of $\Lambda_U$ implies a smaller TGC $\alpha_i$. For each
of them the unitarity relation may be written as~\cite{BaZe88,GoRe94}
\vspace{-15pt}
\bq
\label{TGCunitarity}
|\aW|  \simeq  19~\left ({m_W \over \Lambda_U} \right )^2  , \ \
| \aWphi|   \simeq   15.5\left ({m_W \over \Lambda_U} \right )^2 ,\ \
|\aBphi|  \simeq  49\left ({m_W\over  \Lambda_U} \right )^2  .
\eq
For any given value of $\alpha_i$ the corresponding scale $\Lambda_U$ 
provides an upper bound on the NP scale $\Lambda_{NP}$.
Conversely, a sensitivity to small values of an anomalous coupling constant
is equivalent to a sensitivity to potentially high values of the
corresponding NP scale.
Applying (\ref{TGCunitarity}) for $\Lambda_U=1$~TeV, we get $|\aW | \simeq
0.12$, $|\aWphi| \simeq 0.1$, $|\aBphi|\simeq 0.3$. Since these values are
larger than the expected LEP2 sensitivity by less than a factor 3, it is clear
that LEP2 is sensitive to $\Lambda_{NP}\lsim 1$~TeV. Thus a caveat is in
order: for these low values of $\Lambda_{NP}$ the neglect of dimension 8
operators may no longer be justified, leading to deviations from the
relations (\ref{tgcd6})\cite{ArEiWu94}.
%
%
%
\subsubsection{Nonlinear Realization}\label{TGCnlr}
In the absence of a light Higgs a non-linear approach should be used to
render ${\cal L}^{WWV}_{eff}$ gauge invariant. The SM Lagrangian, deprived of
the Higgs field,
violates unitarity at a scale of 
roughly\footnote{
One should caution that this estimate of $\Lambda_{NP}$ follows
directly from analogy with low energy QCD and 
Chiral perturbation theory~\cite{GaLe85},
where $v \equiv f_\pi$ and $\Lambda \simeq M_P$ are known,
while in the present context $\Lambda_{NP}$ is essentially unknown.
It should be taken as a rough order of magnitude estimate only.}
 $4\pi v
\sim 3$ TeV, so that the new physics should appear at a scale
$\Lambda_{NP} \lsim 4 \pi v$.
Technically the construction of gauge-invariant operators follows
closely the linear case above, except that in place
of the scalar doublet $\Phi$ 
a (unitary, dimensionless) matrix $U \equiv \exp(i
{\vec\omega}\cdot
{\vec\tau}/v)$, where the $\omega_i$ are the WBGBs, and the
appropriate matrix form of the $SU(2)_L\times U(1)_Y$ covariant
derivative are used.
The so-called ``naive dimensional analysis''
(NDA)~\cite{NDA} dictates that the expected order of magnitude of a
specific operator involving $b$ WBGB fields,
$d$ derivatives and $w$ gauge fields is
$ \sim v^2 \Lambda_{NP}^2 \; ( 1/ v)^b \;
( 1/ \Lambda_{NP})^d \;
(  g/ \Lambda_{NP})^w $.
Applying NDA to the terms in Eqs.~(\ref{LeffWWV}),
we see that $\Delta g_1^V$ and  $\Delta\kappa_V$ are of ${\cal
O}(m_W^2/\Lambda_{NP}^2)$. In
other words, just as in the linear realization, these terms are effectively
of dimension 6 (in the sense that there is an explicit factor of
$1/\Lambda_{NP}^2$).  On the other hand, we see that the
$W^{\dagger}_{\rho\mu}{W^{\mu}}_{\nu}V^{\nu\rho}$ term is effectively of
dimension~8, {\it i.e.} the coefficient $\lambda_V$ is expected
to be of order
$m_W^4/\Lambda_{NP}^4$. Thus, within the nonlinear realization scenario, the
$\lambda_V$ terms are expected to be negligible compared to those
proportional to $\Delta g_1^V$ and $\Delta\kappa_V$.
Accordingly there remain three parameters at lowest dimensionality, which
can be taken as $g_1^Z$, $\kappa_Z$ and $\kappa_\gamma$.

\subsubsection{Operators of Higher Dimension and Global Symmetry Arguments}
\label{TGCcustodial}
As mentioned in section \ref{TGCgaugeinv},
one may argue that relations
like in (\ref{tgcd6}) would not even be approximately correct if $\Lambda_{NP}$
is substantially smaller than 1~TeV, since higher dimensional operators
are no longer suppressed, and may even be more important than the $dim=6$
operators~\cite{ArEiWu94}.
In fact, as far as the 5 C and P conserving TGC
in (\ref{LeffWWV}) are concerned, the most general choice can be realized by
invoking two $dim=8$ operators in addition to the 3 terms in (\ref{eq:alphas})
\cite{Haetal93,GoRe93,GrKuSc95}.
Requiring restoration of an $SU(2)$ {\it global} ("custodial")
symmetry for $g^\prime \to 0$ (i.e in the limit of decoupling B
field) implies~\cite{GrKuSc95}
 the coefficient of one of these two operators to vanish,
because it
violates $SU(2)$ global\footnote{Note that there is no contradiction with
the $SU(2)_L \times U(1)_Y$ local invariance of all these operators,
since $SU(2)$ custodial is a different symmetry from the $SU(2)_L$ global
\cite{custodial}.} {\it independently} of the B field.
In that way one recovers the constraints between $\Delta\kappa_\gamma$ and
$\Delta\kappa_Z$ in (\ref{tgcd6}), in both the nonlinear realization and
in the linear realization at the $dim =8$ level.
Nevertheless a second $dim = 8$ operator spoils the
relation, $\lambda_\gamma = \lambda_Z$ in (\ref{tgcd6}).
One may neglect this term (which vanishes in the limit $g^\prime \to 0$)
by imposing exact $SU(2)$ at the scale $\Lambda_{NP}$,
which in our context is similar to neglecting the $\pi^\pm -\pi^0$
mass difference in strong interaction physics.
 
Largely these are simplifying assumptions only, intended to reduce the number
of free parameters. Motivated by the previous discussion we recommend two sets
of three parameters each for full correlation studies between anomalous
couplings at LEP2:
\begin{itemize}
\item{}
set1 = ($\Delta g_1^Z,\; \Delta\kappa_\gamma,\; \Delta\kappa_Z$) with
$\lambda_\gamma=\lambda_Z=0$.
These correspond to the operators of lowest dimensionality in the nonlinear
realization. A reduction to 2 parameters (using $\Delta\kappa_\gamma =
-\frac{c^2_W}{s^2_W}(\Delta\kappa_Z -\Delta g_1^Z )$) is achieved by
assuming~\cite{Bietal93,GrKuSc95}
custodial $SU(2)$ for $g^\prime \to 0$.
\item{}
set2 = ($\Delta g_1^Z,\; \Delta \kappa_\gamma,\; \lambda_\gamma$) 
with $\lambda_Z$ and
$\Delta\kappa_Z$ given by (\ref{tgcd6}). 
It is this set which has been used in this
report for the determination of precisions achievable from $WW$ production at
LEP2, presented in sections 5--9 as limits on the parameters \aBphi, \aWphi\
and \aW\ defined by (\ref{eq:alphas}). 
\end{itemize}
Expressing results in terms of $\Delta g^Z_1$, $\Delta\kappa_\gamma$, etc.
will be useful for ease of comparison with published
hadron collider data~\cite{ref:TGC-CDF,D0}.
 
In addition, it would clearly be of interest to present fits to each of the
parameters in ${\cal L}_{eff}^{WWV}$ in order to reduce the dependence of
the analysis on specific models. However, this can only be achieved bearing
in mind the limited data which will be available from LEP2, and the
correlations inherent in the extraction of many parameters from the data.
We return to this point in sections {3.1, 4.2 and 5.1} below.
\subsection{Present constraints on TGC}\label{TGCpresentbounds}
The errors of present direct measurements, via pair production of
electroweak bosons at the Tevatron, are still fairly large. The latest,
best published 95\% CL bounds by $CDF$ and $D0$ are obtained from studies
of $W\gamma$ events~\cite{ref:TGC-CDF,D0}
$$
-1.6< \Delta\kappa_\gamma < 1.8\; ,\qquad -0.6 < \lambda_\gamma < 0.6
$$
but constraints from the study of $WW,\;WZ\to \ell\nu jj,\;\ell\ell jj$ events
are becoming competitive and should lead to 95\% CL bounds of roughly
$-0.65<\Delta\kappa_\gamma < 0.75,\; |\lambda_\gamma| = |\lambda_Z| < 0.4$,
once the already collected run 1b data are fully analyzed.
Increasing the integrated luminosity to 1~fb$^{-1}$ with the Fermilab main 
injector is expected to improve these limits by another factor 2~\cite{DPF}.
Note that these latter bounds assume the relations between anomalous
couplings as given by (\ref{tgcd6}) with {\aWphi\ =\aBphi}. In addition,
the Tevatron measures these parameters at considerably larger momentum
transfers than LEP2 and, hence, form factor effects could result in different
measured values at the two machines.
 
Alternatively, constraints may be derived also from evaluating {\it
virtual} contributions of TGC to precisely measured quantities such as
$(g-2)_\mu$~\cite{g-2}, the $b\to s\gamma$ decay
rate~\cite{btosgamma,CLEO}, $B\to K^{(*)}\mu^+\mu^-$~\cite{Kstar}, the
$Z\to b\bar b$~\cite{zbb} rate
and oblique corrections\cite{Ruetal92,Haetal93} ({\it i.e.}
corrections to the W, Z, $\gamma$ propagators). Oblique corrections
combine information from the recent LEP/SLD data, neutrino scattering
experiments, atomic parity violation, $\mu$-decay, and the $W$-mass
measurement at hadron colliders.
 
When trying to derive TGC bounds from their virtual contributions one must
make assumptions about other NP contributions to the observable in question.
In the linear realization, for example, Higgs contributions to the oblique
parameters tend to cancel the TGC contributions and as a result the TGC bounds
are relatively weak for a light Higgs boson~\cite{Haetal93}. In general, there
are other higher dimensional operators which contribute directly to the
observable, in addition to the virtual TGC effects. Bounds on the TGC then
require to either specify the underlying model of NP completely or to
{\it assume} that no significant cancellation occurs. The bounds on the TGC
parameters in (\ref{LeffWWV}) due to virtual effets thus depend on the
underlying hypotheses and are of ${\cal O}(0.1)$ to
${\cal O} (1)$~\cite{Ruetal92,Haetal93,HeVe94}.
 
More stringent bounds are obtained~\cite{Ruetal92} by comparing the higher
dimensional operators which induce TGC with those operators which directly
induce oblique effects (see Section 10.1). In simple models the coefficients
of these two sets of operators are of similar size and hence the stringent
LEP1 bounds on the latter~\cite{HagiwaraLEP1} indicate that one should not
expect anomalous TGC above ${\cal O}(0.01)$. One should stress, however,
that no rigourous relation between oblique effects and TGC can be derived
except by going to specific models of NP. Therefore, these stringent bounds
must be verified, by a direct measurement of the TGC at LEP2.
%
%
%
\subsection{Virtual Contributions to TGC in the MSSM
\protect\footnote{A complementary study of virtual MSSM contributions 
to the \eeWW cross-section is done in the New Particle 
chapter of these proceedings.}}\label{TGCmssm}
Definite TGC contributions are certainly present
at the {\it loop level} in any
renormalizable model, although
such loop effects contribute to TGC with a factor of
$(g^2/16\pi^2) \simeq $ 2.7 10$^{-3}$, being therefore too small a priori to
be observed at LEP2. 
For instance, SM one-loop TGC predictions are 
known~\cite{Fletal92,pinch,LaSp94} and give, at $\sqrt s = $ 190 GeV, 
$\Delta \kappa_\gamma $ ($\Delta \kappa_Z$) $\simeq $  4.1--5.7 10$^{-3}$
(3.3--3.1 10$^{-3}$), for $m_{Higgs} =$ 0.065--1 TeV and  
$m_{top}=$ 175 GeV~\cite{ArKnMo95}. 
(Contributions to $\lambda_V$ are about a factor
of 3 smaller). 
One may, however, expect that the ``natural scale" $(g^2/16\pi^2)$  
could be substantially enhanced 
if, for example, some particles in the loop have strong coupling
and/or are close to their production threshold. To obtain a ``reference point''
it is thus important to explore more quantitatively
{\it how far} one is from the LEP2 accuracy limit, within
some well-defined model of NP.
We here use the contributions of the
(MSSM)~\cite{MSSM} as an example. These contributions
were calculated independently by two groups in the
framework of the Workshop. We summarize the main results,
referring for more details to refs.~\cite{LaSp94,ArKnMo95}.
\begin{table}[htb]
\center{
\begin{tabular}{|l|l|}
\hline\hline
SUGRA-GUT MSSM
         & Unconstrained MSSM (maximal effects) \\ \hline
    $A_0, m_0, M_{1/2} =$ 300,300,80 (GeV), 
& $\tan\beta =$1.5; $M^{\chi^+}_{1,2} \simeq$ 95, 130; 
$M^{\chi^0}_i \simeq $ 20--132 (GeV) ; 
\\
 $\tan\beta =$ 2 $\;\;$ ($\mu < 0 $) 
&
$m_{H+} \simeq $ 95; 
$m_{\tilde \nu_l} \simeq$ 45; $m_{\tilde l} \simeq $ 92--110; 
$m_{\tilde q} \simeq $45-800 GeV \\ \hline
$\vert \Delta\kappa_\gamma \vert \! =\!$
 0.44 10$^{-2}$,
$\vert \Delta\kappa_Z \vert \! = \!$
 0.72 10$^{-2}$ \hspace{-3 mm} & \hspace{1 cm} $\Delta\kappa_\gamma =$
 1.75 10$^{-2}$,
$\Delta\kappa_Z =$
 0.84 10$^{-2}$ 
         \\ \hline
         \hline
\end{tabular}
\caption{$\Delta \kappa_{\gamma ,Z}$ (as defined in eq. \ref{LeffWWV}) in 
MSSM at $\protect \sqrt{s} =$ 190 GeV.  
(Contributions to $\lambda_V$ are
about a factor of 2-3 smaller).
   }
\label{tab:tgcmssm}
}
\end{table}

Naively, TGC are obtained by
summing all MSSM contributions to the appropriate parts in eq. (\ref{LeffWWV})
from vertex loops with entering $\gamma$ (or $Z$) and
outgoing  $W^+$, $W^-$.
But as is well-known, the vertex graphs with virtual gauge bosons
need to be combined with parts of box graphs for the full process,
$e^+e^- \to W^+W^-$, to form a gauge-invariant contribution.
The resulting combinations define purely $s$-dependent
\footnote{By definition, $t$ and $u$-dependent
box contributions are left over in this procedure.
We have evaluated~\cite{ArKnMo95} a definite (gauge-invariant)
sample of this remnant part,
the slepton box contributions, and found them negligible,
$\simeq 0.1\: (g^2/16\pi^2) \simeq 3\;10^{-4}$ at most, 
at LEP2 energies.} TGC~\cite{pinch}.
In table \ref{tab:tgcmssm} we illustrate our results for
($s$-dependent) contributions in two different cases.
First, for a representative choice of the free parameters in the more
constrained MSSM spectrum obtained~\cite{LaSp94} from the SUGRA-GUT
scenario~\cite{SUGRAGUT}: the only parameters are the universal
soft terms
$m_0$, $M_{1/2}$, $A_0$ at the GUT scale, 
$\tan\beta $ (and the sign of $\mu$). Second, we give one illustrative
contribution, obtained~\cite{ArKnMo95}
from a rather systematic search of maximal effects in  
the unconstrained MSSM parameter space. 
The largest contributions are mostly due to gauginos 
and/or some of the sleptons
and squarks being practically at threshold. 
One may note, however, that
some individual 
contributions, potentially larger, 
were quite substantially reduced when   
the present constraints on the MSSM parameters are taken into 
account~\cite{ArKnMo95}.
Even these maximal contributions 
hardly reach the level of the most optimistic accuracy limit expected on
TGC (compare section \ref{sec:TGC-gen} below). 
One should also note that radiatively generated TGC 
generically have a complicated $\sqrt{s}$ form factor dependence as well
as contributions from boxes, which are well approximated by an
expansion in $1/\Lambda_{NP}$ only when one probes well below threshold.
%
%
%

\def\diff#1#2          {\mbox{$\frac{{\mathrm d} #1}{{\mathrm d} #2}       $}}
\subsection{TGC from extra $Z^\prime$
\protect\footnote{A complementary study can be found in the
$Z^\prime$ working group chapter of these proceedings}}\label{TGCzprime}
\def\temp{1.37}%
\let\tempp=\relax
\expandafter\ifx\csname psboxversion\endcsname\relax
  \message{PSBOX(\temp)}%
\else
    \ifdim\temp cm>\psboxversion cm
      \message{PSBOX(\temp)}%
    \else
      \message{PSBOX(\psboxversion) is already loaded: I won't load
        PSBOX(\temp)!}%
      \let\temp=\psboxversion
      \let\tempp= 
    \fi
\fi
\tempp
\let\psboxversion=\temp
\catcode`\@=11
%
%
\def\psfortextures{
\def\PSspeci@l##1##2{%
\special{illustration ##1\space scaled ##2}%
}}%
\def\psfordvitops{
\def\PSspeci@l##1##2{%
\special{dvitops: import ##1\space \the\drawingwd \the\drawinght}%
}}%
\def\psfordvips{
\def\PSspeci@l##1##2{%
\d@my=0.1bp \d@mx=\drawingwd \divide\d@mx by\d@my
\includegraphics{##1\space}}}%
\def\psforoztex{
\def\PSspeci@l##1##2{%
\special{##1 \space
      ##2 1000 div dup scale
      \number-\psllx\space\space \number-\pslly\space\space translate
}}}%
\def\psfordvitps{
\def\dvitpsLiter@ldim##1{\dimen0=##1\relax
\special{dvitps: Literal "\number\dimen0\space"}}%
\def\PSspeci@l##1##2{%
\at(0bp;\drawinght){%
\special{dvitps: Include0 "psfig.psr"}
\dvitpsLiter@ldim{\drawingwd}%
\dvitpsLiter@ldim{\drawinght}%
\dvitpsLiter@ldim{\psllx bp}%
\dvitpsLiter@ldim{\pslly bp}%
\dvitpsLiter@ldim{\psurx bp}%
\dvitpsLiter@ldim{\psury bp}%
\special{dvitps: Literal "startTexFig"}%
\special{dvitps: Include1 "##1"}%
\special{dvitps: Literal "endTexFig"}%
}}}%
\def\psfordvialw{
\def\PSspeci@l##1##2{
\special{language "PostScript",
position = "bottom left",
literal "  \psllx\space \pslly\space translate
  ##2 1000 div dup scale
  -\psllx\space -\pslly\space translate",
include "##1"}
}}%
\def\psforptips{
\def\PSspeci@l##1##2{{
\d@mx=\psurx bp
\advance \d@mx by -\psllx bp
\divide \d@mx by 1000\multiply\d@mx by \xscale
\incm{\d@mx}
\let\tmpx\dimincm
\d@my=\psury bp
\advance \d@my by -\pslly bp
\divide \d@my by 1000\multiply\d@my by \xscale
\incm{\d@my}
\let\tmpy\dimincm
\d@mx=-\psllx bp
\divide \d@mx by 1000\multiply\d@mx by \xscale
\d@my=-\pslly bp
\divide \d@my by 1000\multiply\d@my by \xscale
\at(\d@mx;\d@my){\special{ps:##1 x=\tmpx cm, y=\tmpy cm}}
}}}%
\let\psforpsprint=\psforoztex
\def\psonlyboxes{
\def\PSspeci@l##1##2{%
\at(0cm;0cm){\boxit{\vbox to\drawinght
  {\vss\hbox to\drawingwd{\at(0cm;0cm){\hbox{({\tt##1})}}\hss}}}}
}}%
\def\psloc@lerr#1{%
\let\savedPSspeci@l=\PSspeci@l%
\def\PSspeci@l##1##2{%
\at(0cm;0cm){\boxit{\vbox to\drawinght
  {\vss\hbox to\drawingwd{\at(0cm;0cm){\hbox{({\tt##1}) #1}}\hss}}}}
\let\PSspeci@l=\savedPSspeci@l
}}%
%
%
\newread\pst@mpin
\newdimen\drawinght\newdimen\drawingwd
\newdimen\psxoffset\newdimen\psyoffset
\newbox\drawingBox
\newcount\xscale \newcount\yscale \newdimen\pscm\pscm=1cm
\newdimen\d@mx \newdimen\d@my
\newdimen\pswdincr \newdimen\pshtincr
\let\ps@nnotation=\relax
{\catcode`\|=0 |catcode`|\=12 |catcode`|
|catcode`#=12 |catcode`*=14
|xdef|backslashother{\}*
|xdef|percentother{
|xdef|tildeother{~}*
|xdef|sharpother{#}*
}%
\def\R@moveMeaningHeader#1:->{}%
\def\uncatcode#1{%
\edef#1{\expandafter\R@moveMeaningHeader\meaning#1}}%
\def\execute#1{#1}
\def\psm@keother#1{\catcode`#112\relax}
\def\executeinspecs#1{%
\execute{\begingroup\let\do\psm@keother\dospecials\catcode`\^^M=9#1\endgroup}}%
\def\@mpty{}%
\def\matchexpin#1#2{
  \fi%
  \edef\tmpb{{#2}}%
  \expandafter\makem@tchtmp\tmpb%
  \edef\tmpa{#1}\edef\tmpb{#2}%
  \expandafter\expandafter\expandafter\m@tchtmp\expandafter\tmpa\tmpb\endm@tch%
  \if\match%
}%
\def\matchin#1#2{%
  \fi%
  \makem@tchtmp{#2}%
  \m@tchtmp#1#2\endm@tch%
  \if\match%
}%
\def\makem@tchtmp#1{\def\m@tchtmp##1#1##2\endm@tch{%
  \def\tmpa{##1}\def\tmpb{##2}\let\m@tchtmp=\relax%
  \ifx\tmpb\@mpty\def\match{YN}%
  \else\def\match{YY}\fi%
}}%
\def\incm#1{{\psxoffset=1cm\d@my=#1
 \d@mx=\d@my
  \divide\d@mx by \psxoffset
  \xdef\dimincm{\number\d@mx.}
  \advance\d@my by -\number\d@mx cm
  \multiply\d@my by 100
 \d@mx=\d@my
  \divide\d@mx by \psxoffset
  \edef\dimincm{\dimincm\number\d@mx}
  \advance\d@my by -\number\d@mx cm
  \multiply\d@my by 100
 \d@mx=\d@my
  \divide\d@mx by \psxoffset
  \xdef\dimincm{\dimincm\number\d@mx}
}}%
%
\newif\ifNotB@undingBox
\newhelp\PShelp{Proceed: you'll have a 5cm square blank box instead of
your graphics.}%
\def\s@tsize#1 #2 #3 #4\@ndsize{
  \def\psllx{#1}\def\pslly{#2}%
  \def\psurx{#3}\def\psury{#4}
  \ifx\psurx\@mpty\NotB@undingBoxtrue
  \else
    \drawinght=#4bp\advance\drawinght by-#2bp
    \drawingwd=#3bp\advance\drawingwd by-#1bp
  \fi
  }%
\def\sc@nBBline#1:#2\@ndBBline{\edef\p@rameter{#1}\edef\v@lue{#2}}%
\def\g@bblefirstblank#1#2:{\ifx#1 \else#1\fi#2}%
{\catcode`\%=12
\xdef\B@undingBox{
\def\ReadPSize#1{
 \readfilename#1\relax
 \let\PSfilename=\lastreadfilename
 \openin\pst@mpin=#1\relax
 \ifeof\pst@mpin \errhelp=\PShelp
   \errmessage{I haven't found your postscript file (\PSfilename)}%
   \psloc@lerr{was not found}%
   \s@tsize 0 0 142 142\@ndsize
   \closein\pst@mpin
 \else
   \loop
     \executeinspecs{\catcode`\ =10\global\read\pst@mpin to\n@xtline}%
     \ifeof\pst@mpin
       \errhelp=\PShelp
       \errmessage{(\PSfilename) is not an Encapsulated PostScript File:
           I could not find any \B@undingBox: line.}%
       \edef\v@lue{0 0 142 142:}%
       \psloc@lerr{is not an EPSFile}%
       \NotB@undingBoxfalse
     \else
       \expandafter\sc@nBBline\n@xtline:\@ndBBline
       \ifx\p@rameter\B@undingBox\NotB@undingBoxfalse
         \if\matchexpin{\GlobalInputList}{, \lastreadfilename}%
           \else\xdef\GlobalInputList{\GlobalInputList, \lastreadfilename}%
             \immediate\write\psbj@inaux{\lastreadfilename,}%
         \fi%
         \edef\t@mp{%
           \expandafter\g@bblefirstblank\v@lue\space\space\space}%
         \expandafter\s@tsize\t@mp\@ndsize
       \else\NotB@undingBoxtrue
       \fi
     \fi
   \ifNotB@undingBox\repeat
   \closein\pst@mpin
 \fi
\message{#1}%
}%
%
%
\def\psboxto(#1;#2)#3{\vbox{%
   \ReadPSize{#3}%
   \advance\pswdincr by \drawingwd
   \advance\pshtincr by \drawinght
   \divide\pswdincr by 1000
   \divide\pshtincr by 1000
   \d@mx=#1
   \ifdim\d@mx=0pt\xscale=1000
         \else \xscale=\d@mx \divide \xscale by \pswdincr\fi
   \d@my=#2
   \ifdim\d@my=0pt\yscale=1000
         \else \yscale=\d@my \divide \yscale by \pshtincr\fi
   \ifnum\yscale=1000
         \else\ifnum\xscale=1000\xscale=\yscale
                    \else\ifnum\yscale<\xscale\xscale=\yscale\fi
              \fi
   \fi
   \divide\drawingwd by1000 \multiply\drawingwd by\xscale
   \divide\drawinght by1000 \multiply\drawinght by\xscale
   \divide\psxoffset by1000 \multiply\psxoffset by\xscale
   \divide\psyoffset by1000 \multiply\psyoffset by\xscale
   \global\divide\pscm by 1000
   \global\multiply\pscm by\xscale
   \multiply\pswdincr by\xscale \multiply\pshtincr by\xscale
   \ifdim\d@mx=0pt\d@mx=\pswdincr\fi
   \ifdim\d@my=0pt\d@my=\pshtincr\fi
   \message{scaled \the\xscale}%
 \hbox to\d@mx{\hss\vbox to\d@my{\vss
   \global\setbox\drawingBox=\hbox to 0pt{\kern\psxoffset\vbox to 0pt{%
      \kern-\psyoffset
      \PSspeci@l{\PSfilename}{\the\xscale}%
      \vss}\hss\ps@nnotation}%
   \global\wd\drawingBox=\the\pswdincr
   \global\ht\drawingBox=\the\pshtincr
   \global\drawingwd=\pswdincr
   \global\drawinght=\pshtincr
   \baselineskip=0pt
   \copy\drawingBox
 \vss}\hss}%
  \global\psxoffset=0pt
  \global\psyoffset=0pt
  \global\pswdincr=0pt
  \global\pshtincr=0pt 
  \global\pscm=1cm 
}}%
%
%
\def\psboxscaled#1#2{\vbox{%
  \ReadPSize{#2}%
  \xscale=#1
  \message{scaled \the\xscale}%
  \divide\pswdincr by 1000 \multiply\pswdincr by \xscale
  \divide\pshtincr by 1000 \multiply\pshtincr by \xscale
  \divide\psxoffset by1000 \multiply\psxoffset by\xscale
  \divide\psyoffset by1000 \multiply\psyoffset by\xscale
  \divide\drawingwd by1000 \multiply\drawingwd by\xscale
  \divide\drawinght by1000 \multiply\drawinght by\xscale
  \global\divide\pscm by 1000
  \global\multiply\pscm by\xscale
  \global\setbox\drawingBox=\hbox to 0pt{\kern\psxoffset\vbox to 0pt{%
     \kern-\psyoffset
     \PSspeci@l{\PSfilename}{\the\xscale}%
     \vss}\hss\ps@nnotation}%
  \advance\pswdincr by \drawingwd
  \advance\pshtincr by \drawinght
  \global\wd\drawingBox=\the\pswdincr
  \global\ht\drawingBox=\the\pshtincr
  \global\drawingwd=\pswdincr
  \global\drawinght=\pshtincr
  \baselineskip=0pt
  \copy\drawingBox
  \global\psxoffset=0pt
  \global\psyoffset=0pt
  \global\pswdincr=0pt
  \global\pshtincr=0pt 
  \global\pscm=1cm
}}%
%
\def\psbox#1{\psboxscaled{1000}{#1}}%
\newif\ifn@teof\n@teoftrue
\newif\ifc@ntrolline
\newif\ifmatch
\newread\j@insplitin
\newwrite\j@insplitout
\newwrite\psbj@inaux
\immediate\openout\psbj@inaux=psbjoin.aux
\immediate\write\psbj@inaux{\string\joinfiles}%
\immediate\write\psbj@inaux{\jobname,}%
%
%
\def\toother#1{\ifcat\relax#1\else\expandafter%
  \toother@ux\meaning#1\endtoother@ux\fi}%
\def\toother@ux#1 #2#3\endtoother@ux{\def\tmp{#3}%
  \ifx\tmp\@mpty\def\tmp{#2}\let\next=\relax%
  \else\def\next{\toother@ux#2#3\endtoother@ux}\fi%
\next}%
%
%
\let\readfilenamehook=\relax
\def\re@d{\expandafter\re@daux}
\def\re@daux{\futurelet\nextchar\stopre@dtest}%
\def\re@dnext{\xdef\lastreadfilename{\lastreadfilename\nextchar}%
  \afterassignment\re@d\let\nextchar}%
\def\stopre@d{\egroup\readfilenamehook}%
\def\stopre@dtest{%
  \ifcat\nextchar\relax\let\nextread\stopre@d
  \else
    \ifcat\nextchar\space\def\nextread{%
      \afterassignment\stopre@d\chardef\nextchar=`}%
    \else\let\nextread=\re@dnext
      \toother\nextchar
      \edef\nextchar{\tmp}%
    \fi
  \fi\nextread}%
\def\readfilename{\bgroup%
  \let\\=\backslashother \let\%=\percentother \let\~=\tildeother
  \let\#=\sharpother \xdef\lastreadfilename{}%
  \re@d}%
%
%
\xdef\GlobalInputList{\jobname}%
\def\psnewinput{%
  \def\readfilenamehook{
    \if\matchexpin{\GlobalInputList}{, \lastreadfilename}%
    \else\xdef\GlobalInputList{\GlobalInputList, \lastreadfilename}%
      \immediate\write\psbj@inaux{\lastreadfilename,}%
    \fi%
    \let\readfilenamehook=\relax%
    \ps@ldinput\lastreadfilename\relax%
  }\readfilename%
}%
\def\psinputredef{%
\expandafter\ifx\csname @@input\endcsname\relax    
  \immediate\let\ps@ldinput=\input\def\input{\psnewinput}%
\else
  \immediate\let\ps@ldinput=\@@input
  \def\@@input{\psnewinput}%
\fi}%
\def\psinputunredef{%
\expandafter\ifx\csname @@input\endcsname\relax
  \immediate\let\input=\ps@ldinput
\else
  \immediate\let\@@input=\ps@ldinput
\fi
}%
\def\nowarnopenout{%
 \def\warnopenout##1##2{%
   \readfilename##2\relax
   \message{\lastreadfilename}%
   \immediate\openout##1=\lastreadfilename\relax}}%
\def\warnopenout#1#2{%
 \readfilename#2\relax
 \def\t@mp{TrashMe,psbjoin.aux,psbjoint.tex,}\uncatcode\t@mp
 \if\matchexpin{\t@mp}{\lastreadfilename,}%
 \else
   \immediate\openin\pst@mpin=\lastreadfilename\relax
   \ifeof\pst@mpin
     \else
     \edef\tmp{{If the content of this file is precious to you, this
is your last chance to abort (ie press x or e) and rename it before
retexing (\jobname). If you're sure there's no file
(\lastreadfilename) in the directory of (\jobname), then go on: I'm
simply worried because you have another (\lastreadfilename) in some
directory I'm looking in for inputs...}}%
     \errhelp=\tmp
     \errmessage{I may be about to replace your file named \lastreadfilename}%
   \fi
   \immediate\closein\pst@mpin
 \fi
 \message{\lastreadfilename}%
 \immediate\openout#1=\lastreadfilename\relax}%
{\catcode`\%=12\catcode`\*=14
\gdef\splitfile#1{*
 \readfilename#1\relax
 \immediate\openin\j@insplitin=\lastreadfilename\relax
 \ifeof\j@insplitin
   \message{! I couldn't find and split \lastreadfilename!}*
 \else
   \immediate\openout\j@insplitout=TrashMe
   \message{< Splitting \lastreadfilename\space into}*
   \loop
     \ifeof\j@insplitin
       \immediate\closein\j@insplitin\n@teoffalse
     \else
       \n@teoftrue
       \executeinspecs{\global\read\j@insplitin to\spl@tinline\expandafter
         \ch@ckbeginnewfile\spl@tinline
       \ifc@ntrolline
       \else
         \toks0=\expandafter{\spl@tinline}*
         \immediate\write\j@insplitout{\the\toks0}*
       \fi
     \fi
   \ifn@teof\repeat
   \immediate\closeout\j@insplitout
 \fi\message{>}*
}*
\gdef\ch@ckbeginnewfile#1
 \def\t@mp{#1}*
 \ifx\@mpty\t@mp
   \def\t@mp{#3}*
   \ifx\@mpty\t@mp
     \global\c@ntrollinefalse
   \else
     \immediate\closeout\j@insplitout
     \warnopenout\j@insplitout{#2}*
     \global\c@ntrollinetrue
   \fi
 \else
   \global\c@ntrollinefalse
 \fi}*
\gdef\joinfiles#1\into#2{*
 \message{< Joining following files into}*
 \warnopenout\j@insplitout{#2}*
 \message{:}*
 {*
 \edef\w@##1{\immediate\write\j@insplitout{##1}}*
\w@{
\w@{
\w@{
\w@{
\w@{
\w@{
\w@{
\w@{
\w@{
\w@{
\w@{\string\input\space psbox.tex}*
\w@{\string\splitfile{\string\jobname}}*
\w@{\string\let\string\autojoin=\string\relax}*
}*
 \expandafter\tre@tfilelist#1, \endtre@t
 \immediate\closeout\j@insplitout
 \message{>}*
}*
\gdef\tre@tfilelist#1, #2\endtre@t{*
 \readfilename#1\relax
 \ifx\@mpty\lastreadfilename
 \else
   \immediate\openin\j@insplitin=\lastreadfilename\relax
   \ifeof\j@insplitin
     \errmessage{I couldn't find file \lastreadfilename}*
   \else
     \message{\lastreadfilename}*
     \immediate\write\j@insplitout{
     \executeinspecs{\global\read\j@insplitin to\oldj@ininline}*
     \loop
       \ifeof\j@insplitin\immediate\closein\j@insplitin\n@teoffalse
       \else\n@teoftrue
         \executeinspecs{\global\read\j@insplitin to\j@ininline}*
         \toks0=\expandafter{\oldj@ininline}*
         \let\oldj@ininline=\j@ininline
         \immediate\write\j@insplitout{\the\toks0}*
       \fi
     \ifn@teof
     \repeat
   \immediate\closein\j@insplitin
   \fi
   \tre@tfilelist#2, \endtre@t
 \fi}*
}%
\def\autojoin{%
 \immediate\write\psbj@inaux{\string\into{psbjoint.tex}}%
 \immediate\closeout\psbj@inaux
 \expandafter\joinfiles\GlobalInputList\into{psbjoint.tex}%
}%
%
%
%
\def\centinsert#1{\midinsert\line{\hss#1\hss}\endinsert}%
\def\psannotate#1#2{\vbox{%
  \def\ps@nnotation{#2\global\let\ps@nnotation=\relax}#1}}%
\def\pscaption#1#2{\vbox{%
   \setbox\drawingBox=#1
   \copy\drawingBox
   \vskip\baselineskip
   \vbox{\hsize=\wd\drawingBox\setbox0=\hbox{#2}%
     \ifdim\wd0>\hsize
       \noindent\unhbox0\tolerance=5000
    \else\centerline{\box0}%
    \fi
}}}%
%
\def\at(#1;#2)#3{\setbox0=\hbox{#3}\ht0=0pt\dp0=0pt
  \rlap{\kern#1\vbox to0pt{\kern-#2\box0\vss}}}%
%
\newdimen\gridht \newdimen\gridwd
\def\gridfill(#1;#2){%
  \setbox0=\hbox to 1\pscm
  {\vrule height1\pscm width.4pt\leaders\hrule\hfill}%
  \gridht=#1
  \divide\gridht by \ht0
  \multiply\gridht by \ht0
  \gridwd=#2
  \divide\gridwd by \wd0
  \multiply\gridwd by \wd0
  \advance \gridwd by \wd0
  \vbox to \gridht{\leaders\hbox to\gridwd{\leaders\box0\hfill}\vfill}}%
%
\def\fillinggrid{\at(0cm;0cm){\vbox{%
  \gridfill(\drawinght;\drawingwd)}}}%
%
%
\def\textleftof#1:{%
  \setbox1=#1
  \setbox0=\vbox\bgroup
    \advance\hsize by -\wd1 \advance\hsize by -2em}%
\def\textrightof#1:{%
  \setbox0=#1
  \setbox1=\vbox\bgroup
    \advance\hsize by -\wd0 \advance\hsize by -2em}%
\def\endtext{%
  \egroup
  \vskip2pt
  \hbox to \hsize{\valign{\vfil##\vfil\cr%
\box0\cr%
\noalign{\hss}\box1\cr}}}%
%
\def\frameit#1#2#3{\hbox{\vrule width#1\vbox{%
  \hrule height#1\vskip#2\hbox{\hskip#2\vbox{#3}\hskip#2}%
        \vskip#2\hrule height#1}\vrule width#1}}%
\def\boxit#1{\frameit{0.4pt}{0pt}{#1}}%
\catcode`\@=12 
%
\psfordvips   
%

%
%
A light and weakly coupled $Z'$ provides an illustrative example of 
relatively large deviations of the TGC from their SM values and of 
strong form-factor effects~\cite{frere}.
%
%
%
Consider
an extra gauged $U(1)'$ symmetry with associated coupling $g^\prime_1$,
whose vector boson $Z'$ is
relatively light, say $M_{Z'} \simeq 200$GeV.
For such a boson to remain undetected at
LEP1 and CDF, it must have rather small couplings to fermions: $\lambda\equiv
\hbox{sin}(\theta_W) g'_1/g_1<0.2$ or less~\cite{frere}.
However, this new $Z'$
might be only part of the new physics beyond the SM, and we 
parametrize this by {\it gauge invariant} higher dimensional 
operators. For illustration, let us focus on the $dim=6$ operator
\begin{equation}
{\cal L}_{B' W}\equiv {\epsilon \over v^2} O_{B^\prime W}
  = {\epsilon \over v^2}
   \phi^\dagger B^{\prime \mu \nu}\vec{\hat{W}}_{\mu \nu}
\cdot \vec{\tau}\phi
  \label{eIiv}
\end{equation}
where $B'_{\mu\nu}$ is the new
$U'(1)$ field strength.  This operator has 
a part linear in $W_\mu$ inducing unusual mixing through the
kinetic terms, from which LEP1 data put upper bounds on $\lambda$ and
$\epsilon$. The other piece is quadratic in $W_\mu$ and brings anomalous
contributions to $W$-pair production at LEP2, which may be
enhanced at will by approaching the $Z'$ pole. Within a gauge-invariant 
framework, enlarging the symmetry group 
has given us enough freedom to escape
the more stringent LEP1 constraints on the coefficient of the similar
operator ${\cal O}_{BW}$~\cite{Ruetal92,Haetal93} of Eq.~(\ref{nonblind}).
Having such an (admittedly contrived) counter-example to
\cite{Ruetal92} (depending on the 3 parameters $M_{Z'}$,
$\lambda<0.2$ and $|\epsilon|<0.2$), it is instructive to see how it fits into
our TGC parametrization. 

%
%
\textleftof{\pscaption{
\psannotate{\psboxto(0.48\hsize;0cm){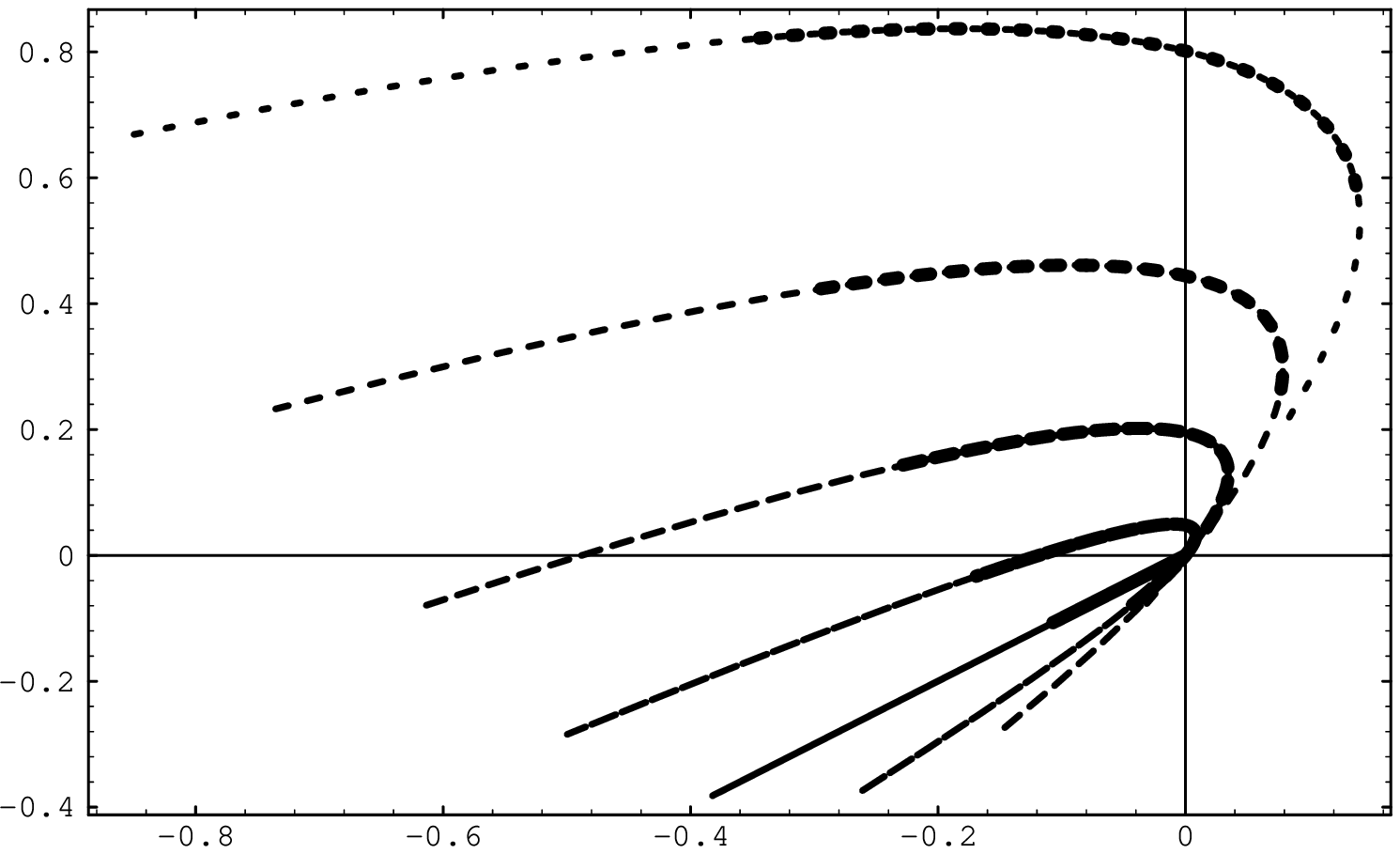}}
  {
   \at(13.5\pscm;-0.5\pscm){$\Delta g_1^Z$}%
   \at(-0.5\pscm; 7.8\pscm){$\Delta g_1^\gamma$}}}
{\it The deviations $\Delta g_1^Z$ vs. $\Delta g_1^\gamma$ for
  $\sqrt{s}=205$GeV and $M_{Z'}=210$GeV. For each $\lambda$ ranging from 0
  (plain curve) to 0.2 (smallest dashes), $\epsilon$ is limited to
  satisfy today's $W$ mass accuracy, 
$|\delta M_W|<160$MeV (LEP2's $|\delta M_W|<45$MeV for the
  thick curves).}}:
The normal way of extracting the predictions of this model for
$W$-pair production would be to add all the amplitudes for $e^+e^-\rightarrow
W^+W^-$, namely the $t$-channel $\nu$ pole, and $s$-channel $\gamma$, $Z$ and
$Z'$ poles, including the contributions of $O_{B'W}$ in the latter. 
Alternatively, the correct angular dependence in \eeWW from such
a $Z^\prime$ is recovered through the introduction of ``process -- dependent" 
TGC form factors: 
the $Z^\prime $ exchange only contributes to the $J=1$ partial
wave and the TGC of Eq. (\ref{LeffWWV}) allow to 
parameterize the most general $J=1$
amplitude. For the case at hand one can always find TGC
$(g^Z_1, \kappa^Z, g^\gamma _1, \kappa^\gamma)$ 
matching the $Z^\prime $ parameter dependence 
{\em in this particular ee-WW
  channel}, but these TGC will depend on the
incoming electron's coupling to the Z and the photon.
\endtext
In general, a non-zero $\Delta g^\gamma _1$ 
is needed to
match the precise $t$-dependence, but in such a process-dependent
approach, this
does not imply any violation of charge conservation. Finally one should note
that the $Z^\prime$ described above would also appear in $e^-e^+\to q\bar q,\;
\ell^-\ell^+$ at LEP2 and thus all channels need to be searched for NP effects.
%
%
%
\section{The $W$ Pair Production Process}\label{TGCWpairpheno}
\subsection{Phenomenology of On-shell $WW$ Production}
Deviations of the TGC's from their SM, tree level form are most directly
observed in vector boson pair production. At LEP2 this is the process
$e^-e^+\to W^-W^+$, which, to lowest order, proceeds via the Feynman graphs
of Fig.~\ref{figww}. We start by describing the
core process, including the $W$ decay into fermion anti-fermion pairs
in the zero-width approximation, since most of the effects of anomalous
couplings can already be understood at this level. A full simulation of
the signal will, of course, need refinements
such as finite width effects, the ensuing contributions from final state
radiation graphs and the inclusion of $t$-channel vector boson exchange graphs
for specific final states such as $e^-\bar\nu u\bar d$.
The simulation of this
full $e^+e^-\to 4$~fermions process will be discussed later.
\begin{figure}[htb]
\epsfxsize=4.5in
\epsfysize=1.0in
\begin{center}
\hspace*{0in}
\epsffile{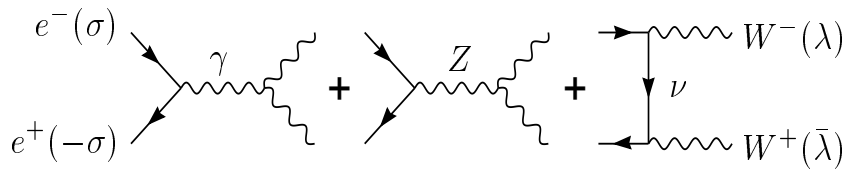}
\vspace*{0.1in}
\caption{\label{figww} Feynman graphs for the process $e^+e^-\to W^+W^-$.}
\vspace*{-0.1in}
\end{center}
\end{figure}
It is instructive to consider first the
individual contributions of $s$-channel photon and $Z$ exchange and of
$t$-channel neutrino exchange to the various helicity amplitudes
for the process $e^-e^+\to W^-W^+$~\cite{Haetal87},
\begin{equation}
{\cal M}(\sigma,\lambda,\bar\lambda) = {\cal M} =
{\cal M}_\gamma + {\cal M}_Z + {\cal M}_\nu \; .
\end{equation}
Here the $e^-$ and $e^+$ helicities are given by $\sigma/2$ and $-\sigma/2$,
and $\lambda$ and $\bar \lambda$ denote the $W^-$ and $W^+$ helicities.
Let us define reduced amplitudes $\tilde{\cal M}$
by splitting off the leading angular dependence in terms of the
$d$-functions~\cite{dfunctionPDG}
$d^{J_0}$ where $J_0= 1,2$ denotes the lowest angular momentum
contributing to a given helicity combination. In the c.m. frame, with the $e^-$
momentum along the $z$-axis and the $W^-$ transverse momentum pointing along
the $x$-axis, the helicity amplitudes are given by\footnote{As compared 
to Ref.~\cite{Haetal87} a phase factor $(-1)^{\bar\lambda}$ is absorbed into 
the definition of the $W^+$ polarization vector.}
\begin{equation}
{\cal M}(\sigma,\lambda,\bar\lambda;\theta) =
\sqrt{2}\;\sigma \; e^2\;
\tilde{\cal M}_{\sigma,\lambda,\bar\lambda}(\theta)\;
d^{J_0}_{\sigma,\lambda-\bar\lambda}(\theta)\; .
\label{Mdf}
\end{equation}
For $(\lambda,\bar\lambda)=(\pm,\mp)$, i.e. $|\lambda-\bar\lambda |=2$,
only $t$-channel neutrino exchange contributes and the incoming electron must
be lefthanded. The corresponding amplitudes are given by
\bq
{\cal M}(-1,\lambda,\bar\lambda=-\lambda;\theta)
= -\sqrt{2}e^2\;\;
{-\sqrt{2}\over {\rm sin}^2\theta_W}\;
{1\over 1+\beta^2-2\beta\,{\rm cos}\,\theta } \;\;
\lambda\; {\rm sin}\theta\; (1-\lambda\; {\rm cos}\theta)/2 \; .
\eq
$s$-channel photon and $Z$ exchange is possible only for
$|\lambda-\bar\lambda| = 0,1$. The corresponding reduced amplitudes can
be written as
\begin{eqnarray}
\tilde{\cal M}_\gamma & = &
-\beta A^\gamma_{\lambda\bar\lambda} \; , \nonumber \\
\tilde{\cal M}_Z & = &
+\beta A^Z_{\lambda\bar\lambda} \left[
1-\delta_{\sigma,-1}{1\over 2\, {\rm sin}^2 \theta_W }
\right] {s \over s-m_Z^2}\; , \nonumber \\
\tilde{\cal M}_\nu & = &
+\delta_{\sigma,-1}{1\over 2\beta\, {\rm sin}^2 \theta_W }
\left[B_{\lambda\bar\lambda} -
{1\over 1+\beta^2-2\beta {\rm cos} \theta} C_{\lambda\bar\lambda} \right]
\; . \label{Mreduced}
\end{eqnarray}
Here $s$ denotes the $e^+e^-$ center of mass energy and
$\beta = \sqrt{1-4m_W^2/s}$ is the $W^\pm$ velocity. The subamplitudes
$A^V$, $B$ and $C$ are given in Table~\ref{tableamp}.
%
\begin{table}[htb]
\caption{
Subamplitudes for $J_0=1$ helicity combinations of the process
$e^-e^+\to W^-W^+$, as defined in Eq.~(\protect\ref{Mreduced}). $\beta$
denotes the $W$ velocity and $\gamma = \protect\sqrt{s}/2m_W$. The
abbreviation $f_3^V = g_1^V+\kappa_V+\lambda_V$ is used.
}
\label{tableamp}
\vspace*{0.2in}
\arraycolsep=2.5em
\begin{tabular}{ccccc}
\vspace*{0.05in}
{$\lambda\bar\lambda$}& $A^V_{\lambda\bar\lambda}$ &
   $B_{\lambda\bar\lambda}$ & $C_{\lambda\bar\lambda}$ &
   $d^{J_0}_{\sigma,\lambda-\bar\lambda}$ \\
\hline
$++$ & $g_1^V +2\gamma^2\lambda_V +
{i\over\beta}(\tilde\kappa_V+\tilde\lambda_V-2\gamma^2\tilde\lambda_V)$
& 1 & $1/ \gamma^2$ & $-\sigma\, {\rm sin}\,\theta\, /\sqrt{2}$ \\
$--$ & $g_1^V +2\gamma^2\lambda_V -
{i\over\beta}(\tilde\kappa_V+\tilde\lambda_V-2\gamma^2\tilde\lambda_V)$
& 1 & ${1/ \gamma^2}$ & $-\sigma\, {\rm sin}\,\theta\, /\sqrt{2}$ \\
$+0$ & $\gamma (f_3^V -ig_4^V+\beta g_5^V +
{i\over \beta}(\tilde\kappa_V-\tilde\lambda_V) )$
& $2\gamma$ & $2(1+\beta)/\gamma$ & $(1+\sigma\,{\rm cos}\,\theta)\,/2$ \\
$0-$ & $\gamma (f_3^V +ig_4^V+\beta g_5^V -
{i\over \beta}(\tilde\kappa_V-\tilde\lambda_V) )$
& $2\gamma$ & $2(1+\beta)/\gamma$ & $(1+\sigma\,{\rm cos}\,\theta)\,/2$ \\
$0+$ & $\gamma (f_3^V +ig_4^V-\beta g_5^V +
{i\over \beta}(\tilde\kappa_V-\tilde\lambda_V) )$
& $2\gamma$ & $2(1-\beta)/\gamma$ & $(1-\sigma\,{\rm cos}\,\theta)\,/2$ \\
$-0$ & $\gamma (f_3^V -ig_4^V-\beta g_5^V -
{i\over \beta}(\tilde\kappa_V-\tilde\lambda_V) )$
& $2\gamma$ & $2(1-\beta)/\gamma$ & $(1-\sigma\,{\rm cos}\,\theta)\,/2$ \\
$00$ & $g_1^V +2\gamma^2\kappa_V$
& $2\gamma^2$ & ${2/ \gamma^2}$ & $-\sigma\, {\rm sin}\,\theta\, /\sqrt{2}$ \\
\end{tabular}
\end{table}

One of the most striking features of the SM are the gauge theory cancellations
between $\gamma$, $Z$ and neutrino exchange graphs at high energies.
Within the SM the only non-vanishing couplings in the table are $g_1 = \kappa
=1$ and $f_3=2$ for both the photon and
the $Z$-exchange graphs. As a result $A^\gamma_{\lambda\bar\lambda}=
A^Z_{\lambda\bar\lambda}$ and the $\beta A^V$ terms in Eq.~(\ref{Mreduced})
cancel, except for the difference between photon and $Z$ propagators.
Similarly, the $B_{\lambda\bar\lambda}$ term in $\tilde{\cal M}_\nu$ and the
$\delta_{\sigma,-1}$ term in $\tilde{\cal M}_Z$ cancel in the high energy
limit for all helicity combinations. While the contributions from individual
Feynman graphs grow with energy for longitudinally polarized $W$'s, this
unacceptable high energy behavior is avoided in the full amplitude due to the
cancellations which can be traced to the gauge theory relations between
fermion--gauge boson vertices and the TGC's.
 
LEP2 will operate 
close to $W$ pair production threshold and these cancellations
are not yet fully operative. For example, at $\sqrt{s}=190$~GeV one has
$\beta=0.54$, $\beta\;s/(s-m_Z^2)=0.70$, and $1/\beta=1.87$ instead of unity.
As a result, the linear combinations of couplings which enter in
$\tilde{\cal M}_\gamma$ and $\tilde{\cal M}_Z$ are quite different from their
asymptotic forms. In particular the $\gamma^2$ enhancement factors
are still small, the $(\pm,\pm)$ and $(0,0)$ amplitudes are not yet dominated
by individual couplings, and interference effects between different TGC
are very important.
 
Table~\ref{tableamp} shows that only seven $W^-W^+$ helicity combinations
contribute to the $J_0 =1$ channel and the various $WWV$ couplings enter in as
many different combinations. This explains why exactly
seven form factors or coupling constants are needed to parameterize the
most general $WWV$ vertex. Since we have both $WWZ$ and $WW\gamma$ couplings
at our disposal, the most general $J=1$ amplitudes
${\cal M}_L={\cal M}(\sigma=-1,\lambda,\bar\lambda)$ and
${\cal M}_R={\cal M}(\sigma=+1,\lambda,\bar\lambda)$ for both
left- and right-handed incoming electrons can be parameterized. Turning the
argument around one concludes that
all 14 helicity amplitudes need to be measured independently for a complete
determination of the most general $WW\gamma$ and $WWZ$ vertex.
 
A first step in this direction is the measurement of the angular distribution
of produced $W$'s, $d\sigma/d\;{\rm cos}\;\theta$. In terms of the reduced
amplitudes $\tilde{\cal M}_{\sigma,\lambda,\bar\lambda}$ of (\ref{Mdf}) this
distribution is given by
\begin{eqnarray}
{d\sigma\over d\;{\rm cos}\;\theta} = {\pi\alpha^2\beta\over 4s} \Bigl\{
\;\;\sum_{\sigma=\pm 1}\;\; \Bigl[ && {{\rm sin}^2\theta\over 2}
\left( |\tilde {\cal M}_{\sigma,++}|^2+|\tilde {\cal M}_{\sigma,--}|^2
+|\tilde {\cal M}_{\sigma,00}|^2 \right)  \nonumber \\
&& + {(1+\sigma\, {\rm cos}\, \theta)^2\over 4}
\left( |\tilde {\cal M}_{\sigma,+0}|^2+|\tilde {\cal M}_{\sigma,0-}|^2 \right)
\nonumber \\
&& + {(1-\sigma\, {\rm cos}\, \theta)^2\over 4}
\left( |\tilde {\cal M}_{\sigma,0+}|^2+|\tilde {\cal M}_{\sigma,-0}|^2 \right)
\;\;\Bigr] \nonumber \\
&+& {1\over 2}(1+{\rm cos}^2\theta)\;{\rm sin}^2\theta \;\;
{2\over {\rm sin}^4\theta_W}
{1\over (1+\beta^2-2\beta\,{\rm cos}\,\theta )^2}\;
\Bigr\}
\; . \label{eq:prodangle}
\end{eqnarray}
Due to the different $d$-function factors amplitudes with different values of
$\lambda-\bar\lambda$ can be
separated in principle. In practice, the additional $\theta$-dependence of
the neutrino exchange graphs (the $C_{\lambda\bar\lambda}$ terms in
Eq.~(\ref{Mreduced})) distorts these angular distributions and
leads to contributions from the individual $W^-W^+$ helicity combinations as
shown in Fig.~\ref{figsighel}. In fact, the interference with the
$\nu$-exchange graphs can be used to further separate the various $s$-channel
helicity amplitudes.

\begin{figure}[htb]
\mbox{\epsfig{file=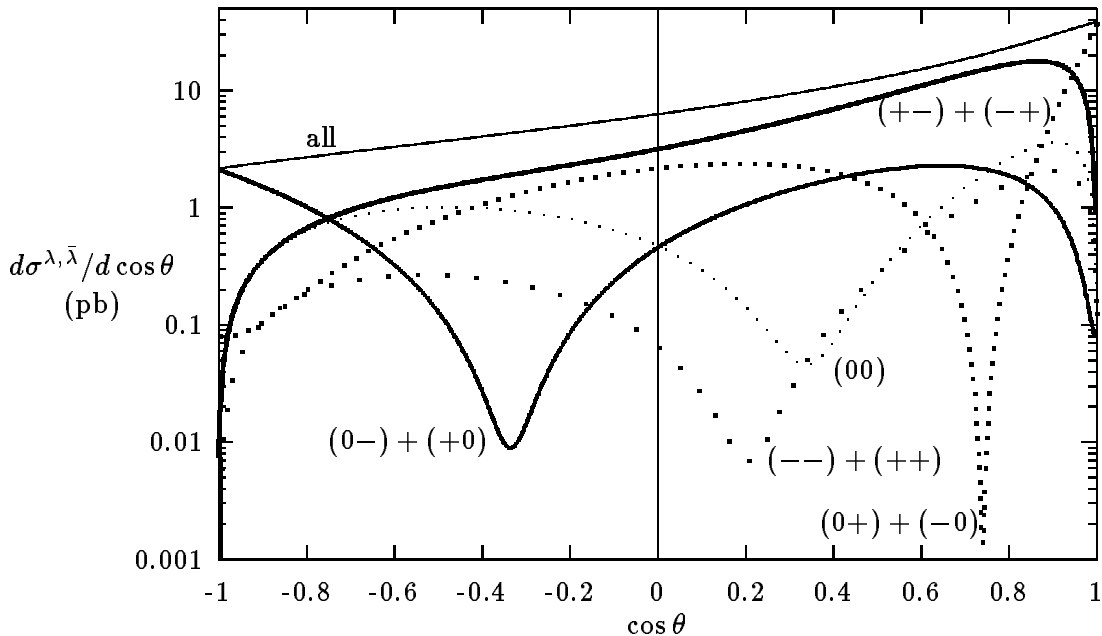,height=8cm,width=10cm,
bbllx=100,bblly=500,
bburx=350,bbury=700}}
\caption{\label{figsighel}
Angular distributions $d\sigma/d\; {\rm cos}\; \theta$ for
$e^-e^+\to W^-W^+$: SM contributions from fixed $W^-W^+$ helicities
$(\lambda\bar\lambda)$ at $\protect\sqrt{s} = 190$~GeV.
}
\end{figure}

Due to the $V-A$ structure of the $W$--fermion vertices the decay angular
distributions of the $W$'s are excellent polarization analyzers and a further
separation of the various $W^+W^-$ helicities can be
obtained~\cite{Haetal87,Bietal93}. These decay distributions are most 
easily given
in the rest frame of the parent $W$. Choose the $e^-e^+\to W^-W^+$ scattering
plane as the $x-z$ plane with the $z$-axis along the $W^-$ direction and
obtain the $W^\pm$ rest frames by boosting along the z-axis. In the $W^-$ frame
we define the momentum of the decay fermion for $W^-\to f_1\bar f_2$ as
\bq
\label{eq:theo-angdef1}
p_1^\mu = {m_W\over 2}\; (1,\;{\rm sin}\theta_1\;{\rm cos}\phi_1,\;
{\rm sin}\theta_1\;{\rm sin}\phi_1,\;{\rm cos}\theta_1)\; ,
\eq
and, similarly, for $W^+\to f_3\bar f_4$, the anti-fermion momentum in the
$W^+$ frame is given by
\bq
\label{eq:theo-angdef2}
p_4^\mu = {m_W\over 2}\; (1,\;{\rm sin}\theta_2\;{\rm cos}\phi_2,\;
-{\rm sin}\theta_2\;{\rm sin}\phi_2,\;-{\rm cos}\theta_2)\; ,
\eq
Thus, $\theta_i=0$ corresponds to the charged lepton or the down-type
(anti)quark being emitted in the direction of the parent $W^\pm$.
 
Neglecting any fermion masses, the $W^-\to \ell^-\bar\nu$ decay amplitude
is given by~\cite{Haetal87}
\bq
{\cal M}_D(\lambda) = {e\;m_W\over \sqrt{2}{\rm sin}\theta_W }\;\;
\ell_\lambda(\theta_1,\phi_1)\;,
\eq
where the angular dependence is contained in the functions
\bq
(\ell_-,\ell_0,\ell_+)(\theta_1,\phi_1) = \left(
{1\over \sqrt{2}}(1+{\rm cos}\theta_1)\;{\rm e}^{-i\phi_1},\;
-{\rm sin}\theta_1,\;
{1\over \sqrt{2}}(1-{\rm cos}\theta_1)\;{\rm e}^{i\phi_1}\right)\;.
\eq
An analogous expression is obtained for the $W^+$ decay amplitude.
 
The production and decay amplitudes can easily be combined to obtain the
five-fold differential angular distribution for the process
$e^-e^+\to W^-W^+\to f_1\bar f_2\; f_3\bar f_4$, in the narrow $W$-width
approximation~\cite{Haetal87,Bietal93},
\ba
\label{eq:dsig5}
{ d^5 \sigma\; (e^-e^+\to W^-W^+\to f_1\bar f_2\; f_3\bar f_4) \over
d\;{\rm cos}\;\theta\;\;
d\;{\rm cos}\;\theta_1\;d\;\phi_1\;\;
d\;{\rm cos}\;\theta_2\;d\;\phi_2 } & = &
{\beta\over 128\pi s}\left({3\over 8\pi}\right)^2
B(W\to f_1\bar f_2)\; B(W\to f_3\bar f_4)\;        \nonumber \\
&& \times
\sum_{\sigma,\lambda,\bar\lambda,\lambda',\bar\lambda'}
{\cal M}(\sigma,\lambda,\bar\lambda){\cal M}^*(\sigma,\lambda',\bar\lambda')
\nonumber \\
&& \times \;\;\;
D_{\lambda,\lambda'}(\theta_1,\phi_1)\;
D_{\bar\lambda,\bar\lambda'}(\pi-\theta_2,\phi_2+\pi)\; .
\ea
Here the production amplitudes ${\cal M}(\sigma,\lambda,\bar\lambda)$ are
given in Eq.~(\ref{Mdf}) and the $D_{\lambda,\lambda'}$ are given
by
\bq
D_{\lambda,\lambda'}(\theta,\phi) = \ell_\lambda (\theta,\phi)\;
\ell_{\lambda'}^* (\theta,\phi)\; .
\eq
The information contained in the five-fold differential distribution
(\ref{eq:dsig5}) can be used to isolate different linear combinations of
$WWV$ couplings and hence reduce the possibility of cancellations between
them. For example, by isolating $W^+W^-$ pairs
which are both transversely polarized (and hence give $1+{\rm cos}^2\theta_i$
decay distributions) the combinations $g_1^V+2\gamma^2\lambda_V$ are
determined which appear in the production amplitudes ${\cal M}_{++}$ and
${\cal M}_{--}$ (see Table~\ref{tableamp}). Similarly, longitudinal $W$'s
produce a characteristic ${\rm sin}^2\theta_i$ decay distribution. The
isolation of LT+TL and of LL polarizations of the two $W$'s allows 
independent measurements of the combinations $f_3^V = g_1^V+\kappa_V+\lambda_V$
and $g_1^V + 2\gamma^2\kappa_V$, respectively,  and thus 
the three $C$- and $P$-conserving anomalous couplings\footnote{Note however
that if relations among TGC such as those in eq.  
(\ref{eq:alphas}) are relaxed, it will not be easy to
distinguish \kg\ from \kz\ (or \lm\ from \lz)
with unpolarized beams, since these both feed the same
helicity amplitudes in table~\ref{tableamp}.} 
may be isolated.  
 
Additional information is obtained from the azimuthal angle distributions of
the decay products. A nontrivial azimuthal angle dependence arises from the
interference between helicity amplitudes for different $W^+$ or different
$W^-$ polarizations. The large ${\cal M}_{+-}$ and ${\cal M}_{-+}$ amplitudes,
which arise solely from neutrino exchange, can thus be put to use:
interference with these large amplitudes can amplify the effects of anomalous
couplings.
 
The observation of azimuthal angular dependence and 
correlations is particularly important for the study of $CP$-violating effects
in $W^-W^+$ production~\cite{Haetal87,CPWW}. 
The methods suggested in section \ref{sec:TGC-Techniques} below
for TGC determination from data can all be used for this purpose, and the
reader is referred to the literature for details of procedures using
density matrix~\cite{CPWW} and optimal observable~\cite{DiNa94} analyses. 
Similarly, the study of rescattering effects between the produced $W$ pairs,
i.e. the presence of nontrivial phases in the production amplitudes, relies
on the interference with the phase factors introduced by the azimuthal angle
dependence of the decay amplitudes. We do not explicitly discuss these
techniques here but rather refer to the literature~\cite{Haetal87,rescatterWW}.
 
\begin{table}[htb]
\center{
\begin{tabular}{c|c|lll}
\hline\hline
WW decay channel
         & Decay fraction & \mco{3}{c}{Available angular information} \\ \hline
   \jjlv & $l=e$: 14\%    &  \ctw     & (\ctl, \phil) & (\ctj, \phij)\fold
                                                                      \\
         & $l=\mu$: 14\%  &           &               &               \\
         & $l=\tau$: 14\% &           &               &               \\ \hline
   \jjjj &    49\%        & $\mid\ctw\mid$
                                      & (\ctja, \phija)\fold
                                                      & (\ctjb, \phijb)\fold
                                                                      \\ \hline
   \lvlv &     9\%        &  \ctw     & (\cta, \phia) & (\ctb, \phib) \\
         &                &  \mco{3}{c}{2 solutions}                  \\ \hline
                                                                         \hline
\end{tabular}
\caption{Availability of angular information in different WW final states. The
         production angle is denoted by $\theta$ and
         $(\theta_{l,j},\phi_{l,j})$ denote decay angles for W $\rightarrow$
         (leptons, jets) respectively. $(\ctj, \phij)\fold$ implies the
         ambiguity $\ctj \leftrightarrow -\ctj$,
         $\phij \leftrightarrow \phij + \pi$ incurred by the inability to
         distinguish quark from antiquark jets.}
\label{tab:anginf}
}
\end{table}

The application of (\ref{eq:dsig5}) to experimental data must take account of
some restrictions in the ability to determine the angles involved: in the case
of hadronic \W\ decays, 
and in the absence of any quark charge or flavour tagging
procedure, the fermion and anti-fermion cannot be distinguished; also, in
the case where both \Ws decay leptonically, a quadratic ambiguity is
encountered.
The ambiguities in each of the three \WW\
 final states \jjlv, \jjjj\ and \lvlv,
where $j$ represents the jet fragmentation of a quark or antiquark and $(l\nu)$
the products of W decay into lepton-antilepton, are summarized in
table~\ref{tab:anginf}. 
%
%
%
\newcommand{\nl}{\nonumber \\}
\newcommand{\lb}{\linebreak}
\newcommand{\fig}[1]{Fig.\ref{#1}}
\newcommand{\bc}{\begin{center}}
\newcommand{\ec}{\end{center}}
\renewcommand{\theenumiii}{\arabic{enumiii}}
\newcommand{\bqa}{\begin{eqnarray}}
\newcommand{\eqa}{\end{eqnarray}}
\newcommand{\ben}{\begin{enumerate}}
\newcommand{\een}{\end{enumerate}}
\renewcommand{\theenumii}{ \arabic{enumii} }
\newcommand{\eqn}[1]{Eq.(\ref{#1})}
\newcommand{\itema}{\item\addtocounter{cit}{1}}
\newcommand{\itemb}{\item\addtocounter{cit}{1}\addtocounter{cita}{1}}
\def\a2n{${\cal A}(2\nolinebreak \to n)$}
\def\an{${\cal A}(1\to n)$}
\def\cgp{C.G.~Pa\-pa\-do\-pou\-los }
\def\sdv{S.D.P.~Vlas\-so\-pu\-los }
\def\ena{E.N.~Ar\-gy\-res }
\def\rk{R.H.P.~Kleiss}
\def\nga{N.G.~Anto\-ni\-ou}
\def\an{{non-standard TGC}}
\def\ano{{non-standard}}
\def\el{{$e^-\bar{\nu}_e u \bar{d}$}}
\def\mul{{$\mu^-\bar{\nu}_{\mu} u \bar{d}$}}
\def\ert{{\tt ERATO}}
\def\exc{{\tt EXCALIBUR}}
\def\exca{{\tt EXCALIBUR1}}
\def\pb{\parbox}
\def\va{\vspace*{-4pt}}
\subsection{Four-fermion production and {\an}
}\label{TGC4f}
Most studies of TGC so far have been made with zero width simulated
data and with an analysis program based on the same assumptions.
This procedure might neglect some important effects, however,
 and the corresponding
physics issues will be discussed in this subsection. These are the influence
of a finite W-width, of background diagrams, i.e. graphs other than
the three W-pair diagrams of Fig.~\ref{figww}, and the influence
of radiative corrections (RC) in particular the dominant QED
initial state radiation (ISR).
 
At the moment there are many Monte Carlo (MC) programs for four fermion
production, but only two of them can at present study the above issues,
namely {\ert}\cite{ert0} and {\exc}\cite{exc0,exc1}.
For a detailed description we refer to the WW event
generator report, but we make a few comments here.
Although the programs can study {\an} effects\cite{ert0,exc-an} for all
the channels of Table \ref{tab:anginf},
we will only consider the $jj\ell\nu$ case in the following. More specifically
we will study {\el} or {\mul} final states.
The amplitude for these final states consists of 20 and 10 diagrams,
respectively, of which 3 are the W-pair diagrams of Fig.~\ref{figww}.
Since the four fermions
are assumed to be massless in the calculations, cuts have to be applied to
avoid singularities in the phase space. Experimental cuts usually have this
effect as well. In the case of only three diagrams such cuts are not
required.
ISR is incorporated following the prescription of Ref.~\cite{exc-isr}.
\begin{table}[htb]
\bc\begin{tabular}{|c|c|c|}
\hline
Standard Model & {\an} & physical assumptions
\\ \hline
 $\sigma_{SM,on}$ & $\sigma_{AN,on}$ & $\Gamma_W=0$
\\ \hline
 \pb[c]{3cm}{\bc \va $\sigma_{SM,off}$ \\$\sigma_{SM,off,cuts}$ \va \ec} &
 \pb[c]{3cm}{\bc \va $\sigma_{AN,off}$ \\$\sigma_{AN,off,cuts}$ \va \ec} &
 3 diagrams
\\ \hline
 $\sigma_{SM,all}$ & $\sigma_{AN,all}$ & 20 diagrams, cuts
\\ \hline
 $\sigma_{SM,ISR}$ & $\sigma_{AN,ISR}$ &
 3 diagrams, ISR
\\ \hline
 $\sigma_{SM,all,ISR}$ & $\sigma_{AN,all,ISR}$ & 20 diagrams, cuts, ISR
\\ \hline
\end{tabular}
\caption[.]{Cross sections and the corresponding physical assumptions
under which they have been calculated. The subscripts $SM,\; AN,\; on, \; off$
refer to Standard Model, {\an}, on-shell and off-shell, respectively.}
\label{EXC-tab1}
\vspace*{-.2in}
\ec\end{table}
In table \ref{EXC-tab1}, we list a number of differential cross-sections which
have been calculated, and correspond to 
different physical asumptions. The first column refers to
the SM and the second one to a {\an} case (usually with only one of the 
CP-conserving couplings being different from its SM value).
For the cross-sections labeled $\sigma_{cuts}$, cuts are applied
mainly to lepton and quark energies and angles in the laboratory frame:
\bq
E_{e^-,u,\bar{d}} > 20\, \mbox{GeV}\;, \qquad
|\cos\theta_{e^-,u,\bar{d}}| <  0.9\;, \qquad
 \cos\theta_{u-\bar{d}}  <  0.9\;, \qquad
 m_{u\bar{d}} > 10\, \mbox{GeV} \; .
\eq
The calculations were performed with input parameters as prescribed in the
WW cross-section Working Group chapter.
Results from the two programs agree within the MC errors.
The particular case of $d\sigma_{AN,off}/d\cos\theta$
(for the full phase space) has also been calculated by M. Bilenky
in a semi-analytical method and full agreement with {\exc} has been
obtained for all CP conserving TGC.

\begin{figure}[htb]
\epsfxsize=6.0in
\epsfysize=2.0in
\begin{center}
\hspace*{0in}
\epsffile{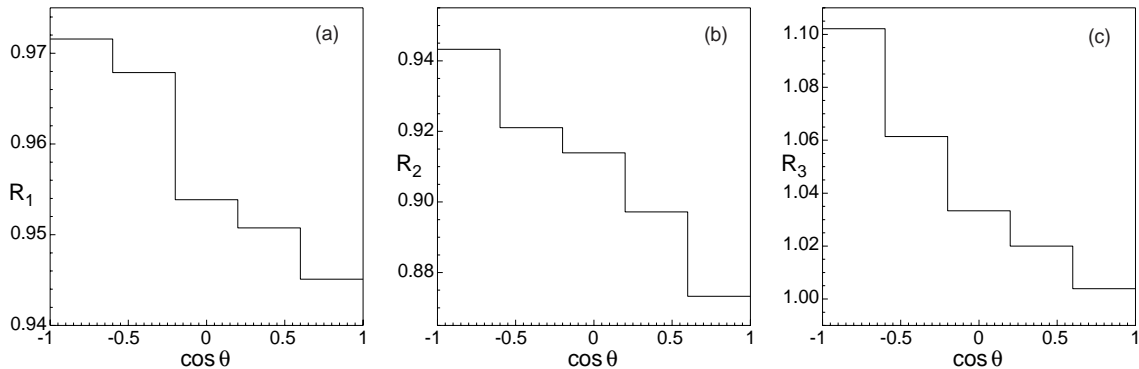}
\vspace*{0.1in}
\caption{\label{EXC-fig2}
Ratios of differential cross-sections
at various levels of the simulation of the 4-fermion processes,
(a) $\mbox{R}_1=\sigma_{SM,off}/\sigma_{SM,on}$,
(b) $\mbox{R}_2=\sigma_{SM,ISR}/\sigma_{SM,off}$
and (c) $\mbox{R}_3=\sigma_{SM,all}/\sigma_{SM,off,cuts}$.}
\end{center}
\end{figure}

Different physical mechanisms could influence the angular distribution
of the produced $W$s and thus simulate the effect of {\an}.
Typical examples are shown in Fig.~\ref{EXC-fig2},
namely the effect of a finite $W$ width, of ISR and of background graphs
on $d\sigma/d\cos\theta$. ISR, for instance, lowers the available $\sqrt{s}$
of the event and thus reduces the forward peak of the $W^-W^+$ production
cross-section. In addition, the recoil of the $W^-W^+$ system against the
emitted photon further smears out the $W$ angular
distribution~\cite{BeDe94}. A similar
effect, relative depletion of forward as compared to
backward produced $W^-$s can also arise from negative TGC parameters.
This is evident from Fig.~\ref{EXC-fig3}, where
ratios of a {\ano} $d\sigma/d\cos\theta$ and
SM cross-sections are presented, both having been calculated under
the same physical assumptions. 
Fig.~\ref{EXC-fig3}(b) demonstrates the quantitative importance
of this phenomenon.  For final state electrons the background
graphs, if not included in the analysis,
could mimic a $\delta_Z$ of the order of $-0.2$.
While the shape of the angular distribution $d\sigma/d\cos\theta$ for
negative TGC parameters shows a trend
similar to that induced by ISR, finite width or background graph effects,
the normalization of the cross-section might provide some
discriminating power, as do the decay angular distributions.
Another very important message coming from Fig.~\ref{EXC-fig3} is that
the sensitivity to the TGC remains the same at the different levels of the
simulation (from on-shell $W$s up to four-fermion production).
Conversely, the influence of the various physics effects on
production and decay angular distributions is largely independent of whether
or not {\an} are present.
 
We conclude that it is clearly important to account for and to correct the
effects considered above in experimental analyses. We return to the effects of
ISR and finite W width in Section {5.2} where their neglect in TGC
determination at LEP2 is quantified. In Section {6.2} we indicate how they
contribute to the overall bias in a typical simulated TGC determination.
%
\begin{figure}[hbt]
\epsfxsize=5.5in
\epsfysize=3.0in
\begin{center}
\vspace{-5pt}
\hspace*{0in}
\epsffile{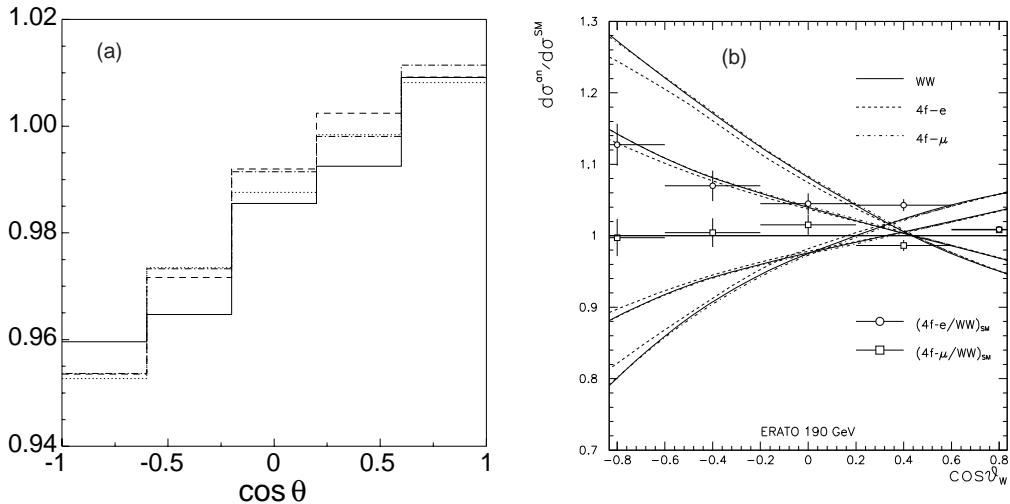}
\vspace*{0.1in}
\caption{\label{EXC-fig3}
Ratio of anomalous to SM differential cross-section. (a)
$\sigma_{AN,off}/\sigma_{SM,off}$ (solid line),
$\sigma_{AN,ISR}/\sigma_{SM,ISR}$ (dotted line),
$\sigma_{AN,all}/\sigma_{SM,all}$ (dashed line), and
$\sigma_{AN,all,ISR}/\sigma_{SM,all,ISR}$ (dash-dotted line) for
$y_\gamma=+0.1$. (b)
$\sigma_{AN,off}/\sigma_{SM,off}$ (solid line),
$\sigma_{AN,all}/\sigma_{SM,all}$ for muons (dash-dotted line) and electrons
(dashed line) for $\alpha_W=0.2,\delta_Z=0.2,\delta_Z=-0.2,\alpha_W=-0.2$
(bottom-top) and $\sigma_{SM,all}/\sigma_{SM,off}$ for muons
(squares) and electrons (circles).}
\end{center}
\end{figure}
%
%
%
%
%
%
%
\newcommand{\note}[1]{\footnote{\tt #1}}  
\newcommand{\eqpt}{\; .}              
\newcommand{\eqcm}{\; ,}
\newcommand{\ra}{/}
\newcommand{\gw}{\mbox{$\Gamma_W$}}
\newcommand{\SM}{Standard Model }
\newcommand{\MC}{Monte Carlo }
\newcommand{\goesto}{\rightarrow}
\newcommand{\order}{\cal}
\newcommand{\ecm}{E_{cm}}
\newcommand {\qqlnu}{q\bar{q'}l\bar{\nu}}
%
\def\wha{$W^+W^-~\rightarrow~q_1\overline{q}_2q_3\overline{q}_4$}
\def\wle{$W^+W^-~\rightarrow~q_1\overline{q}_2 \ell\nu$}
%
\def\diff#1#2          {\mbox{$\frac{{\mathrm d} #1}{{\mathrm d} #2}       $}} 
\section{Statistical techniques for TGC determination\protect\footnote{The
experimental sections, \ref{sec:TGC-Techniques}--\ref{sec:TGC-lvlv}, have
been coordinated by R.~L.~Sekulin}}
\label{sec:TGC-Techniques}
Three different methods have thus far been proposed for the determination of
TGCs at LEP2, --- the density matrix method, the maximum  likelihood method and
the method of optimal observables. These methods are outlined in the following
subsections and their application to common simulated datasets is compared. In
devising these  methods, two considerations have been borne in mind: first, ---
as will be elaborated in the next section --- that it is advantageous to use as
much of the available angular data for each \WW\ event as possible; second, that
the  expected LEP2 data (a total of $\approx 8000$ events for an integrated
luminosity of $500 {\mathrm pb}^{-1}$ at 190 GeV) will not be sufficient, for
instance, to bin the data into the five angular variables appearing in the \WW\
production and decay distribution (\ref{eq:dsig5}) and subsequently to perform a
$\chi^2$ fit. The studies reported in this section have been performed assuming
that the final state momenta of the four partons from \Wm\ and \Wp\ decay have
been successfully reconstructed from the data; the practical difficulties of
doing this are discussed in  section~\ref{sec:TGC-gen}.

\subsection{Density matrix method}
\label{sec:TGC-DM}

In this method, TGC parameters are extracted from the data in a two-stage
analysis.
First, experimental density matrix elements and their statistical
errors are determined from the angular distribution (\ref{eq:dsig5}) in
bins of \ctw; then the predictions of different theoretical models are
fitted to the resulting distributions using a $\chi^{2}$ minimization method. 
The joint \WW\ helicity density matrix elements $\rho_{\lambda \bar{\lambda}
\lambda^{'} \bar{\lambda}^{'}}$ are defined from (\ref{eq:dsig5}) as the sums 
$\sum_{\sigma} {\cal M} (\sigma, \lambda, \bar{\lambda})
               {\cal M}^{*} (\sigma, \lambda^{'}, \bar{\lambda}^{'})$ 
of bilinear products of production amplitudes and the dependence of the
cross-section on the TGC parameters is fully contained in the complete density
matrix thus evaluated. Similarly, by integrating over the observables of one $W
$, single \W\ density matrix elements  $\rho_{\lambda \lambda^{'}}$ and 
$\rho_{\bar{\lambda} \bar{\lambda}^{'}}$ can be defined.

The density matrix elements can be calculated in two ways:
\vspace{-.1in}
\begin{itemize}
\item[-] Using the orthogonality properties of the \W\ decay functions 
      $D_{\lambda \lambda^{'}}$ and $D_{\bar{\lambda} \bar{\lambda}^{'}}$
      in (\ref{eq:dsig5}), density matrix elements can be
      extracted by integrating over the \W\ decay angles with suitable
      projection operators. Thus, unnormalized density matrix elements of the
      leptonically decaying \W\ in \jjlv\ events can be found from the lepton
      spectrum as
\bq
 \rho_{\lambda \lambda^{'}} 
                      \frac{{\mathrm d}\sigma(\eeWW)}{{\mathrm d}\!\cos\theta}
 = \frac{1}{B_{W l\nu}}
   \int{     
   \frac{{\mathrm d}\sigma(\eeWW \rightarrow \jjlv) }
       {{\mathrm d}\!\cos\theta\, {\mathrm d}\!\cos\theta_l\, {\mathrm d}\phi_l}
   \Lambda_{\lambda \lambda^{'}}(\theta_{l},\phi_{l})\,
                          {\mathrm d}\!\cos\theta_{l}\, {\mathrm d}\phi_{l}
   }
\label{eq:TGC-DM-1}
\eq

\noindent where $B_{W l\nu}$ is the branching ratio for the \jjlv\ channel, 
      the angular variables are as defined in (\ref{eq:theo-angdef1}), 
(\ref{eq:theo-angdef2}), with
      the decay angles and helicity indices now referring to the leptonically
      decaying \W.
      Expressions for the normalized  operators $\Lambda_{\lambda \lambda^{'}}$
      are given in~\cite{ref:TGC-Gounaris}; for example, $\Lambda_{00} =
      2-5\cos^2 \theta_l $ projects out the longitudinal cross-section
      $\rho_{00} \frac{{\mathrm d}\sigma}{{\mathrm d}\!\cos\theta }$ of the
      leptonically decaying \W.
     
\item[-] In the second method~\cite{Bietal93}, the production and decay
      angular distribution is expressed in terms of the density matrix elements
      and, in each bin of \ctw\, , they are determined using a maximum
      likelihood fit to the distribution of the decay angles.
\end{itemize}

\noindent Fig~\ref{fig:TGC-DM-1} shows some of the density matrix elements
calculated from a sample of simulated events by the two methods as a function of
\ctw\ and fitted to the prediction of the Standard Model
It can be seen that there is good agreement between the density matrix elements
as calculated by the two methods, and with the fit to the Standard Model.
\begin{figure}[htb]
\mbox{\epsfig
              {file=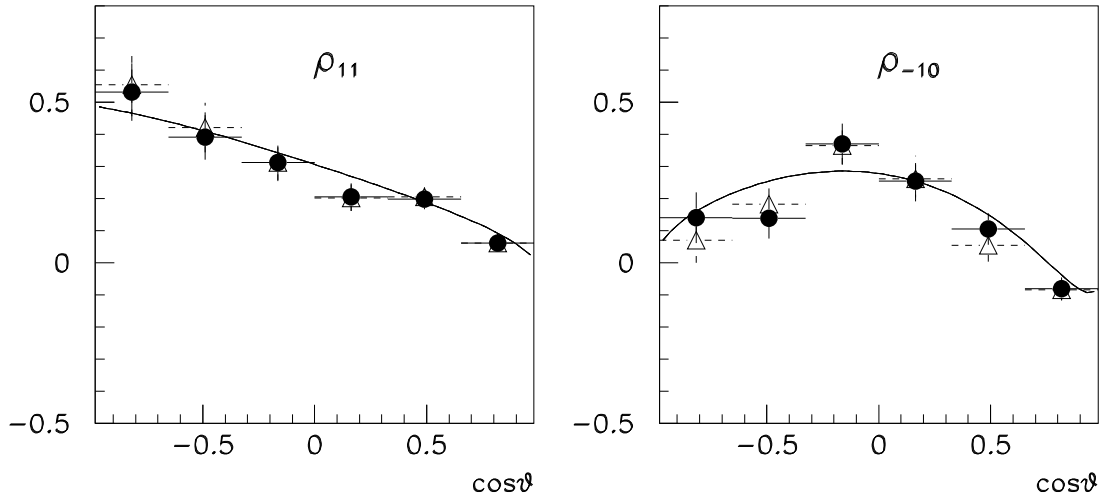,
              width=\textwidth}}
\caption{\ctw\ dependence of density matrix elements $\rho_{11}$ and 
        $\rho_{-10}$ for a sample of 2930 simulated \eeWW\ events at 190 GeV,
        calculated using the projection method (full circles) and the maximum
        likelihood method (triangles) and compared with the prediction of the
        Standard Model (fitted curve).}
\label{fig:TGC-DM-1}
\end{figure}
\subsection{Maximum likelihood method}
\label{sec:TGC-ML}

In this method, the distribution of some or all of the observed angular data is
used directly in an unbinned maximum likelihood fit~\cite{Se94}, in
which parameters ${\bf P}$, denoting one or more of the Lagrangian contributions
(\ref{eq:alphas}), are varied to maximize the quantity 

\bq
 \ln {\cal L}_{ML} = \sum_{i} \ln p({\bf \Omega}_i,{\bf P}) - 
              N_{obs} \ln \left\{\int p({\bf \Omega},{\bf P}) d\Omega \right\},
\label{eq:TGC-ML-1}
\eq

\noindent where the sum is over events in the sample, ${\bf \Omega_i}$
represents, for the i'th event, the angular information being used,
$p({\bf \Omega},{\bf P})$ is derived from the cross-section (\ref{eq:dsig5}),
$N_{obs}$ is the observed number of events, and the integral is over the whole
of phase space. Many of the results shown here have been obtained using the
method of extended maximum likelihood, in which the absolute prediction
for the magnitude of the cross-section is also tested~\cite{ref:TGC-EML}:

\bq
 \ln{\cal L}_{EML} = \sum_{i} \ln p({\bf \Omega}_i,{\bf P}) - N({\bf P}),
\label{eq:TGC-ML-2}
\eq

\noindent where, for integrated luminosity $L$, the predicted number of events 
$N({\bf P})$ in the sample is $ L \int \diff{\sigma}{{\bf \Omega}} ({\bf
\Omega},{\bf P}){\mathrm d}  {\bf \Omega}$.

It may be noted that, while in  the evaluation of $N({\bf P})$ in
(\ref{eq:TGC-ML-2}) the absolute normalization of the cross-section must be used
(as given in  (\ref{eq:dsig5})), constant factors such as the flux factor may be
omitted from the unnormalized expression $\int p({\bf \Omega},{\bf P}) d\Omega$
in  (\ref{eq:TGC-ML-1}). Furthermore, since for any event the probability $p$ is
proportional to the product of a phase space factor, which is independent of
${\bf P}$, and a matrix element squared, ${\mid{\cal M}\mid}^2$, which contains
the dependence on the TGC parameters, the sums over events in
(\ref{eq:TGC-ML-1}) and  (\ref{eq:TGC-ML-2}) may be replaced by  $ \sum_{i} \ln
{\mid{\cal M}\mid}^{2}({\bf \Omega}_i,{\bf P}) $, and the maximum of the
likelihood function will be unchanged. While this replacement is trivial for the
2-body cross-section given by  (\ref{eq:prodangle}), it is essential in the
evaluation of the log-likelihood sum when the reaction is analyzed in terms of 
the 4-fermion processes, in which the phase space factor is different for
every event.

While the maximum likelihood method is able to use all the available angular
information for each event,  it has the disadvantage compared with a $\chi^2$
fit of being unable to provide a goodness of fit criterion.  Nonetheless, the
goodness of fit of a hypothesis represented by the likelihood function 
${\cal L}_1({\bf p})$ can be compared with that of ${\cal L}_2({\bf P})$ if the
parameters ${\bf p}$ of ${\cal L}_1$ satisfy the condition  
${\bf p} \in {\bf P}$. Then the quantity $-2 \ln \left({\cal L}_{1}^{max} /
{\cal L}_{2}^{max}\right)$, derived from the ratio of their likelihood
functions, has a $\chi^2$ distribution~\cite{ref:TGC-Eadie}. This property has
been applied to event samples  generated with non-SM values of one TGC, $P_1$,
and used to distinguish this hypothesis from a wrong one, when a different TGC,
$P_2$, is fitted to the data. --- In general, a fit of $P_2$ produces a result
differing significantly from the SM value. Fig~\ref{fig:TGC-ML-1} shows the
results of applying this test to the correct and wrong models in two alternative
ways. In both cases, ${\cal L}_{1}$ is taken as the likelihood function when
$P_1$ varies; in the ``same family'' case (a), ${\cal L}_{2}$ is the likelihood
function when both $P_1$ and $P_2$ vary, while, in (b), ${\cal L}_{2}$ describes
a ``composite'' hypothesis,

\bq 
{\cal L}_2\left( P_1,P_2;\beta\right) =\prod_{i=1}^{N} \left[
\beta p\left(P_1\right) + \left(1-\beta\right) p\left(P_2\right)\right],
\label{eq:TGC-ML-3}
\eq 

\noindent where $\beta$ is the probability that model 1, represented by
the probability density function $p(P_1)$, is correct, and $P_1$, $P_2$ and
$\beta$ vary in the fit. It can be seen that a simple comparison between the 
values of these probabilities indicates the correct model for the majority of
the cases. In addition, the  absolute probability value indicates the goodness
of the fit.

\begin{figure}[hbt]
\begin{center}\leavevmode
\mbox{\epsfig{clip=,bbllx=1,bblly=240,bburx=560,bbury=550,
      file=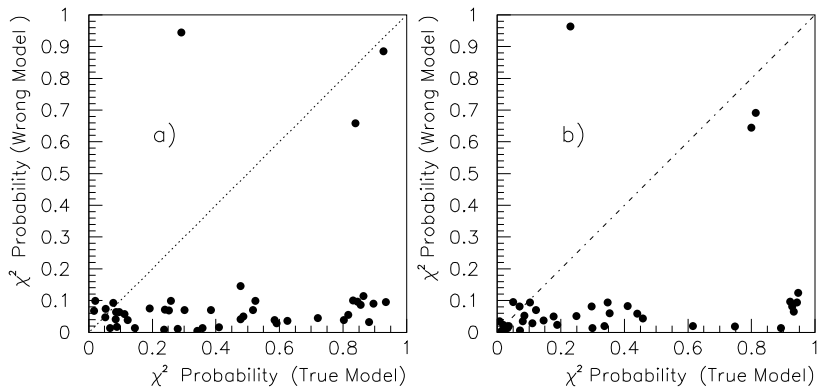,
      width=.7\textwidth}}
\caption{Hypothesis testing using a) the ``same family'' and b) the ``composite
         hypothesis'' methods, for data sets of about 2500~\jjlv~events
         generated with TGC values deviating from the SM values by one to five
         times the expected LEP2 precisions.}
\label{fig:TGC-ML-1}
\end{center}
\end{figure}

\subsection{Optimal observables method}
\label{sec:TGC-OO}

Optimal observables are quantities with maximal sensitivity~\footnote{
This method has been used to search for CP violation in $\tau^+ \tau^-$
production at LEP1, with a clear increase of sensitivity~\cite{
ref:TGC-Akers}.} to the unknown
coupling parameters~\cite{DiNa94,ref:TGC-Atwood}. To construct them, a
particular set of couplings $P_i$ is chosen which are zero at Born level in
the Standard Model (for instance, the TGCs defined by (\ref{eq:alphas})). Then,
recalling that the amplitudes for the four-fermion process are linear in the
couplings,  the differential cross-section may be written

\bq
  \frac{{\mathrm d}\sigma}{{\mathrm d}{\bf \Omega}} = S_0({\bf \Omega}) + 
                             \sum_i S_{1,i}({\bf \Omega}) \, P_i +
                             \sum_{i,j} S_{2,ij}({\bf \Omega}) \, P_i P_j \eqcm
  \label{eq:TGC-OO-1}
\eq 

\noindent where ${\bf \Omega}$ represents the kinematic variables as before. 
Kinematic ambiguities, such as those described in table~\ref{tab:anginf},  can
readily be incorporated into  (\ref{eq:TGC-OO-1}). The distributions of the
functions

\bq 
  {\cal O}_i({\bf \Omega}) = \frac{S_{1,i}({\bf \Omega})}{S_0({\bf \Omega})}
  \label{eq:TGC-OO-2}
\eq 

\noindent are measured, and their mean values $\langle {\cal O}_i \rangle$
evaluated\footnote
       {The functions ${\cal O}_i({\bf \Omega})$ for the TGC parameters used
       in~\cite{Haetal87} are available as a FORTRAN routine~\cite{DiNa94}.
       }.
An example is shown in fig~\ref{fig:TGC-OO-1}. To first order in the $P_i$, the
mean values $\langle {\cal O}_i\rangle $  are given by

\bq
  \langle {\cal O}_i \rangle = {\langle {\cal O}_i \rangle}_0 + \sum_j
  c_{ij} \, P_j \eqcm
  \label{eq:TGC-OO-3}
\eq 

\noindent  from which the couplings $P_j$ can be extracted because 
${\langle {\cal O}_i \rangle}_0$ and $c_{ij}$ are calculable given
(\ref{eq:TGC-OO-1}) and (\ref{eq:TGC-OO-2}).  From the distributions of
the ${\cal O}_i$ the statistical errors on their mean values can be evaluated,
the observables having been constructed to minimize the induced errors on the
$P_j$. If the linear expansion in the couplings is good, the method has the same
statistical sensitivity as a maximum likelihood fit. It can also be
extended to incorporate total cross-section information in a manner analogous to
the use of the extended maximum likelihood method discussed in the previous
section.  

\begin{figure}[htb]
\mbox{\epsfig
              {file=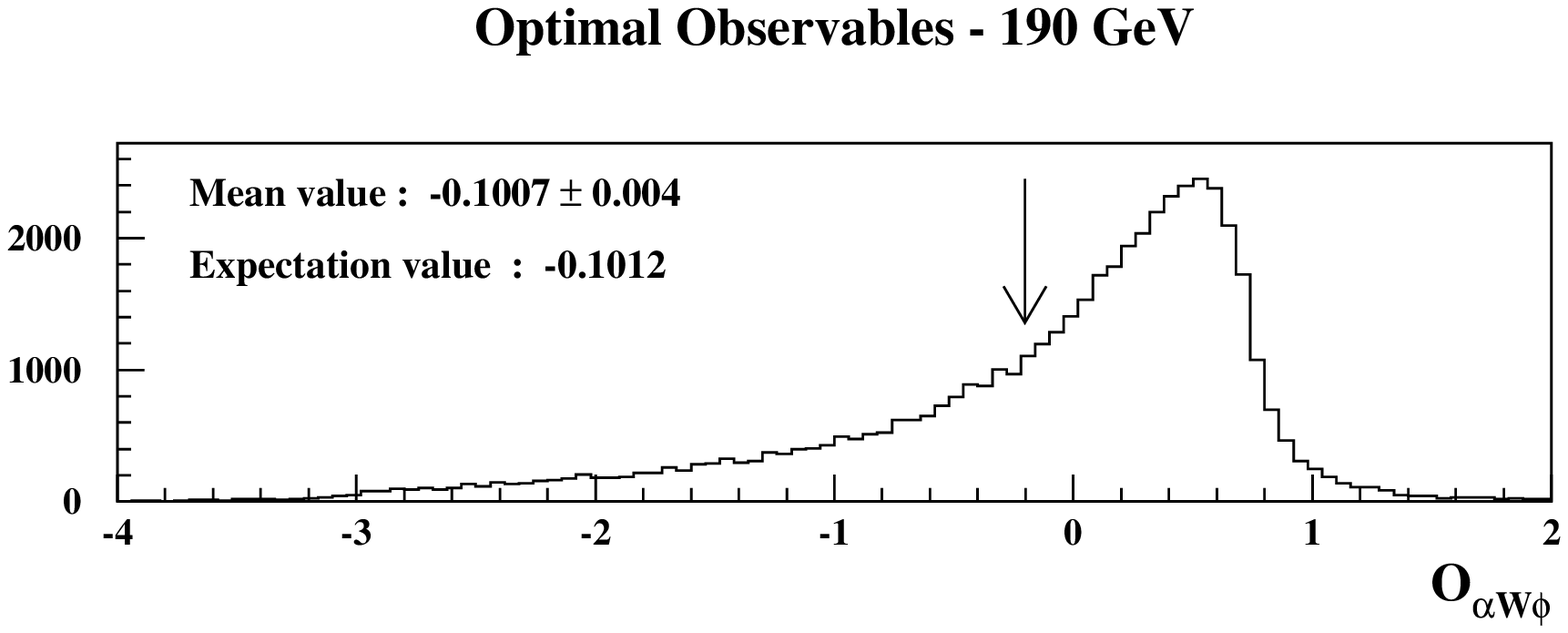,
              width=\textwidth}}
\caption{Distribution of ${\cal O}_{\aWphi}(\ctw, (\ctl, \phil), 
         (\ctj, \phij)\fold)$ for a large sample (50000) of simulated \eeWW\
         events at 190 GeV. The experimentally determined mean value is to be
         compared with the expectation value of this observable in 
the SM, $\aWphi = 0$, used to generate the events.}
\label{fig:TGC-OO-1}
\end{figure}

\subsection{Comparison of methods}
\label{sec:TGC-comp}

In this section a comparison is presented of fits of the TGCs \aWphi, \aBphi\
and \aW, defined in (\ref{eq:alphas}), to common datasets generated with the
PYTHIA\cite{ref:TGC-Sjostrand} Monte Carlo simulation program.

We precede this by mentioning the results of a comparison of the use of the
maximum and extended maximum likelihood (ML and EML) methods, in which both of
these methods were used in fits of the three TGCs to a large sample (50000) of
events using first only the \W\ production angle, and then the complete angular
information (production and decay angles). The extra information contained in
the EML method gave a substantial  improvement (10\%) in precision only in one
case --- the fit of \aBphi, generally the least well determined parameter, to
the production angular distribution. In the other fits the improvement was only
$\sim 1\%$. Similar conclusions have been obtained when applying the optimal
observables method with and without total cross-section information.

In the comparison of the density matrix (DM), EML and optimal observables (OO)
methods, the three analyses were applied to datasets at 175 and 190 GeV
simulating both the expected LEP2 statistics ($\approx$ 2000 events) and much
larger statistics (50000 events). Sample results are given in 
fig~\ref{fig:TGC-comp-1}, in which precisions obtained using the three methods
in 1- and 2-parameter fits to the large dataset at 190 GeV are plotted. In all
cases, the precisions obtained using the three methods are very similar when the
same angular data is used in the fit. This can be seen in the figure, where the
precisions from the EML and OO methods, both of which used angular data \ctw,
(\ctl, \phil) and (\ctj, \phij)\fold, are almost identical. The DM results shown
used the differential cross-section, $\frac{{\mathrm d}\sigma}{{\mathrm
d}\!\cos\theta }$,  density matrix elements $\rho_{00}, \rho_{1\! -1}, \rho_{1\!
0}$ and $\rho_{-1\! 0}$ of the leptonically decaying \W, and the part symmetric
in both polar decay angles of the transverse  element $\rho_{TT} \equiv
\rho_{11,11} + \rho_{-1\!-1,-1\!-1} +  \rho_{11,-1\!-1} + \rho_{-1\!-1,11}$ of
the joint \WW\ density matrix, representing somewhat less than the full 35
(CP-conserving) elements of the full joint density matrix. (Other density matrix
elements can in principle be included in the analysis).

\begin{figure}[htb]
\mbox{\epsfig
              {file=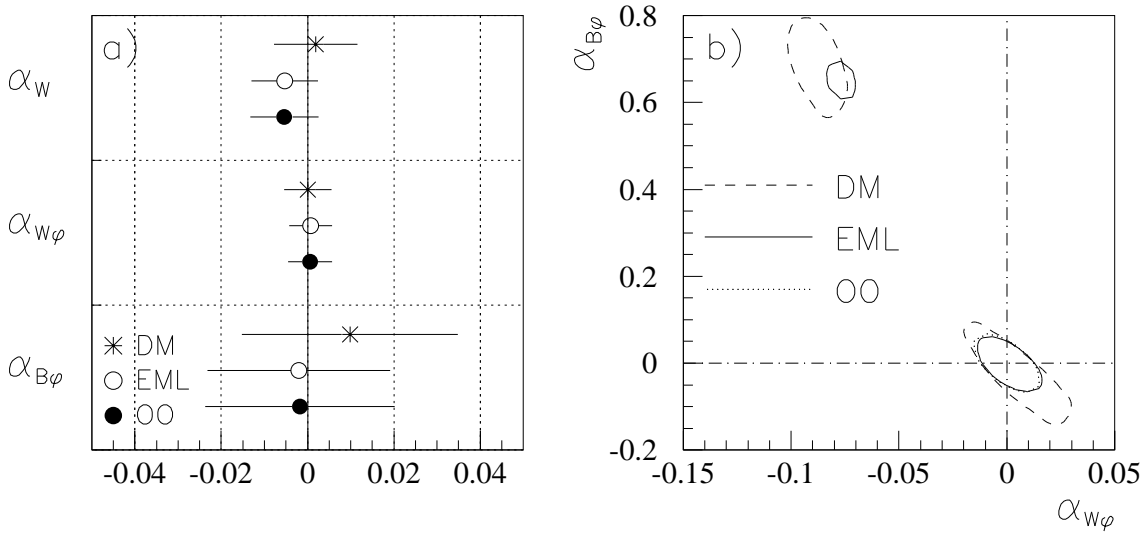,
              width=\textwidth}}
\caption{Comparison of TGC fits to a large sample of simulated events at 190 GeV
         using the density matrix (DM), maximum likelihood (EML) and optimal
         observables (OO) methods. a): 1 s.d. precisions in 1-parameter fits to
         \aW, \aWphi\ and \aBphi.\  b): 95\% confidence contours in
         2-parameter fits to (\aWphi, \aBphi).}
\label{fig:TGC-comp-1}
\end{figure}

A difference between the EML or DM analyses and the OO analysis can be seen  in
the 2-parameter fit shown, where a second allowed region, remote from the SM
region $(\aWphi = 0,\aBphi = 0)$ where the events were generated, is seen by the
EML and DM methods. This effect is discussed in detail in
ref.~\cite{Se94}, where it is shown to arise naturally from the
amplitude structure of \WW\ production, and in particular from the fact that the
helicity amplitudes are linear in the TGCs. It is not seen in the OO results,
because here the {\it cross-section} (\ref{eq:TGC-OO-1}) has been linearized
with respect to the TGCs about their SM values\footnote
{An extension of the OO method to incorporate second order terms in the
parameters is under development.}.

In considering possible extensions to the analyses, two comments may be made.
First, the EML and OO methods could readily be used in a 4-fermion treatment by
replacement of the matrix elements. The DM method does not lend itself to this
adaptation, as the form (\ref{eq:dsig5}) used in the projection of the density
matrix elements assumes $J=1$ for the two final state $f\bar{f}$ pairs. Second,
all three methods can in principle be adapted to the analysis of events with the
experimental and other effects discussed later in this chapter; however, we have
not made an assessment of the relative ease with which this can be done for the
different methods. 

With the above points borne in mind, we can recommend all three methods for
consideration in the analysis of LEP2 data. The studies reported in the
following sections have, except where otherwise indicated, used ML or EML fits
to obtain the results shown.

\section{Precision of TGC determination at LEP2: generator level studies}
\label{sec:TGC-gen}

In this section, the precisions to be expected in TGC determination from the
anticipated LEP2 integrated luminosity are summarized and an estimate of the
biases and systematic errors accessible at generator level is given.

\subsection{TGC precisions in fits to simulated events}
\label{sec:TGC-genresults}

Precisions in TGCs obtained from 1-parameter fits to simulated \eeWW events  at
176 and 190 GeV are shown in table~\ref{tab:TGC-genresults-1}, and confidence 
limits in the planes of two of the three possible combinations of two of the
parameters in eq.~(\ref{eq:alphas}) are shown 
in fig~\ref{fig:TGC-genresults-1}. Results
are shown using various combinations of the angular data appropriate to each of
the three final states \jjlv, \jjjj\ and \lvlv, as indicated in
table~\ref{tab:anginf}, as well as to the ``ideal'' case without angular
ambiguities. For the first two channels (and for the ``ideal'' analysis), 1960
(2600) events were fitted at 176 (190) GeV; for the \lvlv\ channel, 280 (370)
events were used. These figures emulate the statistics anticipated  from an
integrated luminosity of $500 {\mathrm pb}^{-1}$ after experimental efficiency
cuts of $\sim 95\%$, 60\% and 95\% for the three channels respectively, and
excluding leptonic decays into $\tau \nu_{\tau}$.  The extended maximum
likelihood method was used in the fits, and the events were generated and
analyzed in the narrow \W\ width approximation and without initial state
radiation (ISR). In the analysis, the generated values of parton momenta were
used, so that no account has been taken of the subsequent quark fragmentation
nor of possible experimental effects.  No kinematic cuts have been made on the
data. The analysis reported here is therefore to be considered as an idealized
one; the implications of the additional effects mentioned above are considered
in detail in subsequent sections. 

Several conclusions may be drawn from inspection of the table and figure.
As anticipated by the discussion of section~\ref{TGCWpairpheno}, substantial
gains in precision are achievable by running at higher energy. Also, use of as
much as possible of the available angular data serves to increase the precision
and, in 2-parameter fits, to reduce the (quite pronounced) correlations between
the fitted TGCs. The use of the \jjjj\ channel, even with the angular
ambiguities incurred by the inability to distinguish quark from antiquark jets, 
can be seen to provide a modest but worthwhile improvement in the overall
precision attainable. Finally, the occurrence of a second region in the
($\alpha_{W\phi}$, $\alpha_{B\phi}$) plane, remote from the Standard Model
region ($0,0$) at which the events were generated but acceptable at the chosen
significance level, has already been noted in the previous section.

\begin{table}[htb]
\center{
\begin{tabular}{|c|c|l|c|c|}
\hline\hline
Model & Channel   &    Angular data used  & 176 GeV & 190  GeV \\ \hline\hline
$\alpha_{B\phi}$
      & \jjlv     &      \ctw             & 0.222    & 0.109   \\
      &           &  \ctw, (\ctl, \phil)  & 0.182    & 0.082   \\
      &           &  \ctw, (\ctl, \phil),
                     (\ctj, \phij)\fold   & 0.159    & 0.080   \\
      & \jjjj     &  $\mid\ctw\mid$       & 0.376    & 0.149   \\
      &           &  $\mid\ctw\mid$,
                    (\ctja, \phija)\fold,
                   (\ctjb, \phijb)\fold   & 0.328    & 0.123   \\ 
      & \lvlv     &  \ctw, (\cta, \phia),
              (\ctb, \phib), 2 solutions  & 0.323    & 0.188   \\
      & Ideal     &  \ctw, (\cta, \phia),
                     (\ctb, \phib)        & 0.099    & 0.061   \\ \hline\hline
$\alpha_{W\phi}$
      & \jjlv     &   \ctw                & 0.041    & 0.027   \\
      &           &  \ctw, (\ctl, \phil)  & 0.037    & 0.023   \\
      &           &  \ctw, (\ctl, \phil), 
                     (\ctj, \phij)\fold   & 0.034    & 0.022   \\
      & \jjjj     &  $\mid\ctw\mid$       & 0.098    & 0.054   \\
      &           &  $\mid\ctw\mid$,
                    (\ctja, \phija)\fold,
                   (\ctjb, \phijb)\fold   & 0.069    & 0.042   \\ 
      & \lvlv     &  \ctw, (\cta, \phia),
              (\ctb, \phib), 2 solutions  & 0.096    & 0.064   \\
      & Ideal     &  \ctw, (\cta, \phia),
                     (\ctb, \phib)        & 0.028    & 0.018   \\ \hline\hline
$\alpha_W$
      & \jjlv     &      \ctw             & 0.074    & 0.046   \\
      &           &  \ctw, (\ctl, \phil)  & 0.062    & 0.038   \\
      &           &  \ctw, (\ctl, \phil),
                     (\ctj, \phij)\fold   & 0.055    & 0.032   \\
      & \jjjj     &  $\mid\ctw\mid$       & 0.188    & 0.110   \\
      &           &  $\mid\ctw\mid$,
                    (\ctja, \phija)\fold,
                   (\ctjb, \phijb)\fold   & 0.131    & 0.069   \\ 
      & \lvlv     &  \ctw, (\cta, \phia),
              (\ctb, \phib), 2 solutions  & 0.100    & 0.064   \\
      & Ideal     &  \ctw, (\cta, \phia),
                     (\ctb, \phib)        & 0.037    & 0.022   \\ \hline
              
\end{tabular}
\caption{1 s.d. errors in fits of $\alpha_{B\phi}$, $\alpha_{W\phi}$ and
         $\alpha_W$ to various combinations of the angular data at 176 and 190
         GeV. The simulated data corresponds to integrated luminosity of $500
         {\mathrm pb}^{-1}$. Details of the data samples are given in the text.}
\label{tab:TGC-genresults-1}
}
\end{table}

\begin{figure}[p]
\begin{center}
\vspace{-0.5cm}
\mbox{\epsfig
              {file=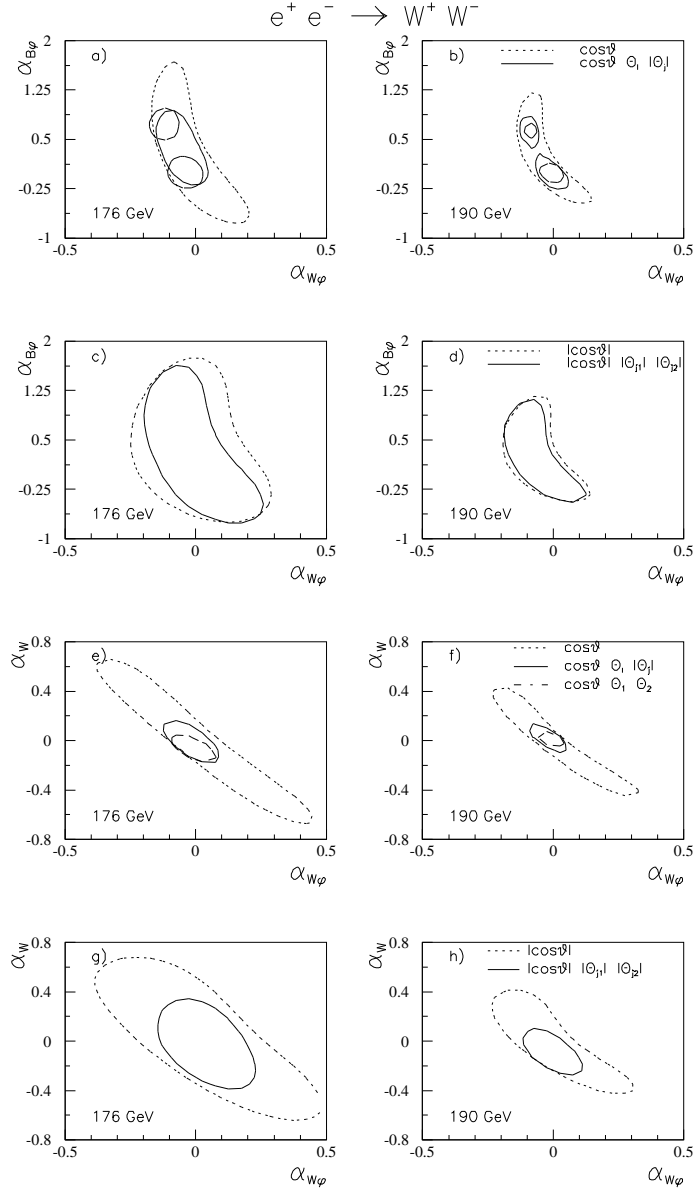,
              width=0.7\textwidth}}
\vspace{-1.0cm}
\caption{95\% confidence limits in the planes of 2-parameter TGC fits at 176 and
         190 GeV, using various combinations of angular data.
         a), b), c), d): Fits to (\aWphi, \aBphi);
         e), f), g), h): Fits to (\aWphi, \aW). 
         In the legend, the  notation $\Theta_{l,j}$ implies a pair of decay
         angles  $(\theta_{l,j},\phi_{l,j})$ for $W \rightarrow$ (leptons, jets)
         respectively, and $\mid\Theta_j\mid$ implies the ambiguity $\ctj
         \leftrightarrow -\ctj$,  $\phij \leftrightarrow \phij + \pi$ incurred
         by the inability to distinguish quark from antiquark jets.
         In plots a), b), e), f), the angular data simulates channel \jjlv\ (and
         the ``ideal'' case, with no ambiguities); in c), d), g), h), it
         simulates channel \jjjj.}
\label{fig:TGC-genresults-1}
\end{center}
\end{figure}

In a first step towards a more realistic simulation of the data, some of the
fits described above have been repeated using calculations corresponding to
4-fermion rather than \WW\ production both in event generation and analysis.  In
so doing,  contributions are included from the complete set of relevant diagrams
and the finite \W\ width  effects ignored in the previous analysis are taken
into account. Using events generated with the ERATO~\cite{ert0}
program corresponding to the expected statistics  at 175 and 190 GeV, similar
precisions to those shown above are obtained in fits of  \aWphi\ and \aW\ to
angular data \ctw, (\ctl, \phil) and (\ctj, \phij)\fold\footnote
      {A computational point may be made here: in the evaluation of the
       differential and total cross-sections needed in the likelihood expression
       (\ref{eq:TGC-ML-2}), time may be saved by noting that, since the
       amplitudes for the process \eeffff\ (or \eeWW) are linear in the TGCs, 
       an exact parametrization of the cross-section dependence  on any one TGC
       may be found from a quadratic fit to its values for any three values of
       the TGC parameter. This procedure can be extended in an obvious way to
       fits of two or more parameters.
       }.
In addition, in fits to a sample of \jjlv\ events  generated at 161 GeV
corresponding to an integrated luminosity of  $100 {\mathrm pb}^{-1}$ (as
suggested for the determination of the \W\ mass from its threshold
excitation~\cite{ref:TGC-Stirling}), 1 s.d. precisions of 0.18 and  0.43 were
obtained in fits of \aWphi\ and \aW\ respectively. It is interesting to note
that these values compare well with current experimental
limits~\cite{ref:TGC-CDF,D0}, implying that TGC measurements from this exposure may
also be of interest. This conclusion, however, remains to be tested when
backgrounds and other experimental effects are included.

\subsection{Biases and systematic errors in TGC determination calculable at
            generator level}
\label{sec:TGC-generrs}

It was pointed out in the previous section that the analyses presented there
are idealized, in the sense that effects due to finite \W\ width (unless a
4-fermion calculation is used), ISR, QCD and experimental reconstruction have
been ignored.  In this section, we consider the biases introduced in TGC
determinations, first, if events generated with a realistic \W\ mass
distribution are nonetheless analyzed in the narrow width approximation, and,
second, if ISR effects are also present, but ignored in the analysis. The
discussion of the overall bias to be expected in TGC determination is pursued in
the next section, where biases arising due to event selection and reconstruction
are added to those discussed here. The systematic errors incurred both in the
assessment of  these biases and from other sources calculable at generator level
are also estimated in this section.

Figs.~\ref{fig:TGC-generrs-1}a) and b) show the effects of ignoring finite \W\
width and ISR in the analysis of events generated with these effects included.
Results are shown for several different generators, all operating in \eeWW\
(CC03) mode. It can be seen, first, that the bias incurred by neglect of ISR is
greater than that from neglect of \W\ width effects, second, that the biases are
smaller when a fit involving more angular data is used, and, third (from b),
that the biases are different for different values of a typical TGC parameter.
Finally, we note that the overall bias is $\lesssim$ the statistical error
expected from LEP2 data.

\begin{figure}[htb]
\mbox{\epsfig
              {file=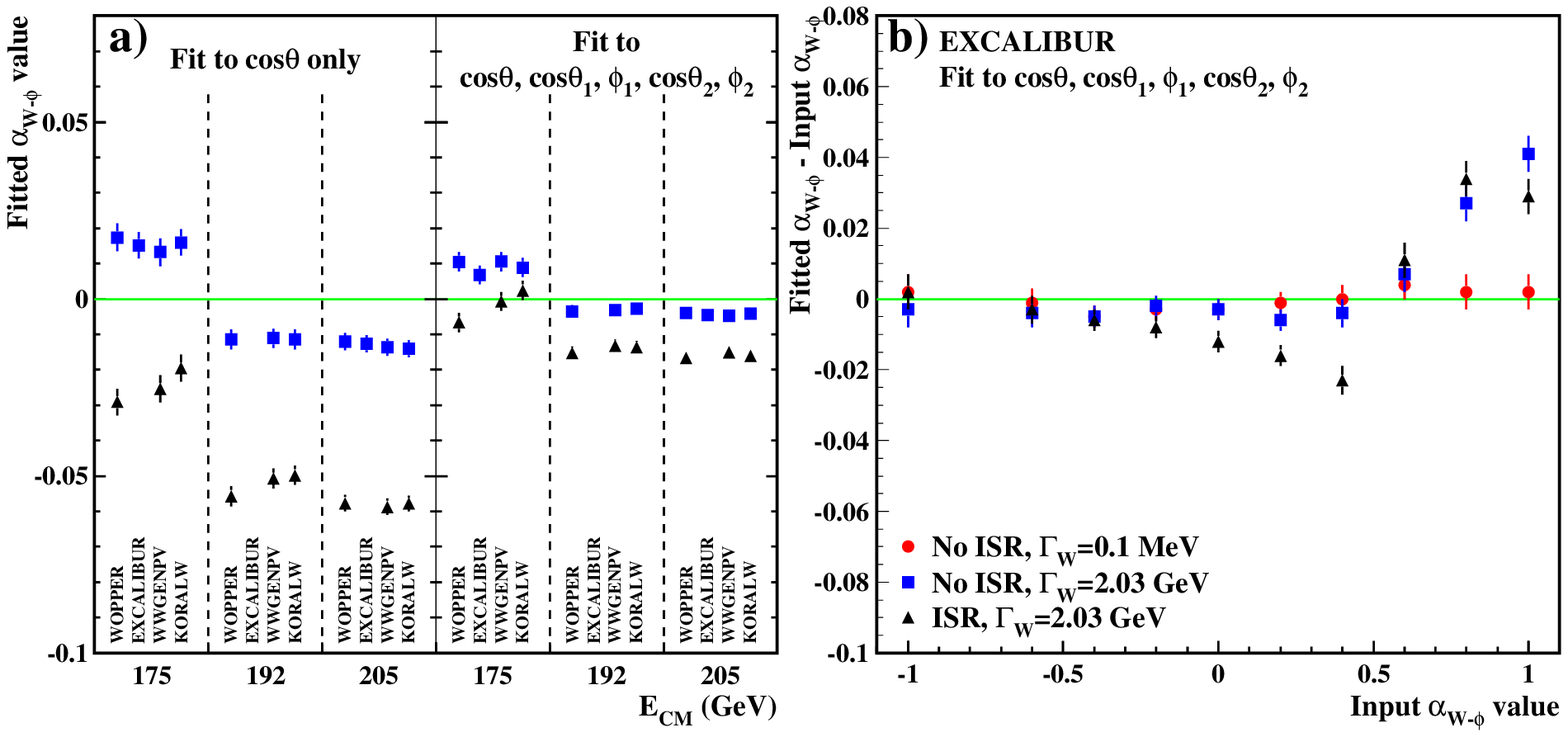,
              width=\textwidth}}
\caption{Effect of ignoring finite \W\ width and ISR in TGC fits. a): Results of
         fits of \aWphi\ to events generated with SM parameters at three 
         energies using various generators. Left-hand plots: fit to \ctw\ only; 
         Right-hand plots: fit to  \ctw, (\cta, \phia), (\ctb, \phib). 
         b): as a), for EXCALIBUR events at 190 GeV, using \ctw, (\cta, \phia),
         (\ctb, \phib), as a function of \aWphi. The legend for both plots is
         shown on b).}
\label{fig:TGC-generrs-1}
\end{figure}

The systematic errors arising from these and other sources calculable at
generator level are summarized, using a particular TGC fit as an example, in
table~\ref{tab:TGC-generrs-1}\footnote
          {The magnitude of some of these errors, in particular those arising
          from finite \W\ width and ISR effects, depend on the angular data
          used in the fit, (c.f. fig~\protect\ref{fig:TGC-generrs-1}).
          }.
The first three entries come from the effects discussed above, the next two
represent two different ways of expressing the uncertainty in the other
electroweak parameters which are important in the evaluation of the matrix
element, and the final pair represent two independent uncertainties coming from
machine and detector considerations. In any analysis which does not compare
total cross-section predictions with the observed data, the second and last
entries will not contribute to the overall uncertainty. It can be seen that,
even when all the relevant entries are added in quadrature, the total is small
compared with the statistical precision expected from LEP2 data, and we expect
the larger component of the systematic error to come from uncertainties in the
experimental effects considered in the next sections.

\begin{table}[htb]
\center{
\begin{tabular}{|l|c|c|}
\hline
Source of uncertainty & Uncertainty & Systematic error          \\
                      &             &     in \aWphi             \\ \hline\hline
           \W\ width  & $\Delta\Gamma_W = \pm 0.07$ GeV 
                                    &      $\pm 0.0004$          \\ \hline
           ISR        & $\Delta\sigma_{tot}/\sigma_{tot}(\eeWW + radiation) 
                        = \pm 1\%$ 
 &      $\pm 0.0013$          \\ \hline
ISR parametrization   & 
    Spread in Monte Carlo estimates 
                                    &      $\pm 0.0020$          \\ \hline\hline
$M_W$                 & $\Delta M_W = \pm 0.18$ GeV 
                                    &      $\mp 0.0021 $         \\ \hline
$\sin^2 \theta_W $    & $\sin^2 \theta_W = 0.226 $ (tree-level) $\rightarrow 
          \sin^2 \theta_W = 0.231 $ &    0.0029                  \\ \hline\hline
Beam energy           & $\Delta\sqrt{s} = \pm 15$ MeV 
                                    &     $\mp 0.0002$           \\ \hline
Absolute normalization
                      &  $\pm 1\%$  &     $\pm 0.0013$           \\ \hline\hline
\end{tabular}
\caption{Systematic errors from various sources incurred in fits of \aWphi\ to
         angular data  \ctw, (\ctl, \phil), (\ctj, \phij)\fold\ at 190 GeV. The
         1 s.d. statistical precision estimate for this fit from LEP2 data 
         ({\it c.f.} table~\protect\ref{tab:TGC-genresults-1}) is $\pm  0.022$.
         }
\label{tab:TGC-generrs-1}
}
\end{table}

In addition to the effects considered above, it is legitimate to ask whether 
colour recombination effects among the two \Ws could affect TGC measurements in
the \jjjj\ channel. It has recently been advocated that such effects may produce
a shift of up to 400 MeV in $M_W$~\cite{ElGe95}.  Therefore, by analogy with the
effects of ISR, it may produce a bias in TGC measurements which would need to
be accounted for, and, if not understood, would have an associated systematic
error. However, a preliminary study~\cite{Geiger} has indicated that the \W\
production angular distribution, reconstructed from the hadronization products
of generated \jjjj\ events, is little affected by application of the colour
recombination models of ref~\cite{ElGe95}, and hence that it is unlikely that
the shift in TGC values determined from the data in this channel will be
significant compared to the expected statistical error.

\section{Analysis of the \jjenu\ and \jjmunu\ final states}
\label{sec:TGC-jjlv}

In the following we address some of the experimental aspects of the analysis of
the $e^+e^- \rightarrow W^+W^- \rightarrow \jjlv$  channel.  In this section, 
we concentrate on the muon and electron channels, these being the cleanest
and very similar in many respects.  The tau channel is considered separately in
the following section. For simplicity, the data are analyzed in terms of the 
five angles describing \WW\ production and decay, by analogy with the
generator-level analysis reported in section~\ref{sec:TGC-genresults}. In its
extension to a four-fermion treatment, also described in that section, the
effect of the experimental selection and reconstruction procedures are expected
to be the same.

In section \ref{sec:TGC-jjlv-kinfit} we describe the efficiencies and purities
obtained after the application of typical selection criteria and  of kinematic
constraints to the events. In the process of reconstructing and analyzing \jjlv\
events,  there are many experimental effects which can potentially bias the
angular distributions, and hence the fitted values of TGC parameters. The scale
of such effects is estimated in section \ref{sec:TGC-jjlv-biases}, and in
section \ref{sec:TGC-jjlv-methods} we discuss briefly some methods proposed to
allow for them in the analysis. The numbers presented result from a comparison
of the work of several different groups and should be regarded as broadly
typical of the four LEP experiments. 
\subsection{Event selection, kinematic reconstruction and residual background}
\label{sec:TGC-jjlv-kinfit} 
The \jjlv\ event selections used typically demand the following:
\begin{itemize}
\item[-] that the event contains a minimum number, typically five or six, of
  charged track clusters;
 
\item[-] that there is an identified electron or muon, or alternatively a high
  energy isolated track;
 
\item[-] that the lepton has a momentum greater than its kinematic minimum,
  $\sim 20$ GeV;
 
\item[-] that the lepton be isolated, by requiring low activity in a cone 
around
  the track (typically that the energy deposited in a cone of 100-200 mrad be
  less than 1-2 GeV).
 
\end{itemize}
\noindent The effect of these cuts corresponds approximately to a fiducial cut
in the centre-of-mass polar angle of the lepton of $|\cos\theta_{lepton}| <
0.95$. The acceptance for jets, which have some angular size, extends further
but with falling efficiency. These numbers vary for specific detectors.
 
The non-lepton system is then split into two (or more) jets using a conventional
jet-finding algorithm. The following kinematic constraints \cite{ref:TGC-kinfit}
can then be applied to impose energy and momentum conservation, and to improve
the measurements using the fact that the system is overconstrained:
 
\begin{itemize}
\item[] 1C fit:\hspace{0.2cm} $\sum E = \ecm$, $\sum \vec{p} = 0$, $m_\nu = 0$;
 
\item[] 3C fit:\hspace{0.2cm} In addition to 1C, $M_{reconstructed} = \mw$ for
  both \W\ candidates;
 
\item[] 3C$^\prime$ fit:\hspace{0.18cm}  In addition to 1C, $M_{reconstructed}$
  for both \W\  candidates is constrained  to a central value of  $\mw$ but is
  allowed to vary approximately within the \W\ width\footnote
       {This is achieved by including either Gaussian approximations or
        true Breit-Wigner constraints in the fit procedure.
       }.
\end{itemize}
\noindent In the above, $m_\nu$ is the neutrino mass and $\mw$ the \W\ mass. A
$\chi^2$ probability cut, typically of 0.1-1\%, is applied to the constrained
fit result.  Typical efficiencies after these stages are shown in table
\ref{tab:TGC-effpur} for centre-of-mass energies $\sqrt{s} = 175$ and 
$192$ GeV.
The main loss is due to geometrical acceptance and lepton identification in the
basic selection.  The kinematic fits themselves are of the order of 90\%
efficient for such a probability cut.
{\small
\begin{table}
\begin{center}
\begin{tabular}{||c||c||c|c|c||c||}\hline \hline
                    & Efficiency \%    & \multicolumn{4}{|c||}{Background \%}
 
\\
                    &                  &   $Z\gamma$ &      \WW\ (non-\jjlv)
     &  $ZZ$      & Total    \\
\hline \hline
\multicolumn{6}{||c||}{$\ecm$ = 175 GeV}    \\
\hline
Basic Selection      & 77          &  8  &  6   &  1         &   15     \\
1C fit               & 75          &  7  &  5   &  1         &   13     \\
3C fit               & 70          &  1  &  2   &  0.5       &   3.5    \\
3C$^\prime$ fit      & 72          &  1  &  4   &  0.5       &   5.5    \\
\hline \hline
\multicolumn{6}{||c||}{$\ecm$ = 192 GeV}     \\
\hline
Basic Selection     &  75          & 7   &  8   & 2          &  17      \\
1C fit              &  73          & 6   &  7   & 2          &  15      \\
3C fit              &  66          & 1   &  2   & 1          &  4       \\
3C$^\prime$ fit     &  71          & 1   &  3   & 1          &  5       \\
\hline \hline
\end{tabular}
\end{center}
\caption{
Efficiencies and purities of the \jjlv\ sample at progressive stages
of selection and kinematic fitting.
}
\label{tab:TGC-effpur}
\end{table}
}

The background estimation was made using event samples, simulated with PYTHIA,
of the final states \WW\ (with neither of the bosons decaying to an electron or
a muon), $Z\gamma$, $ZZ$ and $Zee$. Also, contamination from $\gamma \gamma$
events, generated with TWOGAM \cite{ref:TGC-twogam}, were studied. Backgrounds
from the last two channels were found to be negligible; those from the other
final states are summarized in table \ref{tab:TGC-effpur}. Contributions from
the non-resonant graphs leading to the \jjlv\ final state and containing TGCs
have also been studied. It is found that, taken in isolation and ignoring
interferences, they are rejected by  the selection procedure.  The main
contribution to the $WW$  background comes from events where one of the \Ws\
decays into a tau and then into an electron or muon. Although this channel is
sensitive to the TGCs, it will be seen in section~\ref{sec:TGC-jjlv-biases} 
that
the inclusion of such events into the analysis does not significantly bias the
result.
 
Other approaches can be used instead of the selection procedure described above
.
In particular, if one wishes to avoid the use of the constrained fit, a cut
requiring the missing momentum direction to be away from the beam pipe,
typically $\cos \theta < 0.95$, can be used to reduce the background from the
$ZZ$ and $Z\gamma$ channels. In this case, an algorithm has to be applied  to
impose energy and momentum conservation. Nonetheless, in the rest of this
section we adopt the 3C fits as representative of the efficiency and purity
which can be achieved.
 
The resolutions obtained for the \WW\ production and decay angles before and
after kinematic fitting are shown in table \ref{tab:TGC-resol}. The values 
shown are averages over the whole fiducial region; however, in general, the
resolutions depend  upon the values of the kinematic variables themselves and,
following kinematic fitting, they are correlated.
It can be seen that a modest
improvement in resolution is obtained, the main qualitative effect being due to
the recovery of mis-measured events.
{\small
\begin{table}
\begin{center}
\begin{tabular}{||c||c|c|c|c|c||}\hline \hline
Selection           &\multicolumn{5}{|c||}{Resolution}\\
                    &  $\ctw$    &  $\ctl$  &  $\phil$ (rad)
                              &  $\ctj$  &  $\phij$ (rad)\\
\hline \hline
Before fit             & 0.06-0.13     &  0.11-0.17   &  0.12-0.23   
& 0.13-0.19    & 0.11-0.22 \\
\hline
After 3C$^\prime$ fit  & 0.05-0.12     &  0.07-0.13   &  0.10-0.21   
& 0.10-0.17    & 0.11-0.21 \\
\hline \hline
\end{tabular}
\end{center}
\caption{
Resolutions on \WW\ production and decay angles using simulated events at 192
GeV. The ranges indicate the spread of values obtained from different
experimental simulations.
}
\label{tab:TGC-resol}
\end{table}
}
\subsection{Systematic biases and statistical precision}
\label{sec:TGC-jjlv-biases}
 
We now consider potential  systematic biases, and the degradation of statistical
precision due to experimental effects in the \jjlv\ channel.  In this we include
a) the neglect of ISR and $\gw$, b) experimental acceptance, c) reconstruction
and detector resolution, and d) residual background contamination. The first
item has been discussed in detail in section~\ref{sec:TGC-generrs}; the result
is included here for completeness. We use as example fits to \aWphi\ only.
 
The overall bias due to a)-d)  has been determined using a total of
approximately 20,000  simulated \jjlv\ events at 175 GeV and 30,000 events at
192 GeV. A maximum likelihood or extended maximum likelihood fit was used,
assuming in the analysis that the events originate from \WW\  production with
narrow \W\ width and without initial state radiation. We  emphasize that, since
the purpose of this study is to  show explicitly the scale of the biases, no
corrections for the effects listed above have been applied in the analysis.
{\small
\begin{table}
\begin{center}
\begin{tabular}{||c||c|c||c|c||}\hline \hline
                          & \multicolumn{2}{|c||}{175 GeV} & \multicolumn{2}{|c
||}{192 GeV} \\
                          & 1-D               & 5-D             &  1-D
      & 5-D           \\
\hline \hline
Statistical Precision     &  $\pm 0.041$         &    $\pm 0.034$      &   $\pm
 0.027$          &    $\pm 0.022$          \\
(from table~\protect\ref{tab:TGC-genresults-1})
                          &                      &                     &
                 &                         \\
\hline
Biases to Measurement:    &                      &                     &
                 &                         \\
~~ISR and $\gw$         & $-0.03$              & $-0.01$             &   $-0.
05$              &   $-0.02$               \\
~~Selection/Acceptance  & $-0.06$              & $-0.02            $ &   $-0.
03         $     &   $-0.03              $ \\
~~Reconstruction/resoln.& $-0.05$              & $-0.01            $ &   $-0.
03            $  &   $-0.01            $   \\
                          &                      &                     &
                 &                         \\
Total                     & $-0.14$              & $-0.04$             &  $-0.1
1$               &   $-0.06$               \\
                          &                      &                     &
                 &                         \\
\hline
Approximate additional    &                      &                     &
                 &                         \\
bias due to backgrounds   &                      &                     &
                 &                         \\
~~\WW\                     & $ -0.005           $ &$   -0.002          $&  $
 -0.003           $ &  $  -0.002           $   \\
~~$Z\gamma$              & $+0.003$            &  $+0.008$           &  $ -0.
003$             &  $+0.002$                \\
~~$ZZ$                   & $-0.003$            &  $-0.001$           &  $-0.0
12$              &  $-0.002$               \\
                          &                      &                     &
                 &                         \\
\hline \hline
\end{tabular}
\end{center}
\caption{
Biases in the measurement of $\aWphi$ estimated from studies of large samples 
of
fully simulated events. In the last part of the table the additional biases due
to residual  backgrounds are shown.}
\label{tab:TGC-biases}
\end{table}
}

The results are shown in table~\ref{tab:TGC-biases}.   The column labelled 1D
refers to fits using only the production angle \ctw. The column labelled 5D
refers to fits using the production and decay angles (with the  angles of the
hadronically decaying \W\ folded to take account of the ambiguity described in
table~\ref{tab:anginf}).  The bias due to ISR and $\gw$ is derived as 
described 
earlier. The bias due to event selection and acceptance was determined by
comparing fits to the generated angles before and after event selection, and 
the
bias due to reconstruction and resolution was determined by comparing fits to
generated angles with fits to fully reconstructed angles.  In the last part of
the table the additional biases due to background are shown. However the reader
should be aware that these were measured by adding small numbers of events to
the sample, and in the absence of a systematic study should be considered to be
very approximate.
 
We conclude that the size of the biases from ISR and $\gw$, acceptance and
reconstruction are up to a few  times the expected statistical error in the 
case
of 1D fits, and somewhat smaller when all the angular information is used. In
order that these effects do not present a serious source of systematic error
compared to the statistical error, they will eventually  have to be understood
and corrected for, incurring an error of less than $\sim 10\%$ of their values.
 
Finally, we investigate the extent to which the statistical precision in TGC
determination is degraded due to the effects mentioned above.  The large
simulated sample was divided into subsamples corresponding to the expected LEP2
statistics. The TGC parameter fit was performed on each sample, and the standard
deviation of the spread of the results calculated.  The precisions given for
fits to generator level data for the \jjlv\ channel in
table~\ref{tab:TGC-genresults-1} assume an efficiency of 95\%; thus the ideal
precision in this channel is better by a factor $\sqrt{0.95} = 0.97$. Taking
this and the estimated experimental efficiency of 70\% shown in
table~\ref{tab:TGC-effpur} into account, we expect a statistical degradation of
$\sim \pm 20$\% with respect to this ideal case. This is indeed observed,
together with an additional degradation  of $\pm 10$\% to $\pm 20$\% after
application of the analysis procedure described above, showing the effect of 
the
extra randomization from ISR, $\gw$ and experimental effects.
\subsection{Strategies for allowing for systematic biases}
\label{sec:TGC-jjlv-methods}

In the previous section the scale of the potential systematic bias due to
detector and other effects was quantified.  The simplest method of correction
for such a  bias is to determine its value for many simulated samples, subtract
the mean bias from the experimentally measured TGC value and assign a systematic
error on the basis of the width of the bias  distribution and the experimental
number of events. If the spread on the bias is large compared with the
statistical error, this procedure will clearly  be far from optimal. A second
method is to use a \MC simulation to produce a correction function  to map
between ``true'' and ``measured'' values. This can easily be applied when
fitting to a small number of variables, for instance to the \ctw\ distribution
alone, but is more difficult to apply in 5 dimensions simply because of the
number of events required to  characterize a 5D function in several bins per
variable (unless corrections for each variable can be assumed to factorize). It
has previously been shown at generator level that the precision is maximized by
using all variables; it may however be that when systematic errors are taken
into account the best overall precision is obtained by using a different
strategy. 

It is nonetheless possible to formulate methods which take resolution effects
into account in fits using all the kinematic variables. For instance, if the
resolution/acceptance function for the variables ${\bf \Omega}$  is known, then
the probability function $p({\bf \Omega},{\bf P})$ used for each event in the
maximum likelihood expressions (\ref{eq:TGC-ML-1}) and (\ref{eq:TGC-ML-2}) given
in section~\ref{sec:TGC-ML} can be replaced by 
 
\bq 
  p_{eff}({\bf \Omega}_{meas},{\bf P})= 
         \int p({\bf \Omega}_{true},{\bf P})\times 
                \rho({\bf \Omega}_{true}\rightarrow{\bf \Omega}_{meas}) 
                                                  {\mathrm d}{\bf \Omega}_{true}
\label{eq:TGC-jjlv-biases-1}
\eq 

\noindent (where ${\bf P}$ represents the TGC parameters of the fit).  The
resolution/acceptance function $\rho$  gives the probability that the true value
${\bf \Omega}_{true}$ would be  reconstructed as ${\bf \Omega}_{meas}$.

There are several potential problems with the application of 
(\ref{eq:TGC-jjlv-biases-1}):  (i) a 5-D integration is required; (ii) the
resolution and acceptance functions will almost certainly  not be simple, nor
will they factorize;  (iii) the correlations between angles must be known and
included (in particular if kinematically fitted quantities are used).  One
suggested  method \cite{ref:TGC-box} uses fully simulated  Monte Carlo events  
which are  passed through the same events selection as data, in order  to
calculate the effective likelihood function. The variation of the TGC parameters
is performed by reweighting the Monte Carlo events at their generated
coordinates, while the comparison with data is performed at the reconstructed
coordinates. This  method  can be applied for  any fit dimension and can in
principle take into account the effect of acceptance cuts, experimental
resolution, any kinematic fitting procedure and background contamination in the
data.

\section{Analysis of the \jjtaunu\ final state}
\label{sec:TGC-jjtaunu}

This channel requires special attention for two reasons. First, it comprises a
sizeable part of the semileptonic \WW\ decays and therefore could provide a
useful addition to the available statistics and, second, it is a background
mainly for the hadronic channel and therefore methods are required to reject it.

In this study we consider only the hadronic decays of the $\tau$ and describe
criteria to select this final state. The resulting efficiency and purity
expected for the sample and the resolution expected in the angular variables are
presented. We find that an increase in the overall number of events selected for
analysis in the \jjlv\ channel of between $10-20\%$ can be expected.

\subsection {Selection and reconstruction of \jjtaunu\ events}
\label{sec:TGC-jjtaunu-sel}

To select \jjtaunu\ events, we make use of the characteristics of the $\tau$
jet, namely small jet opening angle and low jet-charge multiplicity and of the
global characteristics of the event, mainly missing energy and event
acoplanarity. 

The signal for the \jjtaunu\ final state has a 3-jet topology, while the main
sources of background ($\WW \rightarrow \jjjj $ and  $\WW \rightarrow Z\gamma(s)
\rightarrow q\bar{q}\gamma$) fall into the 4-jet  and 2-jet topologies
respectively. Thus the choice of the resolution parameter in a jet-clustering
algorithm is quite significant. Requiring at least 3 jets in the  event, we find
a $\tau$-reconstruction efficiency of $70-80\%$  while only $30-40\%$ of
$Z\gamma$ events survive. The clustering algorithm itself ensures isolation for
the $\tau$ jet.

Jets from $\tau$ decays can be distinguished from quark and gluon jets by the
distribution of quantities such as the track multiplicity (total or charged),
the maximum angle of any charged track in the jet to the jet axis, and the
fractional energy of the jet contained within a cone of a specified angle (say,
0.1 rad) about the jet axis. A likelihood function based on such parameters  has
been constructed, giving a typical  efficiency of about $70\%$ with a rejection
factor for quark and gluon jets close to 50.  The charge of the $\tau$ lepton
can be estimated rather reliably from the  total charge of the tracks in the jet
(excluding those with momenta $<$ 1  GeV/c from the sum in order to  reduce the
contribution from soft tracks  from neighbouring jets). 

The $\tau$ signal can be further enhanced by requiring that the event contains
less than five jets and that the sum of the missing energy and the energy of the
reconstructed $\tau$ candidate should exceed $\sqrt{s}/2$. This results in a
selection efficiency for $\tau$ events of about 90\% with a rejection factor
against the \WW\ $\rightarrow$ \jjjj\ channel and against $ZZ$ events of greater
than 10. In addition, constraints on the polar angle of the missing momentum 
and the  acoplanarity of the event can be imposed to reduce further the
background from $Z \gamma$  events. A rejection factor  of 10 is obtained while
about $20\%$ of the signal is lost. Finally,  the very forward electromagnetic
calorimetry can be used to detect ISR photon(s) in cases where they have not
escaped in the beam pipe.

It may be noted from the above that missing energy and missing momentum are key
variables for the rejection of all types of background, and therefore the
hermiticity of the detector is an  important factor.

The efficiencies and purities obtained for \jjtaunu\ events from a sample of
simulated events at 192 GeV are shown in  table~\ref{tab:TGC-jjtaunu-1}. The
background from the \jjlv\ channel stems mainly from inefficiencies in muon
detection in the simulation used, and some improvement may be possible here.
The application of a 2-constraint kinematic fit\footnote
   {The 2C fit imposes energy and momentum conservation and constrains
    the $jj$ and $\tau\nu_{\tau}$ systems to have the \W\ mass,  leaving  the
    momentum of the neutrino from \W\ decay and  the $\tau$ energy as free
    variables  (with a lower limit on $E_{\tau}$ given by the visible energy of
    the $\tau$ decay products).
   }
can also be seen to provide background rejection, with a small decrease in the
$\tau$ selection efficiency.

\begin{table}[htb]
\begin{center}
\begin{tabular}{||c||c||c|c|c|c|c||}\hline 
Selection & Efficiency \% &\mco{5}{||c||}{Background \%}          \\ \cline{3-7}
          &    &  $Z\gamma$ & $WW \rightarrow \jjjj$ & $WW \rightarrow \jjlv$ 
                                                 & $ZZ$ & total \\ \hline \hline
 No fit      & 35 - 45  &  4 - 6  & 4 - 8  &  5 - 8  & 0 - 2  & 13 - 24 \\
 2C fit      & 32 - 42  &  0 - 2  & 2 - 5  &  5 - 8  & 0 - 0  &  7 - 15 \\
\hline \hline
\end{tabular}
\end{center}
\caption{Typical efficiencies and purities for the \jjtaunu\ channel
        with no kinematic fit and with a 2-constraint kinematic fit.}
\label{tab:TGC-jjtaunu-1}
\end{table}

\noindent An improvement to the kinematic fit may result by constraining  the
$\tau$ momentum, using the fact that the direction of the $\tau$ can be
accurately estimated from the combined momentum of its visible decay products,
so that the $\tau$ energy can then be computed from the \W\ decay
kinematics~\cite{ref:TGC-Rylko}.

\subsection{Resolution in reconstructed quantities}
\label{sec:TGC-jjtaunu-resol}

The resolution in the centre-of-mass polar and azimuthal angular  variables of
the $\tau$, evaluated using 2-Gaussian fits to the differences  between
reconstructed  and generated values, is of the order of 5 mrad in 75\% of the
events, and is not changed much by the kinematic fit.  The energy of the
original  $\tau$ can only be estimated at a level of $\Delta E/ E = 0.15$
with no kinematic fit, but after the fit has a resolution  $\Delta E/ E = 0.05$
in 80\% of events.  The resolutions in the \W\ production and decay angles,
evaluated after the kinematic fit, are found to be $\Delta cos\theta = 0.11$, 
$\Delta cos\theta_{\tau} = 0.13$ and $\Delta \phi_{\tau} = 250$ mrad
respectively.

\subsection{TGC determination from \jjtaunu\ events}
\label{sec:TGC-jjtaunu-tgc}

The precision with which TGCs can be determined from \jjtaunu\ events has been
investigated using a sample of 937 fully simulated events, generated at 192
GeV with finite \W\ width and ISR, corresponding to an integrated luminosity of 
$500 {\mathrm pb}^{-1}$. Of these events, 390 survived the selection and
reconstruction procedures described above.  The parameter \aWphi\ was fitted to
the cross-section (\ref{eq:dsig5}) ({\it i.e.} in the narrow width, no ISR
approximation),  using the  extended maximum likelihood method described in 
section~\ref{sec:TGC-ML} and folding over the 2 ambiguous solutions.  The 1 s.d.
precision in \aWphi\  was  found to be $\pm 0.06$ with estimated biases of
$-0.04$ from the neglect of \W\ width and ISR, $-0.025$ from  the effects of
reconstruction, and $+0.03$ from the presence of background events.

\section{Analysis of the \jjjj\ final state}
\label{sec:TGC-jjjj}

The advantage of the high branching fraction of this channel is somewhat reduced
by experimental difficulties associated with the purely hadronic nature of the
final state. Background rejection in the four-jet channel is difficult, since no
high-energy charged lepton is present to tag one \W\ as in the semileptonic
case. The largest background is expected from the high cross-section channel
$\mathrm{e^+e^-\rightarrow Z/\gamma^*\rightarrow q\bar{q}(\gamma)}$ leading to
multi-jet final states. Also, since the decay modes of the two \Ws are both
hadronic,  the problem arises of selecting the correct pairs of jets to form the
two \Ws and of assigning their charges.

In the following we suggest an analysis of the \jjjj\ channel, including event
selection, jet reconstruction and kinematic fitting, and indicate the expected
efficiency  and background levels. A section is devoted to jet- and
W-charge tagging. We then discuss the determination of TGCs from the selected
events, and draw conclusions on the sensitivity of the four-jet channel.

\subsection{Selection of events and reconstruction of 4 jets in the final state}
\label{sec:TGC-jjjj-sel}

The general criteria for the selection of \jjjj\ events are based on the fact
that the hadronization of four quarks gives rise to a high multiplicity of
particles in the final state, and to a large visible energy. Other types of
events with hadrons in the final state can have similar characteristics, mainly
the \jjlv\ channel and the reactions  $e^+e^-\rightarrow q\overline{q}\gamma$
with $M_{q\overline{q}}>$120~GeV\footnote
     {Events with a lower invariant mass of the $q\overline{q}$ system
      correspond to radiative return to the $Z^0$ peak and can be
      easily rejected either because the photon radiated from
      the initial state is detected or because the missing momentum
      associated with it is very high.
     }
and $e^+e^-\rightarrow \ZZ \rightarrow q\overline{q}q'\overline{q}'$.
The first two reactions can mimic 4 jets when gluon radiation has occurred.

The following variables were typically used to select \jjjj\ events:
\begin{itemize}
\item[-] A large multiplicity of particles in the detector ($N_{charged} > 25$,
  or $N_{charged}+N_{neutr}>25-40$). This cut helps to reject \jjlv\  and QCD
  background, where the observed hadrons originate from a smaller number of
  initial quarks;
\item[-] Small thrust and/or large sphericity ($T<0.9-0.97$ or $S>0.1$). The QCD
  background generally consists of two back-to-back jets ($T \rightarrow 1$,
  $S\rightarrow 0$), while the \WW\ hadronic decays are more isotropic. However,
  the discriminating power of these variables  becomes smaller as $\sqrt{s}$
  increases;
\item[-] Large total visible energy (charged + neutral);
\item[-] Small missing energy ($E_{miss}<40-50$ GeV). Large missing energy and
  momentum are associated with the neutrino in leptonic \W\ decays, and with an
  undetected high energy photon in $q\overline{q}\gamma$ events.
\end{itemize}

\noindent Events from the \jjlv\ channel can also be suppressed by requiring
that no energetic isolated track or high energy identified lepton be present.
The efficiency of the selection criteria at this stage is typically around 80\%
and the purity of the surviving sample is around 60\%.

After the cuts described above, the final state particles are grouped into four
jets. For this purpose, several clustering algorithms have been tried, which
fall into two categories, namely transverse momentum-based clustering, such as
LUCLUS \cite{ref:TGC-luclu}, PUJET4 \cite{ref:TGC-pujet}, DURHAM
\cite{ref:TGC-durha} or GENEVA \cite{ref:TGC-genev}, and scaled invariant mass
squared clustering, such as JADE \cite{ref:TGC-jade}. Comparative studies have
shown that differences are contained to within about 3\%, with the algorithms
based on transverse momentum reproducing the initial parton directions somewhat
better, leading to  better jet definition and hence better resolution in
invariant mass.

Further rejection of background can be achieved by application of the following
additional cuts to the reconstructed jets, leading to a \jjjj\ purity of $\sim
80\%$:
\newline
- Minimum number of particles inside each jet (2 to 5);
\newline
- Minimum angular separation between jets ($20^o$);
\newline
- Minimum energy of the 2 least energetic jets (15-20 GeV);
\newline
- Minimum jet-jet invariant mass ( $Y_{cut}=0.002\sqrt{s}$).

\subsection{Kinematic fitting}
\label{sec:TGC-jjjj-kinfit}

The kinematic fit is used as a tool to improve the resolution on measured
quantities by imposing external constraints. For the \jjjj\ channel, the
measured quantities are the energies and polar and azimuthal angles of the four
reconstructed jets (and, for massive jets, their invariant masses).
The external constraints which can be imposed are as follows:
\newline
1) energy-momentum conservation (4C),
\newline
2) as 1), plus equality of the two reconstructed invariant jet-jet masses
     (5C), or
\newline
3) as 1), plus equality of the two reconstructed invariant jet-jet masses
      with $\mathrm{M_W}$ (6C).

The importance and the limits of kinematic fitting have been discussed in
previous sections of this report, and technical details can be found  in
references \cite{ref:TGC-pujet,ref:TGC-paira1}. As in their application to TGC
determination in the \jjlv\ channel (see section~\ref{sec:TGC-jjlv}) the second
and third constraints can be imposed without fear of introducing biases, as they
would if applied to \W\ mass determination. Nonetheless, a comparison of results
using different constraints has shown that there is negligible gain in going
from the 4C fit to the 5C or 6C fits, and the results given below have used a 4C
fit, followed by cuts on the invariant masses of the jets assigned to each \W.
In order to choose the best pairing of the four jets into two \Ws, kinematic
fits are made to each of the three pairings, and that with the  largest $\chi^2$
probability is taken as the correct combination.

\subsection{Results in efficiency and resolution}
\label{sec:TGC-jjjj-effres}

After additional cuts on the fitted quantities to reduce background
contamination, the efficiencies, purities and remaining background content of
selected event samples generated with different detector simulations and at two
centre-of-mass energies are as shown in table~\ref{TGC-jjjj-selec}.

\begin{table}[htb]
\centerline{\begin{tabular}{|c|c|c|} \hline
                ~~ & $\surd s=175$~GeV &$\surd s=192$~GeV
\\ \hline\hline
 Efficiency (\%)   &    54 - 59         &      52
\\ \hline
 Purity (\%)       &       92           &      90
\\ \hline
 Background(\%)    &                    &
\\
$e^+e^-\rightarrow q\bar{q}\gamma$  
                   &  8                 &  8
\\
$e^+e^-\rightarrow \ZZ \rightarrow q\bar{q}q'\bar{q}'$ 
                   & 0                  &   2
\\
$e^+e^-\rightarrow \WW \rightarrow \jjlv$
                   & 0                  &  0
\\ \hline
\end{tabular}}
\caption{Efficiency and purity of samples of events selected with the cuts
         described in the text at two centre-of-mass energies.}
\label{TGC-jjjj-selec}
\end{table}

The resolutions in the radial and azimuthal jet angles $\theta_{jet},
\phi_{jet}$ and the resolution $\Delta E_{jet}/E_{jet}^{true}$  in the jet
energy can be estimated by comparing each reconstructed jet with the closest
generated quark direction. They show little dependence on the centre-of-mass
energy and on the different detector simulations. Results for the resolutions in
jet energy and in the reconstructed \W\ production angle \ctw\ for simulations
at 192 GeV are shown in table~\ref{tab:TGC-resol2}. It can be seen that the
kinematic fit substantially improves the resolutions  in the variables shown (by
a factor of between 30 and 50\%). However, it has less effect on the jet angular
resolutions, which are typically between 20 and 30 mrad for about 2/3 of the
selected events.

\begin{table}[htb]
\centerline
{\begin{tabular}{|c|c|c|} \hline
~~                   & Before kinematic fit & After kinematic fit     \\ \hline
 $\Delta E_{jet}\ /\ E_{jet}^{true}$ & 0.12 & 0.08                    \\  \hline
 $\Delta\ctw $ (mrad)                & 50.0 & 40.0                    \\ \hline
\end{tabular}}
\caption{Resolutions in jet energy and \W\ production angle
         before and after the kinematic fit at 192 GeV.}
\label{tab:TGC-resol2}
\end{table}

\subsection{W charge assignment}
\label{sec:TGC-jjjj-chtag}

The ambiguities in angular data arising from the inability to distinguish quark
from antiquark jets in \W\ decay have been listed in table~\ref{tab:anginf},
and the generator level studies simulating the \jjjj\ channel described in
section~\ref{sec:TGC-genresults} were made using distributions folded in both
production and decay angles. In order to attempt to resolve the ambiguity on the
production angle, a jet charge can be defined by weighting the charge $Q_i$ of
each particle assigned to the jet with some function of its momentum, 
\bq
   Q_{jet}=\frac{\sum_{i\in jet} Q_i\cdot F(p_i)}
                {\sum_{i\in jet} F(p_i)} \eqpt
\eq

\noindent Different weight functions have been tried, based on transverse
momentum, on rapidity, and on a power of the momentum
\cite{ref:TGC-rapid,ref:TGC-powed,ref:TGC-poweo,ref:TGC-powea}. It appears very
difficult to identify the charges of each individual jet. But, since the
separation between the charges of the two \Ws is equal to 2, one can more
easily distinguish the \Wm\  from the \Wp\ and therefore determine the
production angle in the lab frame. The charges of the two jets assigned to one
$W$ on the basis of the kinematic fit are therefore added together to evaluate
the charge of the \W. The fraction of selected events where the charge is
correctly assigned is found to be 80\%. No significant difference among the
various weight functions was found. The \W\ charge identification implies a gain
in sensitivity in TGC determinations.

\subsection{TGC determination from \jjjj\ events}
\label{sec:TGC-jjjj-tgc}

The precision obtained in TGC determination after application of the procedures
outlined above has been estimated using a fully simulated sample of 2292 events
at $\sqrt{s}=192$ GeV, corresponding to an integrated luminosity of 500
${\mathrm pb}^{-1}$.  Two types of fit were performed to the observed angular
distributions, namely,  a $\chi^2$ fit to the production angle \ctw\ only, and 
an unbinned maximum likelihood fit (as described in section~\ref{sec:TGC-ML}) to
the production angle and folded \W\ decay angles (\ctja, \phija)\fold, (\ctjb,
\phijb)\fold. In both fits, the ambiguity in production angle was resolved using
the jet charge assignment. Precisions obtained in fits to the TGC parameters
\aWphi\ and \aW\ are shown in table~\ref{tab:TGC-jjjj-tgc}.

\begin{table}[htb]
\centerline
{\begin{tabular}{|c|c|c|} \hline
 Fitted parameter         & \mco{2}{|c|}{Fitting method}         \\ \cline{2-3}
    value                 &  $  \chi^2 $ method  & Maximum likelihood method
\\ \hline\hline
   \aWphi                 &    0.04        &     0.02           \\ \hline
   \aW                    &    0.07        &     0.04           \\ \hline
\end{tabular}}
\caption{1 s.d. errors in fitted values of parameters \aWphi\ and \aW\ 
         to a sample of 2292 fully simulated
         \jjjj\ events at 192 GeV. $\chi^2$ fits were made to the production
         angle only and maximum likelihood fits to production and folded decay
         angles.}
\label{tab:TGC-jjjj-tgc}
\end{table}

The data used in the fits were generated according to the Standard Model using
PYTHIA, with $\Gamma_W =2.1$ GeV and with ISR.  The theoretical
expectations~\cite{ref:TGC-Gounaris,Bietal93} were calculated with
$\Gamma_W = 0$ and without ISR. In these conditions, a biased result is
expected, as indicated from the results shown in fig~\ref{fig:TGC-generrs-1}
(section~\ref{sec:TGC-generrs}). In addition, experimental biases due to the
selection and reconstruction procedures are to be expected, as found in the
analysis of the \jjlv\ channel and discussed in section~\ref{sec:TGC-jjlv}. In
the case of the \jjjj\ channel, the angular distributions are quite severely
distorted by bad association of pairs of jets to the parent \W\ and by wrong \W\
charge assignment, and the resulting biases can easily simulate an anomalous
TGC. The results shown for the fit to the production angle only include the
effect of the application of a procedure to correct for the bias. Although based
at present on the use of very limited Monte Carlo statistics, the fitted central
values are found to remain within  $\sim 1\sigma$ of the SM values after
application of the correction. However, a full study of the biases in this
channel and the development of methods to correct for them in fits using several
angular variables have yet to be carried out.

\section{Analysis of the \lvlv\ final state}
\label{sec:TGC-lvlv}

The analysis of the channel  in which both \Ws decay leptonically presents
particular problems. It is the least statistically significant final state (with
branching ratio $\sim 11\%$ for $l = e, \mu, \tau$), 
the missing neutrino momenta imply that the \W\
direction cannot be determined unambiguously, and, if one or both of the \Ws\
decay into $ \tau \overline{\nu}_{\tau}$ or its charge conjugate, the presence
of the extra neutrino from $\tau$ decay makes it impossible to reconstruct the
event, reducing the useful branching ratio of such events to around 5\%. On the
other hand, the knowledge of the \W\ charge and  the small reconstruction errors
of leptons favour this channel in contrast to the 4 jet channel. In this section
the usefulness of the purely leptonic decay channel for TGC determination is
discussed in the light of these considerations.

\subsection{Selection of \lvlv\ events}
\label{sec:TGC-lvlv-sel}

The \lvlv\ event signature is very simple: two leptons and large  missing
energy. This makes the channel easy to identify, but the background
contributions, chiefly from $Z\gamma$, are high. Also, \lvlv\ events
with one or two leptonic $\tau$ decays ($BR_{\tau \rightarrow e,\mu} \approx
35\%$) constitute a possible background of about 1.8\% of the total number of
\WW\ events. The typical selection criteria used for \lvlv\ events aim at
reducing these backgrounds by requiring large missing transverse momentum and,
for equal flavours, that the mass of the lepton-lepton system should not be
close to $M_Z$. In addition it is also necessary that physical solutions to the
reconstructed neutrino directions exist -- this turns out to give the strongest
background rejection.

For purely leptonic \WW\ events the momenta of the 2 neutrinos are unknown.
However, in the absence of ISR and for fixed $M_W$, we have six constraints
allowing the momenta of the neutrinos to be reconstructed~\cite{Haetal87}.
The quadratic nature of these constraints results in a two-fold ambiguity,
corresponding to flipping both neutrinos with respect to the lepton-antilepton
plane, hence only affecting \ctw , \phia , and \phib , while leaving \cta\ and
\ctb\ unchanged. 

The  efficiencies and purities after each stage in event selection and
reconstruction are shown in table~\ref{tab:TGC-lvlv-eff} for fully simulated
events generated with ISR and finite width. It can be seen that the required
existence of solutions to the six constraints provides a very strong background
suppression. However, it is important to note that the solution of these
equations requires the use of all the kinematic information available in the
event, leaving no possibility, for instance, of  accounting for ISR or finite
$W$ width effects. Thus, with these effects included, no solution is found at
generator level for about 20\% of the events.

\begin{table}
\begin{center}
\begin{tabular}{|c||c|c|c|c|c|c|c|c|}                                    \hline\hline
Cut & \multicolumn{2}{c}{\tabcolsep=0mm \begin{tabular}{c} $N_{leptons} = 2$\\$leptons\in
\{ e,\mu\}$\end{tabular}} & \multicolumn{2}{|c|}{$P_T^{miss} > 1.5$ GeV} & \multicolumn{2}{c|}{$l=l'$: \tabcolsep = 0mm
\begin{tabular}{c} $M_{l\overline{l'}} < M_Z-\Gamma_Z$,              \\
$M_{l\overline{l'}} > M_Z+\Gamma_Z$
\end{tabular}} & \multicolumn{2}{c|}{\tabcolsep=0mm \begin{tabular}{c} Recon-\\ struction \end{tabular}}
                                                                \\ \hline
$\sqrt{s}$ (GeV)&\makebox[12mm]{175}&190&\makebox[12mm]{175}&190&\makebox[20mm]{175}&190&175&190 \\ \hline\hline
Efficiency (\%) & $82.7$ &$80.1$&$82.6$ &$79.9$&$79.2$ &$77.8$& $58.7$ &$58.5$\\ \hline
Purity (\%) &$9.70$ &$9.35$&$25.2$ &$24.5$&$31.9$ &$30.9$&$88.4$ &$80.8$\\ \hline \hline
Background (\%) &&&&&&&&\\
$Z\gamma$ &$86.0$ &$87.0$&$63.6$ &$65.9$&$53.4$ &$57.7$&$6.33$ &$14.1$\\
$ZZ$ & $0.4$ &$0.5$ & $1.22$ &$1.32$& $1.60$ &$0.89$& $0.32$ &$0.28$\\
$WW \rightarrow l\nu \tau\nu $
& $3.83$ &$3.16$&$9.99$ &$8.28$&$13.1$ &$10.5$&$4.93$ &$4.75$\\ \hline
\end{tabular}
\end{center}
\caption{Efficiencies and purities in selection of \lvlv\ events at 175 and 190 GeV.}
\label{tab:TGC-lvlv-eff}
\end{table}

\subsection{TGC measurements from \lvlv\ events}
\label{sec:TGC-lvlv-tgc}

The precision with which TGCs can be determined from \lvlv\ events has been
investigated using samples of fully simulated events, generated at 175 and 190
GeV with finite \W\ width and ISR, corresponding to an integrated luminosity of 
$500 {\mathrm pb}^{-1}$. The parameter \aWphi\ was fitted to the cross-section
(\ref{eq:dsig5}) ({\it i.e.} in the narrow width, no ISR approximation),  using
the  extended maximum likelihood method described in  section~\ref{sec:TGC-ML}
and folding over the 2 ambiguous solutions.  The 1 s.d. precision in \aWphi\ 
was  found to be $\pm 0.15$ at 175 GeV, with estimated biases\footnote
    {Due to limited statistics the statistical errors on the results from
     which the biases are estimated are of the same order as the biases
     themselves, but since the samples are correlated the statistical error of
     the biases are expected to be smaller.
    }
of $-0.04$ from the neglect of \W\ width and ISR, $-0.05$ from the same sources
plus the effects of reconstruction, and a combined bias of $-0.07$ when, in
addition, background events are added. At 190 GeV the precision in
\aWphi\ was found to be $\pm 0.09$ and the same biases $-0.04$, $-0.13$ and
$-0.21$, respectively.

Taking into account the small number of \lvlv\ events ($\approx 220$) in the
sample,  it is clear that the sensitivity to TGCs  is highly preserved in this
channel, despite the two-fold ambiguity. However, it is clear  that, due to the
very limited statistics, they will have to be used in combination with events
from other decay channels.


%
%
%
%

%
%
%
\section{Other Anomalous Couplings and Other Channels}
\subsection{Constraints on  $WW\gamma$ Coupling from
                   $e^{+}e^{-}\rightarrow
\overline{\nu}\nu\gamma$
}\label{TGCsinglephot}
%
%
The $W^{+}W^{-}$  production process
suffers from the drawback
that both $W^{+}W^{-} \gamma$ and $W^{+}W^{-} Z$ couplings contribute
and it is not easy to disentangle the various contributions.
However, there does exist a process, namely
$e^{+}e^{-}\rightarrow \overline{\nu}\nu\gamma$, which allows us to
concentrate solely on the $W^{+}W^{-} \gamma$ vertex.
The matrix-element
for $\overline{\nu}\nu\gamma$ production in terms of the
$WW\gamma$ TGCs $\kappa_\gamma$, $\lambda_\gamma$ in (\ref{LeffWWV})
has been calculated in Ref.\cite{aks}.
In the numerical analysis we set acceptance cuts of
a minimum photon angle  of $20^\circ$ and transverse momentum of 10 GeV.
To increase the sensitivity to
anomalous couplings the background from the $Z$
exchange graphs, $e^+e^- \rightarrow Z\gamma \rightarrow
\overline{\nu}\nu\gamma$, is eliminated by requiring the energy
of the photon to be  at least
$5 \Gamma_Z$ away from the energy corresponding to the
$Z\gamma$ final state which essentially amounts to an upper limit on
photon energy of 53 GeV.
With these cuts  the cross-section\footnote{We have not included effects
of initial state radiation.}
for the
standard model is 1 pb at $\sqrt{s}=175$ GeV, which still leads to
an appreciable number of events at design luminosity of 500 pb$^{-1}$.
Cross-sections for non-standard TGC, within these cuts,
differ by less than 0.1 pb
for $|\Delta\kappa|$ and/or $|\lambda|$ $<2$, so not
much sensitivity is expected from the total cross-sections alone.
Looking, however,  at the deviations of the differential cross-sections
from the standard model predictions one can set some limits on the parameters.
We consider
a $\chi^2$ fit to SM data,
adding in quadrature a relative systematic error of $\varepsilon=0.02$.
In Fig.~\ref{fig:single_phot}
we show the contour plots for the $\chi^2$ distributions as
functions of $\Delta\kappa_\gamma$ and $\lambda_\gamma$
as extracted from a) the energy,
b) the transverse momentum distributions of the photon\footnote{The angular
distributions are less sensitive to the anomalous couplings.}. We used 
 equal size binning with 17 and 16 bins for the two cases respectively.
\begin{figure}[htb]  
\vskip 3.9in\relax\noindent\hskip -1.6in\relax{\includegraphics{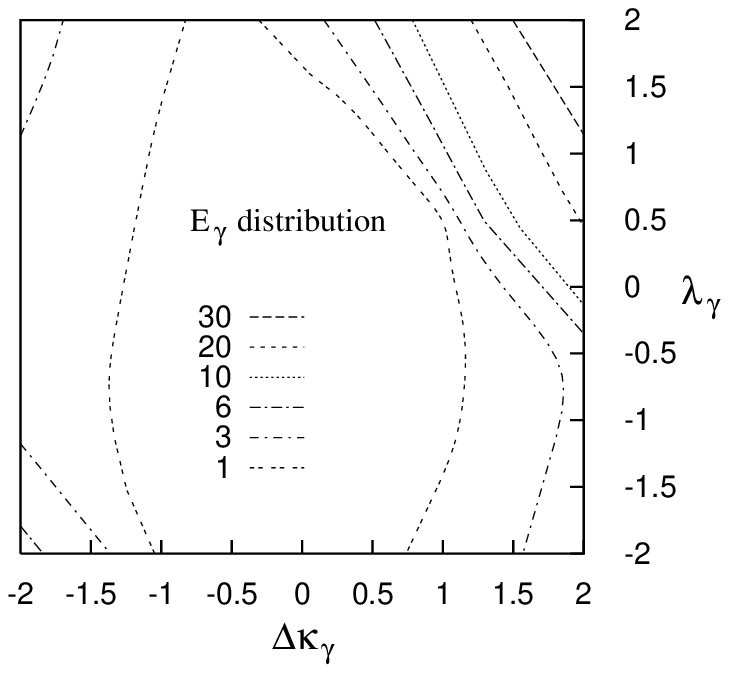}}
          \relax\noindent\hskip 3.0in\relax{\includegraphics{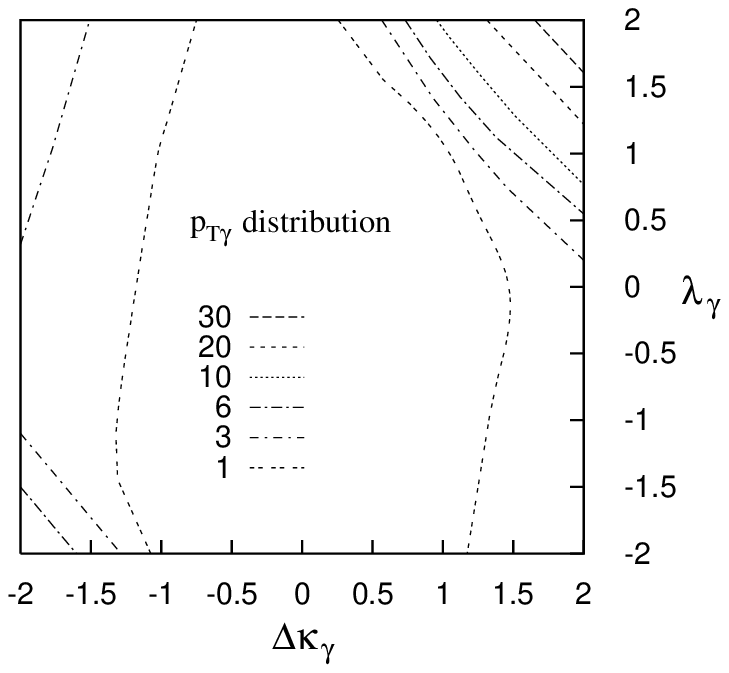}}
\vspace{-22ex}
\caption{\label{fig:single_phot}
$\chi^2$ contours in the $\Delta \kappa_\gamma$--$\lambda_\gamma$
plane as derived from ({\em a}) energy and ({\em b})
transverse momentum distributions, for $\protect \sqrt{s}=$ 175 GeV and
integrated
luminosity of 500 pb$^{-1}$.}
\end{figure}
This process is, in general,
more sensitive to $\Delta\kappa_\gamma$ than to $\lambda_\gamma$. It is thus
somewhat complementary to Tevatron bounds which are derived from $W\gamma$
production. While quantitative improvements
on the constraints may be made by considering two-variable distributions or
by adopting maximum likelihood methods, these would still not be competitive
with those deduced from $W^+ W^-$ production. However, the
$\overline{\nu}\nu\gamma$ channel isolates the $WW\gamma$ couplings and
probes them in a different $q^2$ region. Therefore it 
complements the information obtained from $W$-pair production.

%
%
%
\subsection{Anomalous $Z \gamma$ couplings
\protect\footnote{
We are grateful to Ulrich Baur
for making his $Z\gamma$ event generator available to us.}}\label{TGCneutral}
While the measurement of $WW \gamma$ and $WWZ$
couplings at LEP2 has deservedly received considerable attention,
it is also important to search for couplings between the
neutral gauge bosons\cite{zgrb,zgch}.
For the trilinear $ZV\gamma$ vertex ($V=Z,\gamma$)
the most general vertex function
invariant under Lorentz and electromagnetic gauge transformations
can be described in terms of four independent~\footnote{
As for the WWV TGCs of Eq.~(\ref{LeffWWV}),
constraints on the different $h^V_i$ can be obtained from restriction
to the lowest terms of a gauge-invariant expansion in $1/\Lambda_{NP}$.}
 dimensionless form
factors\cite{zgbaur},
denoted by $h ^V _{i}$, i=1,2,3,4.
The parts of the vertex function proportional to
$h ^V _1$ and $h ^V _2$ are CP--violating while those involving the other
pair of form factors are CP--conserving.  As is well known, all $Z \gamma$ form
factors are zero at the tree level in the SM. 
At the one--loop
level, $h ^V _1$ and $h ^V _2$ are zero while the CP--conserving form factors
are nonzero but too small
to lead to observable effects at any present
or planned experiment.  Observation of $Z \gamma$ couplings
would, therefore, signal physics beyond the SM.
 
We have carried out a generator--level study to estimate  the
sensitivity at LEP2 to anomalous $Z \gamma$ couplings.
Following reference \cite{zgbaur}, the form factors
were parameterized as
$h ^V _i  ~ = ~
h _{i0} ^V / (1 +  (P ^2 /\Lambda _V ^2 )) ^{n _i}$
where P is the effective center--of--mass energy, and h$^V _{i0}$,
$\Lambda _V$, and $n _i$ are parameters.
For comparison with present limits on $Z \gamma$ couplings, we chose
$n _1 = n _3 \equiv n _6= 3.0$ and $n _2 = n _4 \equiv n _8=4.0$.
Two channels,
$e ^+  e ^- \rightarrow \mu ^+ \mu ^- \gamma$
($\mu \mu \gamma$) and
$e ^+  e ^- \rightarrow \nu \bar{\nu} \gamma$ ($1\gamma$),
have been studied in detail.  At LEP2 energies it turns out
that the $1 \gamma$ channel is much more sensitive to anomalous $Z \gamma$
couplings than the $\mu \mu \gamma$ channel.  This is due mainly to
anomalous couplings being dominated by the case where the detected photon
recoils against a resonant Z and that
$\Gamma (Z\rightarrow \nu \bar{\nu}) \cong 6
\Gamma (Z\rightarrow \mu ^+ \mu ^-)$.
Below we thus report on the sensitivity expected from the $1 \gamma$ channel
alone.
 
Experimentally, anomalous couplings in the $1 \gamma$ channel would populate
the same energy range as ``radiative return'' to the $Z$ pole through
initial--state radiation (ISR), namely, the reflection of the $Z$ pole
centered on
$E _0 \equiv (s - m _Z ^2)/(2 \sqrt{s})$.
Unlike photons from ISR, however, photons from anomalous couplings are
distributed almost
uniformly in solid angle.  In our
sensitivity analysis, which employed event counting rather than fits to
distributions,
we therefore required (a) the photon energy to
be within 10 GeV of $E _0$ and
(b) $\mid \! \! cos \theta _{\gamma} \! \! \mid < 0.8$ in order to
maintain good acceptance for anomalous couplings while suppressing the
background from ISR.  For
$1 \gamma$ events passing these cuts, a combined trigger and reconstruction
efficiency of 90\% was assumed.
 
Figure \ref{zgfig1}(a) shows the $ZZ\gamma$ couplings that would be excluded
at the 95\%
C.L. for $\sqrt{s}$=175 GeV and 500 pb$^{-1}$
assuming that the SM expectation is observed\protect\footnote{
The effects of QED corrections on LEP2 sensitivities are not
reflected in Fig. \ref{zgfig1}.  These
corrections reduce the sensitivity to anomalous $Z \gamma$ couplings but by
less than 10\%.}\label{TGCneutralho}.  The limits are shown for two
different values of $\Lambda _Z$ to provide some indication of how much they
depend on the particular choice of parameter values.  Limits on these
couplings have been published recently by L3\cite{zgl3}, CDF\cite{zgcdf},
and D0\cite{zgd0}; the L3 and CDF limits are also plotted.
It is evident that the expected
sensitivity of LEP2 is comparable to the combined sensitivity of searches
by LEP1 and Tevatron experiments.
Figure \ref{zgfig1}(b) shows the corresponding estimated sensitivity to
anomalous
$Z \gamma \gamma$ couplings.  As can be seen from comparison with the
limits from CDF\cite{zgcdf} (competitive limits have also been published
by D0\cite{zgd0}), LEP2 is expected to extend considerably the sensitivity
to $Z \gamma \gamma$ couplings.
\begin{figure}[htb]
\vspace{0.1cm}
\epsfig{figure=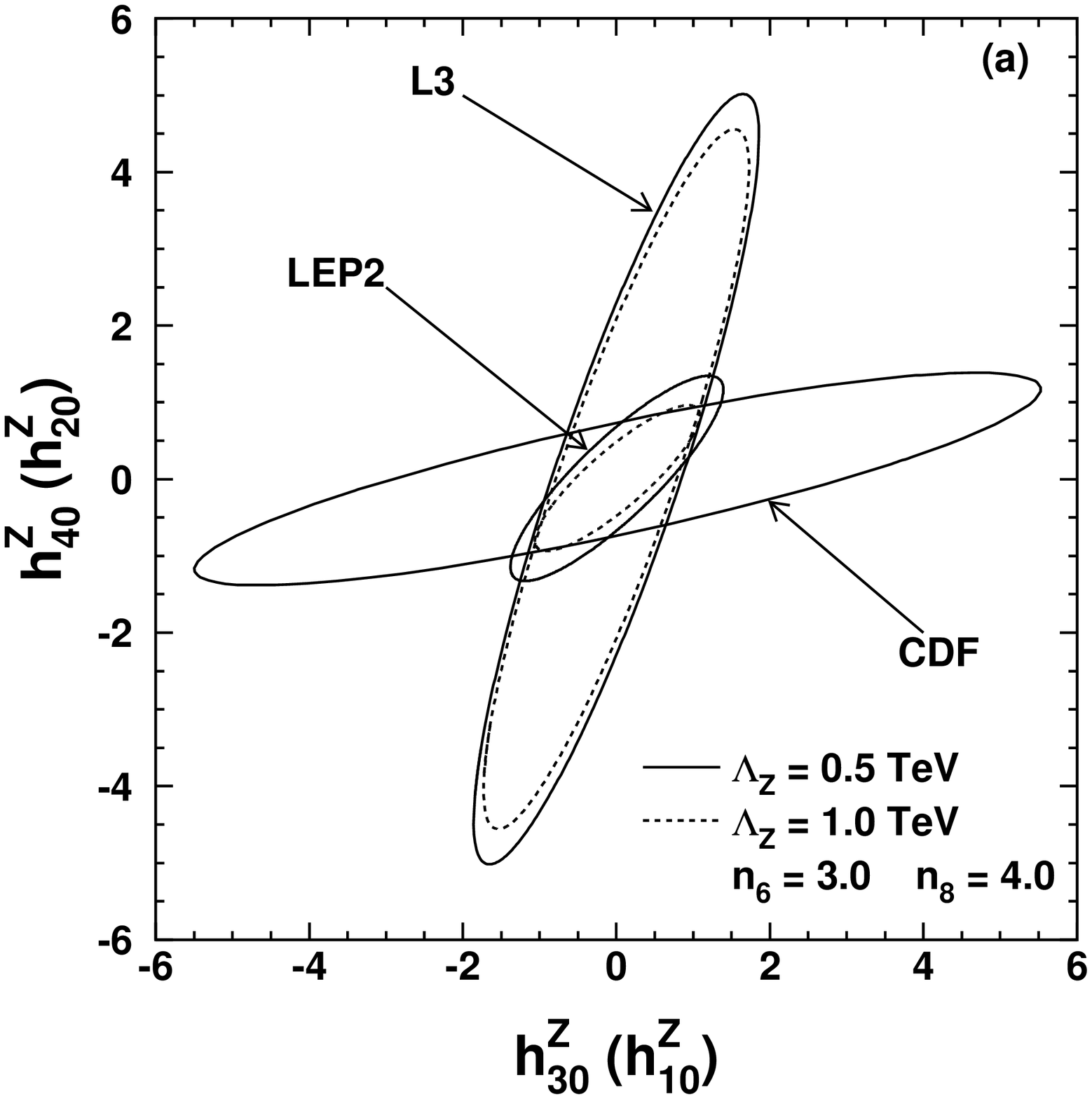,width=8cm,height=8cm,angle=0}
\epsfig{figure=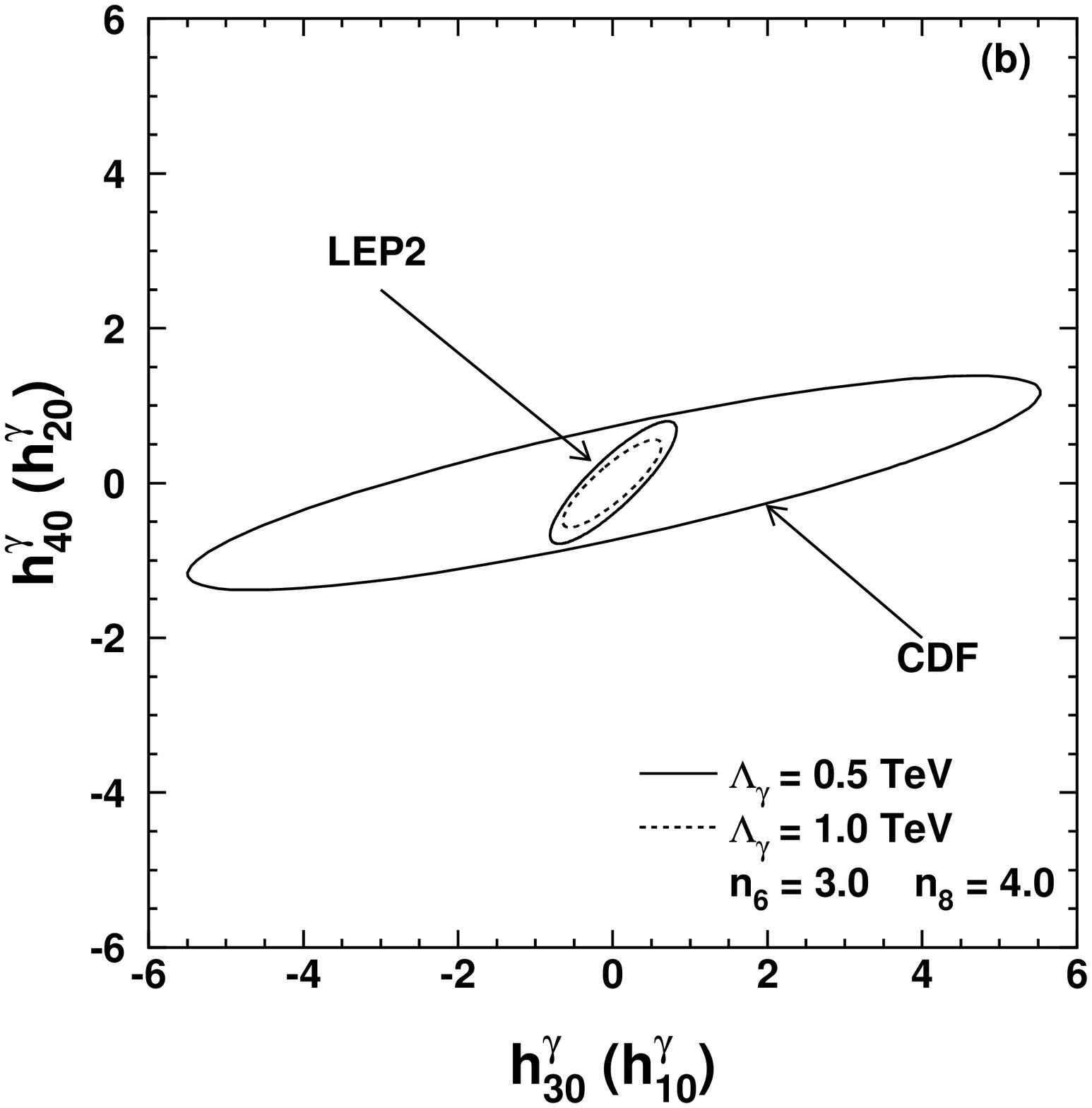,width=8cm,height=8cm,angle=0}
\caption{\label{zgfig1}
Estimated LEP2 sensitivity limits (95\% C.L.) to anomalous $Z \gamma$
couplings and 95\% C.L. limits from present experiments.  The LEP2 estimate
is for $\protect\sqrt{s}$=175 GeV and 500 pb$^{-1}$. See text for 
explanation of the parameters.}
\end{figure}
 
The sensitivity to anomalous $Z \gamma$ couplings increases rapidly with
center--of--mass energy,
the effect being more pronounced for
sensitivity to $h _2 ^V$ and $h _4 ^V$, which correspond to dimension--8
operators compared to dimension--6 operators in the case of $h _1 ^V$ and
$h _3 ^V$.
For example, sensitivity to $h^\gamma_{40}$
($h^\gamma_{20}$) is improved by about 25\% at 192 GeV, even with a smaller
integrated luminosity of 300 pb$^{-1}$.
Although backgrounds are expected
to be more severe, analysis of the event sample consisting of hadrons and an
isolated,
energetic photon may provide another way of
significantly increasing sensitivity
to $Z \gamma$ couplings.
%
%
%
\subsection{Constraints on gauge boson interactions from
$e^-e^+ \to q \bar q ~, l \bar l$
}\label{TGCqqll}
The description of  NP for $e^-e^+ \to q \bar q ~, l \bar l$ in terms of
dimension 6, purely bosonic, $SU(2)\times U(1)$ gauge invariant operators
necessitates the consideration of the interactions
\ba
\label{nonblind}
\L_{NP}  & = &
- ~ \frac{f_{DW} g^2}{2 \Lambda_{NP}^2}
(D_{\mu}\vec{\hat{W}}_{\nu \rho})\cdot
(D^{\mu} \vec{\hat{W}}^{\nu \rho})
~ - ~\frac{f_{DB} {g\prime}^2}{2 \Lambda_{NP}^2}
(\partial_{\mu}B_{\nu \rho})
(\partial^\mu B^{\nu \rho})
\nonumber \\
\null & \null &
- ~ \frac{f_{BW} g g\prime }{4 \Lambda_{NP}^2}
~\Phi^\dagger
B_{\mu \nu} \vec \tau \cdot \vec{\hat{W}}^{\mu \nu}
 \Phi ~ + ~
\frac{f_{\Phi 1}}{\Lambda_{NP}^2}
 (D_\mu \Phi)^\dagger \Phi \Phi^\dagger
(D^\mu \Phi) \ ,  \
\ea
in addition to the ones mentioned in Section
\ref{TGClr}. Such interactions
affect the gauge boson propagator at the tree level and are
thus rather strongly constrained by LEP1 measurements.
Nevertheless LEP2 can significantly improve these constraints,
particularly for the first two terms in
(\ref{nonblind}) which give a $q^4$ contribution to the gauge
boson propagator \cite{HagiwaraLEP1}.
It has been remarked in \cite{RenardZpeak}, that if the
physical quantities measurable in $e^-e^+ \to q \bar q ~,~
l \bar l$ at LEP2 are expressed in terms of Z-peak observables,
then the  aforementioned $q^4$ contribution allows 
the remaining  anomalous dependence of the results to be described
in terms of only the two couplings $f_{DW}$ and $f_{DB}$.
Thus by looking at $\sigma_{\makebox {hadrons}}$ ,
$\sigma_{\mu+\tau}$,
very strong constraints on these couplings should be possible;
(see Fig.~\ref{llqq-verz}).
\begin{figure}[htb]
\begin{center}
\mbox{\epsfig{file=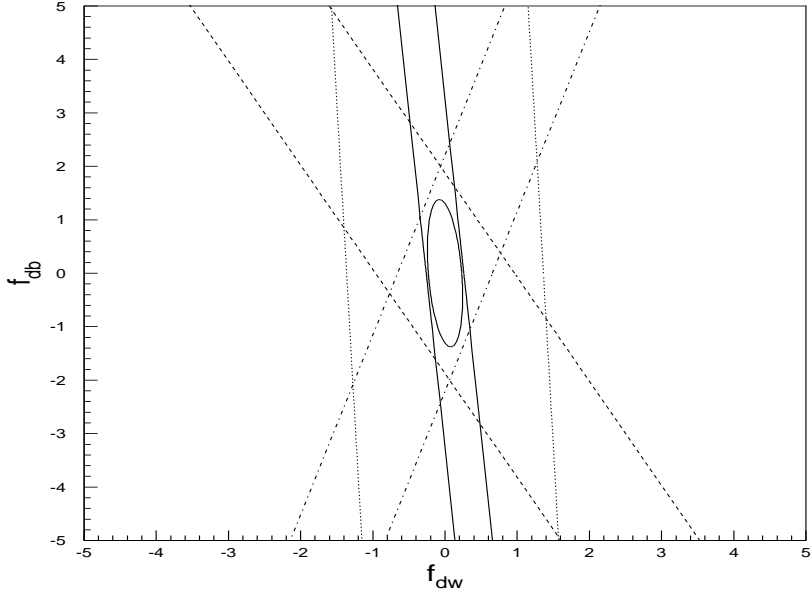,width=12cm,height=8cm}}
%
\caption{\label{llqq-verz}
Sensitivity of LEP2 to $f_{DW}$ and $f_{DB}$ from
$e^-e^+ \to q \bar q ~, l \bar l$ at $\protect \sqrt s =$ 175 GeV
and 500 pb$^{-1}$ (\protect{\it one} experiment).
Constraints from
$\sigma_{hadrons}$ (solid lines);
$\sigma_{\mu +\tau}$ (dashed lines);
$A^{\mu ,\tau}_{FB}$ (dash-dotted lines);
$\sigma_b$ (dotted lines);
global fit (solid ellipse). $\Lambda_{NP}=1$~TeV is assumed.
}
\end{center}
\end{figure}
%
%
\subsection{Higgs anomalous couplings
}\label{TGChghz}
\def\bv#1{{\bf #1}}
\def\scr#1{{\cal #1}}
\def\op#1{{\widehat #1}}
 \def\twi{\widetilde}
\def\tw#1{\tilde{#1}}
\def\ol#1{\overline{#1}}
\def\L{ {\cal L }}
\def\O{ {\cal O }}
\def\Sw{s_W}
\def\Cw{c_W}
\def\swd{s^2_W}
\def\cwd{c^2_W}
\def\ed{e^2}
\def\mwd{M_W^2}
\def\mw{M_W}
\def\mzd{M_Z^2}
\def\mz{M_Z}
\def\lw{\lambda_W}
\def\rd{\sqrt2}
\def\mh{m_H}
\def\dh#1{ {1\over D_H({#1})} }
\def\nh#1{  D_H({#1}) }
Anomalous couplings could also arise for the Higgs interactions with
itself and the gauge bosons. In fact, dynamical considerations indicate that
it is easier to generate anomalous couplings for the Higgs rather than for the
gauge bosons \cite{dyn1,dyn2}. The dimension~6, $SU(2)\times U(1)$ invariant,
CP conserving interaction is
\bq
\label{LeffHiggs}
\L_{NP}  = \frac{1}{v^2}\, (\Phi^\dagger \Phi - \frac{v^2}{2})
(d~ \vec{\hat{W}}^{\mu\nu} \cdot \vec{\hat{W}}_{\mu\nu}
+d_B~ B^{\mu\nu} \ B_{\mu\nu}) +
\frac{4 f_{\Phi2}}{v^2}~
\partial_\mu(\Phi^\dagger \Phi)\partial^\mu(\Phi^\dagger \Phi)
\  . \
\eq
The first two of the above terms generate
Higgs-gauge boson interactions while the last one
induces anomalous  Higgs interactions through a renormalization
of the Higgs field. \par
 
As in section \ref{TGClr},
unitarity can be used to associate to any given value
of each of these anomalous couplings the largest allowed scale $\Lambda_{U}$
where New Physics generates it. For the first two operators
these relations are
%
%
%
\begin{eqnarray}
d  \simeq  \frac{104.5~\left ({\frac{M_W}{\Lambda_{U}}}
\right )^2} {1+6.5 \left ({\frac{M_W}{\Lambda_{U}}}\right )} \
\mbox{ for } d>0 & , &
d  \simeq  -~ \frac{104.5~\left ({\frac{M_W}{\Lambda_{U}}}
\right )^2} {1- 4 \left ({\frac{M_W}{\Lambda_{U}}}\right )} \
\mbox{ for }  d<0 \ ,   \\
d_B  \simeq  \frac{195.8 ~\left ({\frac{M_W}{\Lambda_{U}}}
\right )^2} {1+200 \left  ({\frac{M_W}{\Lambda_{U}}}
\right )^2}\
\mbox{ for } d_B>0 & , &
d_B  \simeq  -~ \frac{195.8 ~\left ({\frac{M_W}{\Lambda_{U}}}
\right )^2} {1 +50 \left  ({\frac{M_W}{\Lambda_{U}}}
\right )^2}
\mbox{ for }  d_B<0 \ . \
\ea
Thus, for $\Lambda_{U}=1$~TeV, the largest allowed values  are
$ d \simeq 0.4 $ or $-1$ and
$ d_B \simeq 0.6 $ or $-1$.
 
The above anomalous Higgs couplings may be studied at LEP2 through
the processes  $e^{-}e^{+}\to ZH,$ provided $m_H<\sqrt{s}-M_Z$,
or via  $e^-e^+ \to \gamma H$ if  $m_H<\sqrt{s}$.
Considering tree level anomalous
contributions and restricting to
cases where only one of the operators above is active
\cite{Hag1, Barger, HagCP, HZHgamma}, we get the results
given in the figures below.
%
%
\begin{figure}[htb]
\mbox{\epsfig{file=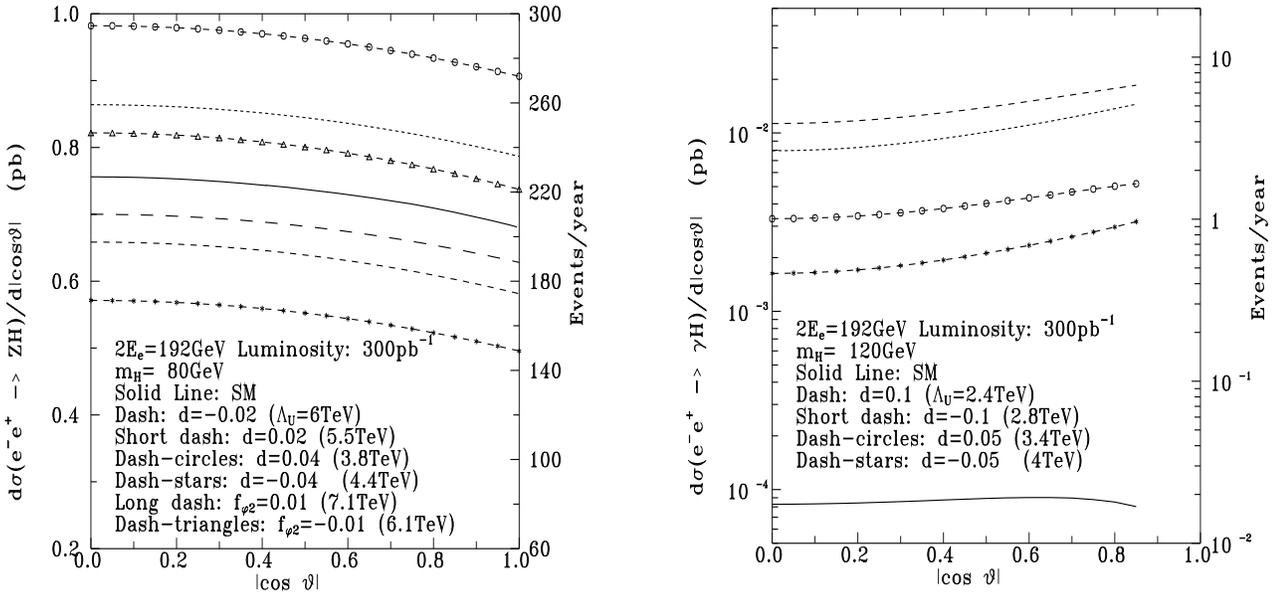,height=8cm,width=17cm}}
\caption{\label{fig:TGC-HZ}
Distribution of Higgs production angle, $d\sigma/d{\rm cos}\theta$,
for (a) $H Z$ production at $m_H=80$~GeV and (b) $H\gamma$ production
at $m_H=120$~GeV.}
\end{figure}
Thus, from Fig.~\ref{fig:TGC-HZ}a, presenting $e^-e^+ \to ZH$,  
we deduce observability limits $|f_{\phi 2}| \simeq 0.01$
and $|d|\simeq 0.015$ ($|d_B|\simeq 0.05$)
corresponding to $\Lambda _{U}\simeq 6-7$~TeV
( $\Lambda _{U}\simeq 5$~TeV) for $m_H\simeq 80$~GeV.\par
 
More striking is the process $e^{-}e^{+}\to \gamma H$ which
is unobservable at LEP2 in the SM
\cite{Barroso,DicusHg},
but may become observable in the presence of  NP interactions
for the Higgs. A sensitivity to $|d|\simeq 0.05$ or
$|d_B|\simeq 0.025$ should be possible from this process for
$m_H\sim 80$~GeV, which means testing  NP scales up to
3 and 7 TeV, respectively\cite{HZHgamma}.
%
%
\section{Conclusions}
\label{sec:conclusions}
Experiments at LEP2 will allow a precise direct measurement of the most
immediate  consequence of the non-Abelian character of the electroweak
bosons, the TGC of the $W$ to the photon and the $Z$. Various channels
can provide information on non-standard interactions in the bosonic sector.
The process $e^-e^+\to f\bar f$ determines oblique parameters
which are complementary to LEP1 results. $Z\gamma$, $HZ$ and $H\gamma$
production allow one to search for non-standard boson couplings in the neutral
sector. $e^-e^+\to \nu\bar\nu\gamma$ is marginally sensitive to the $WW\gamma$
coupling in isolation. However, the most important process is clearly
$e^-e^+\to W^-W^+$ or its generalization, 4~fermion production.
 
Of the various decay channels, the semileptonic modes
$W^-W^+ \to jje\nu,\; jj\mu\nu$ will provide the most precise individual
measurements of TGCs, since high statistics and almost complete information
on the decay distributions are combined. Of particular importance is the
identification of the $W$ charge which is needed to measure the full production
angle distribution $d\sigma/d {\rm cos}\theta$. Also,
the decay angular distributions and their correlations with each other and
with the $W$ production angle are needed to resolve 
the correlations between different TGCs to a maximal extent.
 
A priori, the $jjjj$ final state provides incomplete information on the $W$
charges. However, correct charge assignments at the 80\% level can be obtained
by determining weighted jet charges, providing potentially valuable additional
information in TGC determination. While more limited in statistics, the 
leptonic channel, \lvlv, is particularly clean, and the $jj\tau\nu$ channel
will also be of use in TGC analyses.
 
Using \jjenu\ and \jjmunu\ data alone, measurements of particular TGC
parameters at $\sqrt{s}=192$~GeV appear possible at
generator level with a precision of $\approx\pm 0.02$ for an integrated
luminosity of 500~pb$^{-1}$.
The effects of ISR and finite $W$-width and the application of experimental
selection, acceptance and reconstruction procedures lead to a degradation
estimated at  $\approx 30 - 40\% $ in the precision, and to a systematic
shift which is a factor 3 larger than the statistical error, but our studies
indicate that this bias can be corrected.
For more general TGCs, considerable cancellation
between different parameters is possible, resulting in weaker bounds.
It is for this case that information from the full
five-fold angular distribution of $W^-W^+$ production and decay angles
or its generalization to 4-fermion final states becomes particularly
important.
%
%
%
%

\end{document}